\documentclass{aa}  
\usepackage{graphicx}
\usepackage{txfonts}
\usepackage[breaklinks=true]{hyperref}
\usepackage{natbib}
\bibpunct{(}{)}{;}{a}{}{,} 

\usepackage{placeins}
\usepackage{subcaption}

\newcommand\varTitle{NA-SODINN: A deep learning algorithm for exoplanet image detection based on residual noise regimes}
\newcommand\varADI{ADI}
\newcommand\varSODINN{SODINN}
\newcommand\varLambdaOverD{$\lambda/D$}
\newcommand\varPmaps{PCA-pmap}
\newcommand\varNullHypothesis{$H_{0}$}

\usepackage{xspace}

\newcommand{\eg}{\emph{e.g.}\xspace}

\global\long\def\harmonicMean#1{\bar{#1}}%

\begin{document} 

    \title{\varTitle{}}
    
    \author{C.~Cantero\inst{1,2},
            O.~Absil\inst{2}\fnmsep\thanks{F.R.S.-FNRS Senior Research Associate},
            C.-H.~Dahlqvist\inst{2},
            M.~Van Droogenbroeck\inst{1}
           }

    \institute{Montefiore Institute, Université de Liège, 4000 Liège, Belgium
    \and
    STAR Institute, Université de Liège, Allée du Six Août 19C, 4000 Liège, Belgium
             }
    \date{Received XXX; accepted YYY}
   
    \offprints{C.Cantero,
    \email{ccantero@uliege.be}}

    \abstract
    {Supervised deep learning was recently introduced in high-contrast imaging (HCI) through the SODINN algorithm, a convolutional neural network designed for exoplanet detection in angular differential imaging (ADI) datasets. The benchmarking of HCI algorithms within the Exoplanet Imaging Data Challenge (EIDC) showed that (i) SODINN can produce a high number of false positives in the final detection maps, and (ii) algorithms processing images in a more local manner perform better.}
    {This work aims to improve the SODINN detection performance by introducing new local processing approaches and adapting its learning process accordingly.}
    {We propose NA-SODINN, a new deep learning binary classifier based on a convolutional neural network (CNN) that better captures image noise correlations in ADI-processed frames by identifying noise regimes. The identification of these noise regimes is based on a novel technique, named PCA-pmaps, which allowed us to estimate the distance from the star in the image from which background noise started to dominate over residual speckle noise. NA-SODINN was also fed with local discriminators, such as signal-to-noise ratio (S/N) curves, which complement spatio-temporal feature maps during the model's training.} 
    {Our new approach was tested against its predecessor, as well as two SODINN-based hybrid models and a more standard annular-PCA approach, through local receiving operating characteristics (ROC) analysis of ADI sequences from the VLT/SPHERE and Keck/NIRC-2 instruments. Results show that NA-SODINN enhances SODINN in both sensitivity and specificity, especially in the speckle-dominated noise regime. NA-SODINN is also benchmarked against the complete set of submitted detection algorithms in EIDC, in which we show that its final detection score matches or outperforms the most powerful detection algorithms.}
    {Throughout the supervised machine learning case, this study illustrates and reinforces the importance of adapting the task of detection to the local content of processed images.}

   \keywords{methods: data analysis -- methods: statistical -- techniques: image processing -- techniques: high angular resolution -- planets and satellites: detection}
   
   \titlerunning{NA-SODINN: a deep learning algorithm for exoplanet image detection based on residual noise regimes}
   \authorrunning{C.~Cantero et al.}
   \maketitle
   
    \section{Introduction} 
    The direct imaging of exoplanets through 10-m class ground-based telescopes is now a reality of modern astrophysics \citep[e.g.][]{Bohn_2021,Chauvin_2017_HIP65426,Keppler2018Discovery,Marois_2008_HR8799, Marois_2010_HR8799,Rameau2013Discovery,Wagner_2016_HD131399}. Reaching this milestone is the result of significant advances in the field of high-contrast imaging (HCI). For instance, extreme adaptive optics (AO) is routinely used during observations to correct image degradation caused by the Earth's atmosphere \citep{snik_2018}. In the same way, dedicated HCI instruments, such as Subaru/SCExAO \citep{Lozi2018SCExAO} or VLT/SPHERE \citep{Beuzit2019SPHERE}, make use of state-of-the-art coronagraphs \citep{Soummer2005Lyot,Mawet2009_VORTEX} in order to block out the starlight and mitigate the huge flux ratio (or contrast) between a host star and its companions. Despite all of these approaches, a high-contrast image is still affected by different additive sources of noise, such as photon noise associated with residual stellar light and thermal background emission, speckle noise associated with residual atmospheric turbulence, or residual aberrations that arise in the optical train of the telescope and instrument \citep{Males_2021}. Speckles are scattered starlight blobs in the image that can mimic the expected signal of an exoplanet in both shape and contrast. Therefore, beyond dedicated instrumental developments, powerful image post-processing algorithms are needed to disentangle true companions from speckles. In order to help algorithms achieve this goal, different observing strategies have been proposed, the most popular being angular differential imaging \citep[\varADI{},][]{Marois2006Angular}. An \varADI{} dataset consists of a sequence of high-contrast images acquired in pupil-stabilized mode, where the instrument de-rotator tracks the telescope pupil instead of the field, in such a way that the instrument and optics in the telescope stay aligned while the image rotates in time due to the Earth's rotation. As a result, speckles associated with the telescope and instrument optical train remain mostly fixed in the focal plane while the astrophysical signals rotate around the star as a function of the parallactic angle. 

    Currently, there exist a plethora of post-processing detection algorithms that work on \varADI{} sequences. Most of these algorithms belong to the point spread function (PSF) subtraction family, which aims to model the speckle field and subtract it from each frame in the \varADI{} sequence, derotate the residual images according to the parallactic angles, and finally collapse them into a final frame \citep{Marois2008Confidence}, commonly referred to as a processed frame. Examples of these techniques are the locally optimized combination of images \citep[LOCI,][]{Lafrenierex_2007ANewAlgorithm} and its variants TLOCI \citep{Marois_2014} and MLOCI \citep{wahhaj_2015}, principal component analysis \citep[PCA,][]{Soummer2012Detection,Amara2012Pynpoint}, the low-rank plus sparse decomposition \citep[LLSG,][]{Gomez2016LowRank}, and the non-negative matrix factorization \citep[NMF,][]{Ren2018NonNegative}. PSF subtraction is usually followed by a detection algorithm, which can be either based on an signal-to-noise ratio (S/N) map \citep{Mawet2014Fundamental} or on a more recent technique, such as the standardized trajectory intensity mean \citep[STIM,][]{Pairet2019STIM} or the regime-switching model \citep[RSM,][]{Dahlqvist2020Regime}. Another family of algorithms, based on an inverse problem approach, relies on directly modelling the expected planetary signal and tracking it along the \varADI{} sequence. This is typically done by estimating the contrast of the potential planetary signal via a maximum likelihood estimation. Examples of these methods include ANDROMEDA \citep{Cantalloube2015_andromeda}, the forward model matched filter \citep[FMMF,][]{Ruffio2017_FMMF}, the exoplanet detection based on patch covariances \citep[PACO,][]{Flasseur2018_PACO}, or the temporal reference analysis of planets \citep[TRAP,][]{Samland2021TRAP}. Recently, post-processing approaches based on supervised machine learning have emerged in HCI. \citet{Gomez2018Supervised} introduced the SODIRF and SODINN machine learning models, which are two binary classifiers that use a random forest and a convolutional neural network (CNN), respectively, to distinguish between companion signatures and residual noise in processed frames. \citet{Yip2020Pushing} trained a generative adversarial network with real data from the NICMOS camera (Hubble Space Telescope) to obtain a suitable dataset for training a CNN discriminative model to image companions. More recently, \citet{Gebhard_2022} proposed a modified version of the half-sibling regression \citep{Scholkopf2016_HSR} using a ridge regression with generalized cross-validation. Also, \citet{Flasseur2023deepPACO-arxiv} presented deep PACO, an adaptation of the PACO algorithm to supervised learning through a CNN architecture, which resulted in an improvement on both the detection and characterization of exoplanets.

    A large fraction of these techniques was benchmarked in the context of the Exoplanet Imaging Data Challenge \citep[EIDC,][]{Cantalloube2020Exoplanet}, the first platform designed for a fair and common comparison of post-processing algorithms for exoplanet detection and characterization in HCI. From the whole set of conclusions provided by the first EIDC phase \citep{Cantalloube2020Exoplanet}, we relied on two of them to motivate this paper. First, we observed that detection algorithms that exploit the local behaviour of image noise obtained the highest detection score in the challenge leaderboard. Second, we found that supervised machine learning algorithms produced a relatively high number of false positives, compared with more standard algorithms. Thereby, with the aim of enhancing the supervised machine learning models, for this study, we explored a new stratified noise approach, through which they can better exploit noise statistics in the \varADI{} dataset. This approach relies on the existence of two noise regimes in the processed frame: a speckle-dominated residual noise regime close to the star, and a background-dominated noise regime further away. Our goal is to spatially identify these regimes in the processed frame through the study of their statistical properties, and then adapt the SODINN neural network to work separately in each of them in order to improve its detection performance.  Therefore, in Sect.~\ref{sec:regimes} we first revisit  noise statistics in HCI and present a novel statistical method that allowed us to empirically delimit noise regimes in processed frames. Then, in Sect.~\ref{sec:implementation}, we introduce the noise-adaptive \varSODINN{} (or NA-SODINN) detection algorithm, a neural network architecture optimized to work on noise regimes. Our deep learning method was also fed with local discriminators, such as S/N curves, that contain additional physical-motivated features and help the trained model to better disentangle an exoplanet signature from speckle noise. In Sect.~\ref{sec:evaluation}, NA-\varSODINN{} is evaluated through local ROC analysis using a series of \varADI{} datasets obtained with various instruments. During the evaluation, NA-\varSODINN{} is benchmarked against other state-of-the-art HCI detection algorithms. Section~\ref{sec:conclusions} concludes the paper.

    \section{Noise regimes in processed ADI images} \label{sec:regimes}
    The term local is often used in image processing to describe a process applicable to a smaller portion of the image, such as the neighbourhood of a pixel, in which pixel values exhibit a certain amount of correlation.  In HCI, defining image locality implies a good understanding of the physical information captured in the image. A common way to define locality is linked to the noise distribution along the image field of view. For example, after some pre-processing steps (including background subtraction), a high-contrast image is composed of three independent components: (1) residual starlight under the form of speckles; (2) the signal of possible companions; and (3) the statistical noise associated with all light sources within the field of view, generally dominated by background noise in infrared observations. In these raw images, exoplanets are hidden because starlight speckles and/or background residuals dominate at all angular separations, and act as a noise source for the detection task. According to their origin, starlight speckles can be classified as instrumental speckles \citep{Hinkley2007Temporal,Goebel2016Evolutionay}, which are generally long-lived and therefore referred to as quasi-static speckles, and atmospheric speckles, which have a much shorter lifetime \citep{Males_2021}. Speckle intensity is known to follow a modified Rician probability distribution \citep{Soummer2007Speckle}. Here, the locality of the noise is driven by the distance to the host star \citep{Marois2008Confidence}, which already gives an indication on how local noise will be defined in a processed image. Consequently, a large fraction of post-processing algorithms currently work and process noise on concentric annuli around the star. For example, the annular-PCA algorithm \citep{Absil2013Searching,Gomez2016LowRank} performs PSF subtraction with PCA on concentric annuli.  Nevertheless, more sophisticated local approaches have recently been proposed in the literature. For instance, both the TRAP algorithm \citep{Samland2021TRAP} and the half-sibling regression algorithm \citep{Gebhard_2022} take into account the symmetrical behaviour of speckles around the star when defining pixel predictors for the model. 

    In this section, we aim to introduce an alternative local processing, well-suited for the SODINN framework, as explained later in the paper, based on the spatial stratification of the processed frame into (at least) two noise regimes. For illustrative purposes, we make use, in this section, of two ADI sequences chosen from the set of nine \varADI{} sequences used in the EIDC \citep{Cantalloube2020Exoplanet}; see Table \ref{Table:datasets} for more information about the EIDC datasets. Our two \varADI{} sequences, referred to as \textit{sph2} and \textit{nrc3}, were respectively obtained with the VLT/SPHERE instrument \citep{Beuzit2019SPHERE} and the Keck/NIRC-2 instrument \citep{Serabyn2017TheWW}. They have the advantage of not containing any confirmed or injected companions, which makes them appropriate for algorithm development and tests that rely on the injection of exoplanet signatures in the image.

    \subsection{Spatial noise structure after \varADI{} processing \label{section:residual_noise}}
    
    Performing PSF subtraction on each high-contrast image in an \varADI{} sequence generates a sequence of residual images where speckle noise is significantly reduced and partly whitened \citep{Mawet2014Fundamental}. After derotating these residual images based on their parallactic angle and combining them into a final frame, the remaining speckles are further
    attenuated and whitened. This final frame is commonly referred to as processed frame. Because of the different post-processing steps and the whitening operator that removes correlation effects, the S/N map technique \citep{Mawet2014Fundamental}, the industry standard for exoplanet detection in processed frames, makes use of the central limit theorem to state that residual noise in processed frames follows a Gaussian distribution, an assumption that even today has not been proven experimentally. From practice, it is known that this Gaussian assumption leads to high false positive detection rates \citep{Marois2008Confidence,Mawet2014Fundamental} since residual speckle noise in processed frames is never perfectly Gaussian, and still dominates at small angular separations. \cite{Pairet2019STIM} found experimentally that the tail decay of residual noise close to the star is better explained by a Laplacian distribution than a Gaussian distribution. Later, \cite{Dahlqvist2020Regime} reached the same conclusion by applying a Gaussian and a Laplacian fit to the residuals of PCA-, NMF-, and LLSG-processed frames. These experimental results suggest the presence of two residual noise regimes in the processed frame: a non-Gaussian noise regime close to the star, dominated by residual speckle noise, and a Gaussian regime further away, dominated by background noise.
    
     \begin{figure}
         \centering
         \includegraphics[width=0.98\columnwidth]{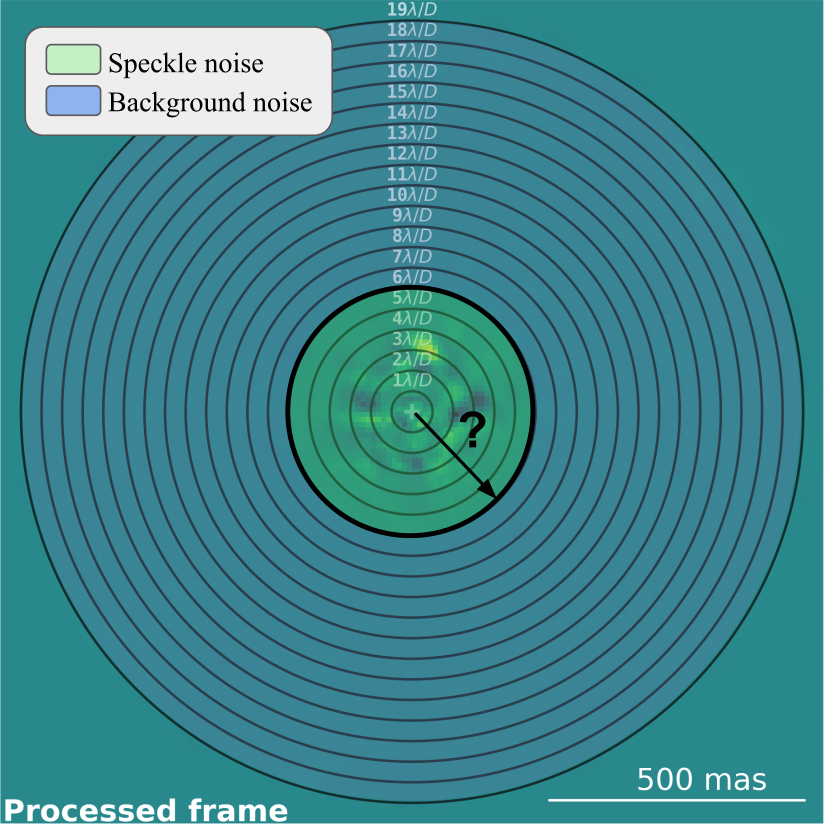}
         \caption{Processed frame from \textit{sph2} dataset with both speckle-dominated and background-dominated residual noise regimes and their annular split (black circle).  The best approximation of this split is what we aim to find in this section.}
         \label{fig:split_assumption}
     \end{figure}
    
    \subsection{Identification of noise regimes}
    
    Based on our understanding of the local statistics of noise in a processed frame, we aim now to spatially delimit both noise regimes in the image. To do so, we try to find the best radial distance approximation from the star where residual speckle noise starts to become negligible compared to background noise (Fig.~\ref{fig:split_assumption}), which is assumed to be uniform over the whole field of view. 
    
     \begin{figure}
         \centering
         \includegraphics[width=\columnwidth]{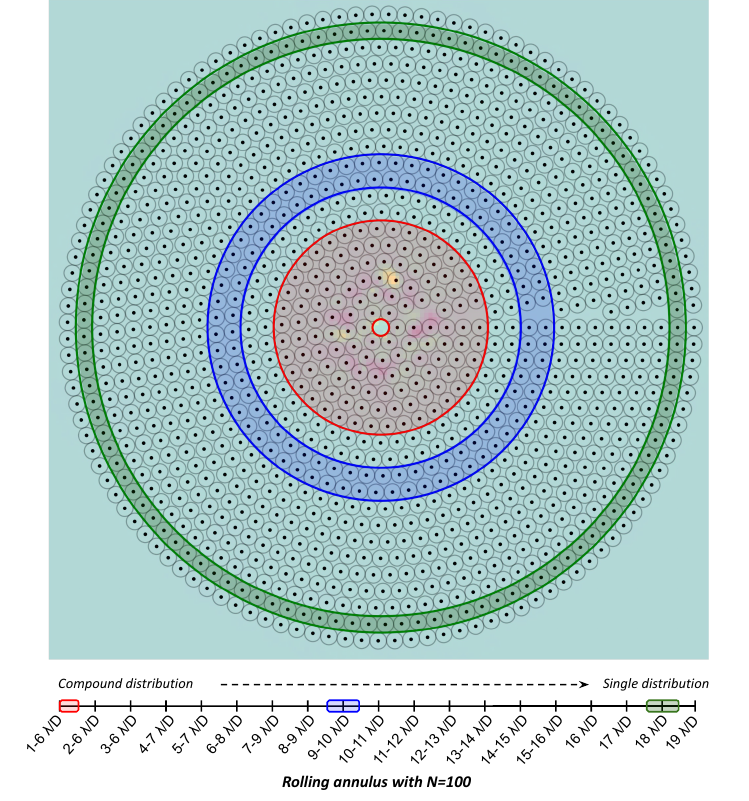}
         \caption{Rolling annulus with $N=100$ over the processed frame of Fig.~\ref{fig:split_assumption}. \textit{Top:} Examples of the first rolling one (in red), the ninth rolling one (in blue) and the eighteenth rolling one (in green) displayed over the central pixel pavement in the image. \textit{Bottom:} A complete set of rolling annuli shown in a straight line that represents the distance from the star. The three rolling annuli shown in the top figure are displayed with the same colours.}
         \label{fig:rolling_annulus}
     \end{figure}
    
      \begin{figure*}[t]
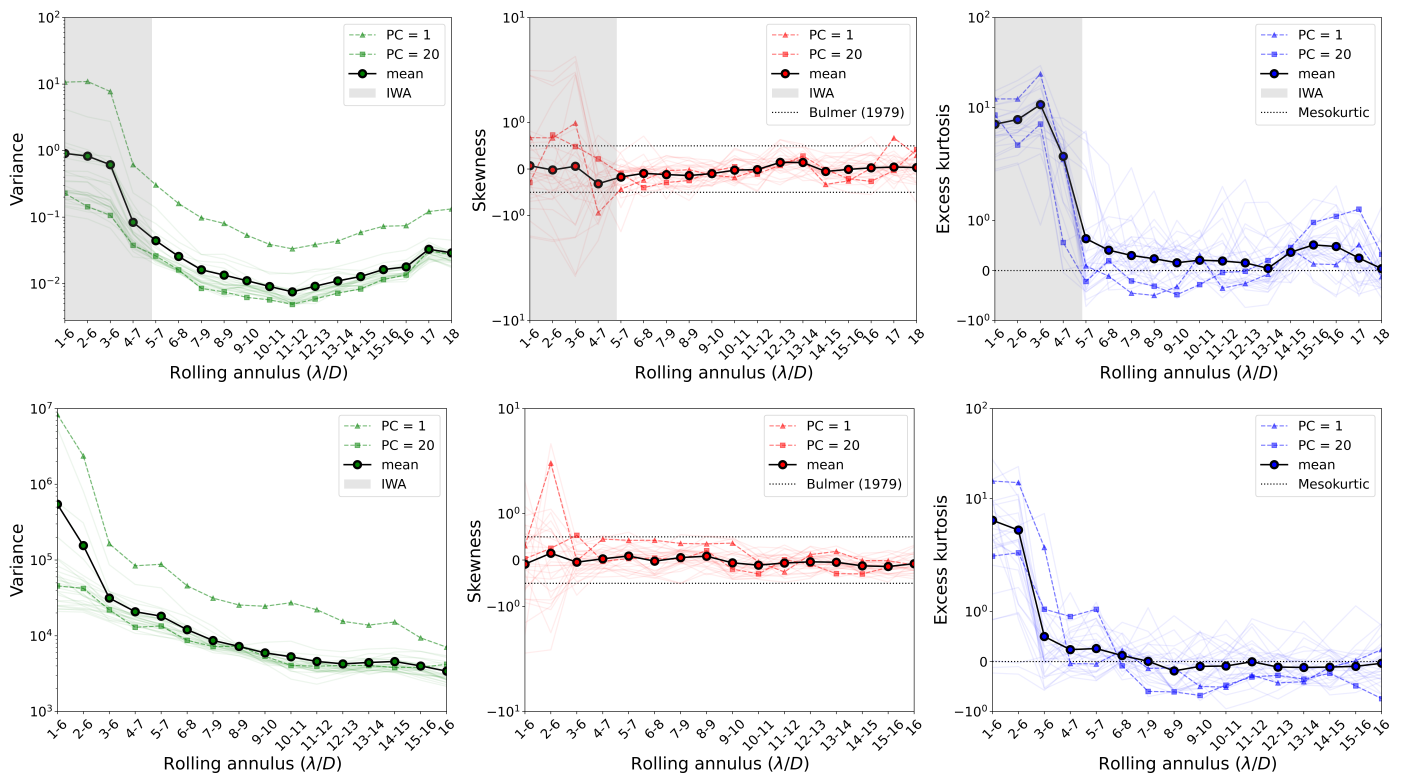

      \includegraphics[width=\textwidth]{sph2_moment_nogrid.png}
      \includegraphics[width=\textwidth]{nirc3_moment_nogrid.png}
    
      \caption{Statistical moments evolution based on a rolling annulus which paves the full annular-PCA processed frame. The top and bottom rows refer, respectively, to the \textit{sph2} and \textit{nrc3} \varADI{} sequences. Colour curves on each subplot refer to a different principal component ranging from one to thirty. The bold curve on top of each subplot indicates the average of the thirty PCs, and PC=1 and PC=20 are illustrated with specific symbols. In the case of \textit{sph2}, grey areas highlight the inner working angle (IWA).    \label{Figure:statistical_moments}}
    \end{figure*}
    
    \subsubsection{Paving the image field of view}
    
    In order to find the radius at which background noise starts to dominate in the image, we study the evolution of noise statistics as a function of angular separation. We first pave the full field of view with concentric annuli of \varLambdaOverD{} width (Fig.~\ref{fig:split_assumption}). Each annulus contains pixels that are expected to be drawn from the same parent population \citep{Marois2008Confidence}, although we acknowledge that this working hypothesis cannot be completely fulfilled when diffraction patterns associated with the spiders of the telescope or the wind-driven halo are present in the image \citep{Cantalloube2019Peering}. We note that, in the presence of residual speckles, pixels that contain information from the same speckle are all spatially correlated. When background noise dominates over residual speckle noise, we can instead assume that all pixels in an annulus are independent, since photon noise occurs on a pixel-wise basis. However, this assumption of independence can be non-optimal when bad pixels are interpolated, since it can still leave spatially correlated footprints. In HCI, a common procedure to guarantee the independence of samples when performing statistical analysis is to work by integrating pixel intensities on non-overlapping circular apertures of \varLambdaOverD{} diameter within a given annulus \citep{Mawet2014Fundamental}, as shown in Fig.~\ref{fig:rolling_annulus}. This procedure is based on the characteristic spatial scale of residual speckles ($\sim\lambda/D$ size). However, \cite{Bonse_2022} have recently showed that, in the presence of speckle noise, this independence assumption on non-overlapping apertures is incorrect. Instead, they propose to (i) only consider the central pixel value in each circular aperture to produce a more statistically independent set of pixels and (ii) possibly repeat the experiment with various spatial arrangements of the non-overlapping apertures to reduce statistical noise in the measured quantities. We follow this recommendation, and therefore, for the rest of this study, we define our annulus samples of both speckle- and background-dominated noise regimes by only taking the central pixel value for each non-overlapping circular aperture (Fig.~\ref{fig:rolling_annulus}). This approach also minimizes the possible effect of bad pixel interpolation. 
    
    One limitation in using non-overlapping apertures is the small sample statistics problem, especially at small angular distances \citep{Mawet2014Fundamental}. Small samples generally lead to conclusions that are not strong enough statistically speaking. In order to avoid this issue, we propose to use the concept of a rolling annulus (Fig.~\ref{fig:rolling_annulus}) that always contains a minimum number of independent pixels $N$. These $N$ pixels are the central pixels of apertures that pave the field of view and are included in the rolling annulus. It can be understood as an annular window around the star for which the inner boundary moves in 1\varLambdaOverD{} steps, while the outer boundary is set to achieve the criterion on the minimum number of independent pixels. An example of this process with $N=100$ pixels is shown in Fig.~\ref{fig:rolling_annulus}, where the first rolling annulus that achieves the condition, composed of all central pixels of the non-overlapping apertures between 1 and 6\varLambdaOverD{}, is displayed in red over the processed frame. Then, the rolling annulus moves away from the star changing its boundaries, as illustrated with the black line at the bottom of Fig.~\ref{fig:rolling_annulus}. For example, the ninth rolling annulus (in blue) with $N=100$ is located between 9 and 10 \varLambdaOverD{}, and the eighteenth rolling annulus (in green) is at 18\varLambdaOverD{} distance, achieving the $N=100$ condition without the need to expand the region to another annulus. In this paper, we select $N=100$ minimum samples, considered to be the minimum number of samples required to reach reliable statistical power and significance for our statistical analysis.  
    
    \subsubsection{Statistical moments \label{section:statistical_moments}}
    
    Once the processed frame is paved, we first study the evolution of different statistical moments, such as the variance (amount of energy), the skewness (distribution symmetry) and the excess kurtosis (distribution tails), as a function of the angular separation from the star for different number of principal components (PCs), ranging from component one to thirty. Figure~\ref{Figure:statistical_moments} shows this evolution for the case of the \textit{sph2} (top row) and \textit{nrc3} (bottom row) datasets, on which we applied annular-PCA to produce the processed frames. We observe that the variance decreases as the rolling annulus moves away from the star. This trend is common to both datasets and is what we would expect in physical terms, as the intensity of residual speckles varies rapidly with angular separation, especially at short distances. We also see that this behaviour is dampened when using a larger number of principal components, which leads to more effective speckle subtraction. Regarding the skewness analysis, we adopt the convention of \cite{Bulmer1979Principles}, which states that a distribution is symmetrical when its skewness ranges from $-0.5$ to $0.5$. For both datasets, we clearly observe a loss of symmetry at small angular separations. The presence of speckles can provoke this distribution asymmetry due to their higher intensity values in comparison with the background. Looking now at the excess kurtosis in Fig.~\ref{Figure:statistical_moments},  we observe a strong leptokurtic\footnote{In statistics, a leptokurtic distribution has a kurtosis greater than the kurtosis of a normal distribution (mesokurtic), and it is associated in HCI to increase the false alarm rate.} trend for the entire set of PCs at small angular separations and for both datasets. This perfectly matches the fact that a Laplacian distribution fits better the tail decay of residual noise \citep{Pairet2019STIM}, since it is, by definition, leptokurtic. At higher angular separations, instead, we observe differences between both datasets. In the \textit{sph2} processed frames, we detect one mesokurtic regime approximately between 6-13\varLambdaOverD{} followed by a weaker leptokurtic regime approximately between 14-18\varLambdaOverD{}. For \textit{nrc3}, we only observe one mesokurtic regime at a large distance from the star, beyond the third rolling annuli (Fig.~\ref{Figure:statistical_moments}).

    \subsubsection{Combined normality test analysis \label{sec:pca_pmap}}
    
    Another way to explore the spatial distribution of noise is to use hypothesis testing. Assuming that residual speckle noise is non-Gaussian by nature, while background noise is Gaussian (see Sect.~\ref{section:residual_noise}), we can assess the probability of the null hypothesis \varNullHypothesis{} that data are normally distributed, that is, explained solely by background noise. We rely on a combination of a series of normality tests, making use of four of the most powerful tests: the Shapiro-Wilk test \citep[$sw$,][]{Shapiro1965AnAnalysis}, the Anderson-Darling test \citep[$ad$,][]{Anderson1952Asymptotic}, the D'Agostino-K2 test \citep[$ak$,][]{Dagostino1973Tests}, and the Lilliefors test \citep[$li$,][]{Lilliefors1967OnTheKolmogorov}. This choice is motivated by the fact that they have been well-tested in many studies, including Monte-Carlo simulations \citep{Yap2011Comparisons, MarmolejoRamos2013APower, Ahmad2015APower, Patricio2017Comparing, Wijekularathna2019Power, Uhm2021ACompparison}. It is worthwhile to remark  that the goal is not to benchmark the robustness of all these tests. Our purpose, instead, is to collect a larger amount of statistical evidence for the same hypothesis, that can then be combined to increase the statistical power when making a decision regarding the null hypothesis. Moreover, regarding the statistical requirements, the only constraints to be verified before using these tests are the independence and sufficient size of the sample. In terms of sample size, \cite{Jensen2017ANewStandard} shows that normality tests can exhibit lower statistical power with sample sizes under 100 observations. Here, the independence and size constraints are met by the proposed approach to pave the field of view, using the central pixels of non-overlapping apertures within rolling annuli of $N=100$ apertures. Additionally, we follow for this analysis the recommendation of \cite{Bonse_2022} to perform our statistical tests with various spatial arrangements for the non-overlapping apertures. We leverage the fact that different aperture arrangements within the same annulus contain valuable noise diversity that can directly benefit the analysis when making a decision about the null hypothesis.
    
    \begin{figure*}[htbp]
      \centering
      \begin{subfigure}[b]{0.6\textwidth}
        \includegraphics[width=\textwidth]{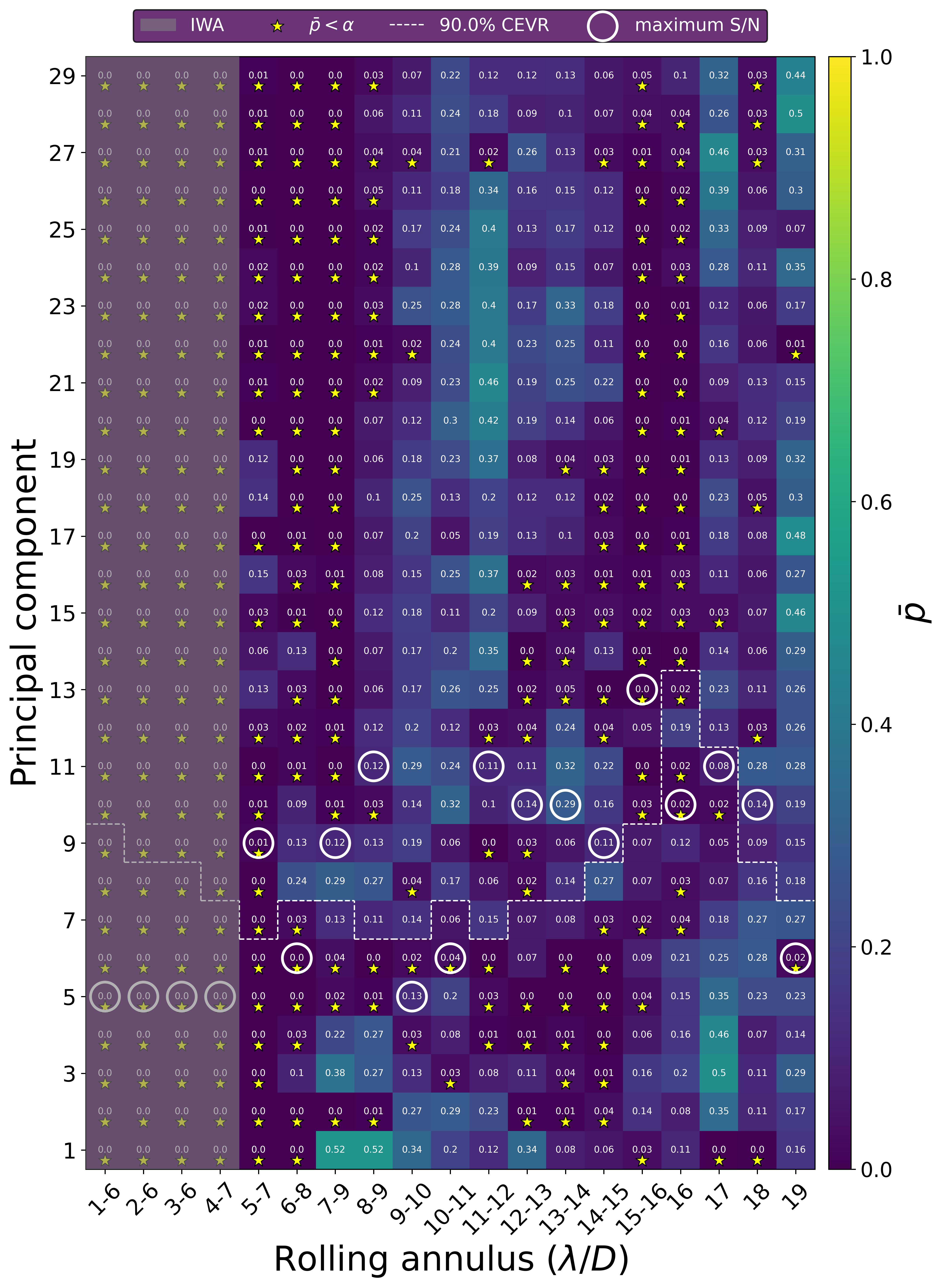}
        \caption{}
        \label{fig4:pcapmap}
      \end{subfigure}
      \hfill
      \begin{subfigure}[b]{0.39\textwidth}
        \includegraphics[width=\textwidth]{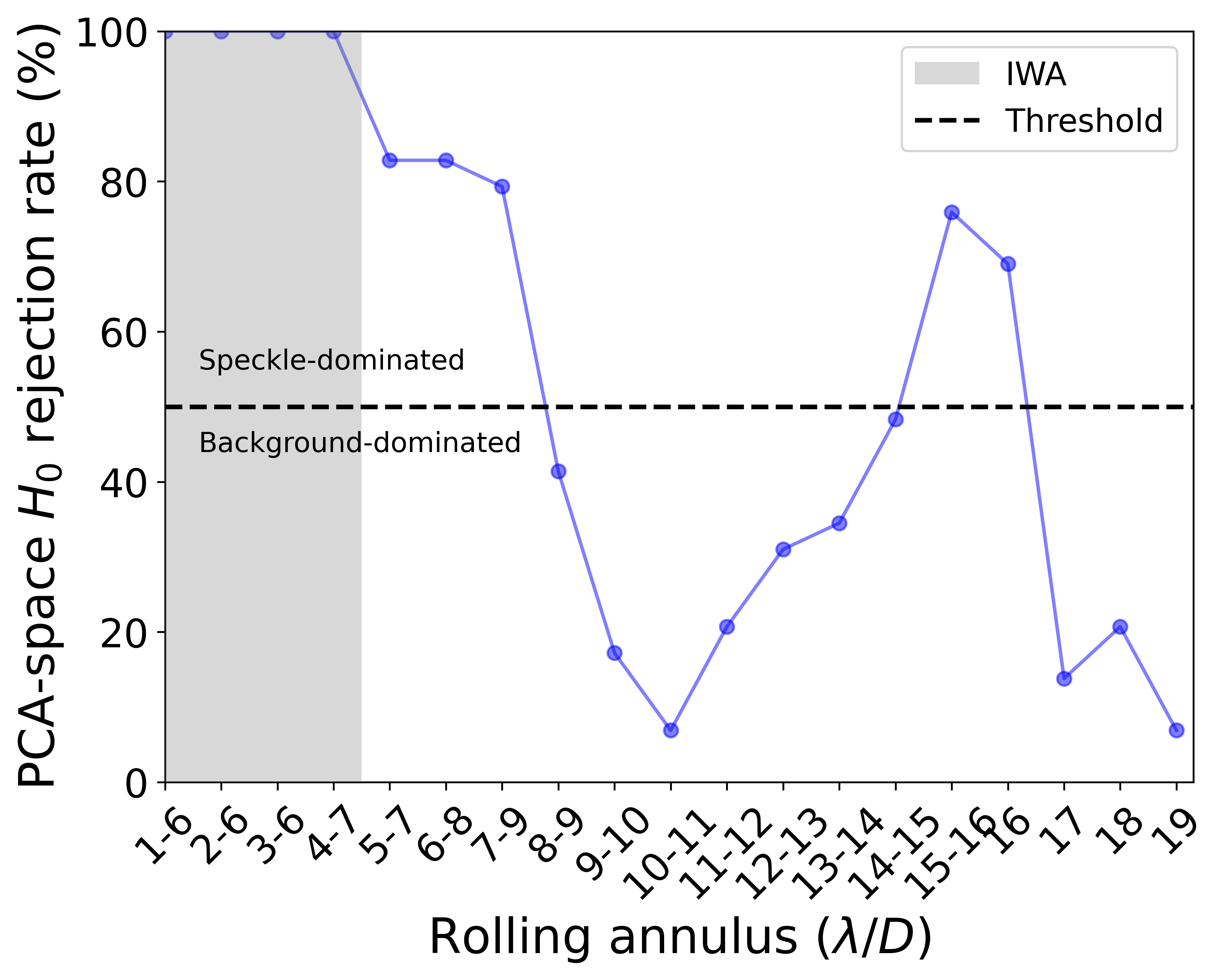}
        \caption{}
        \label{fig4:critera}
        \vspace{0.6 cm}
        \includegraphics[width=\textwidth]{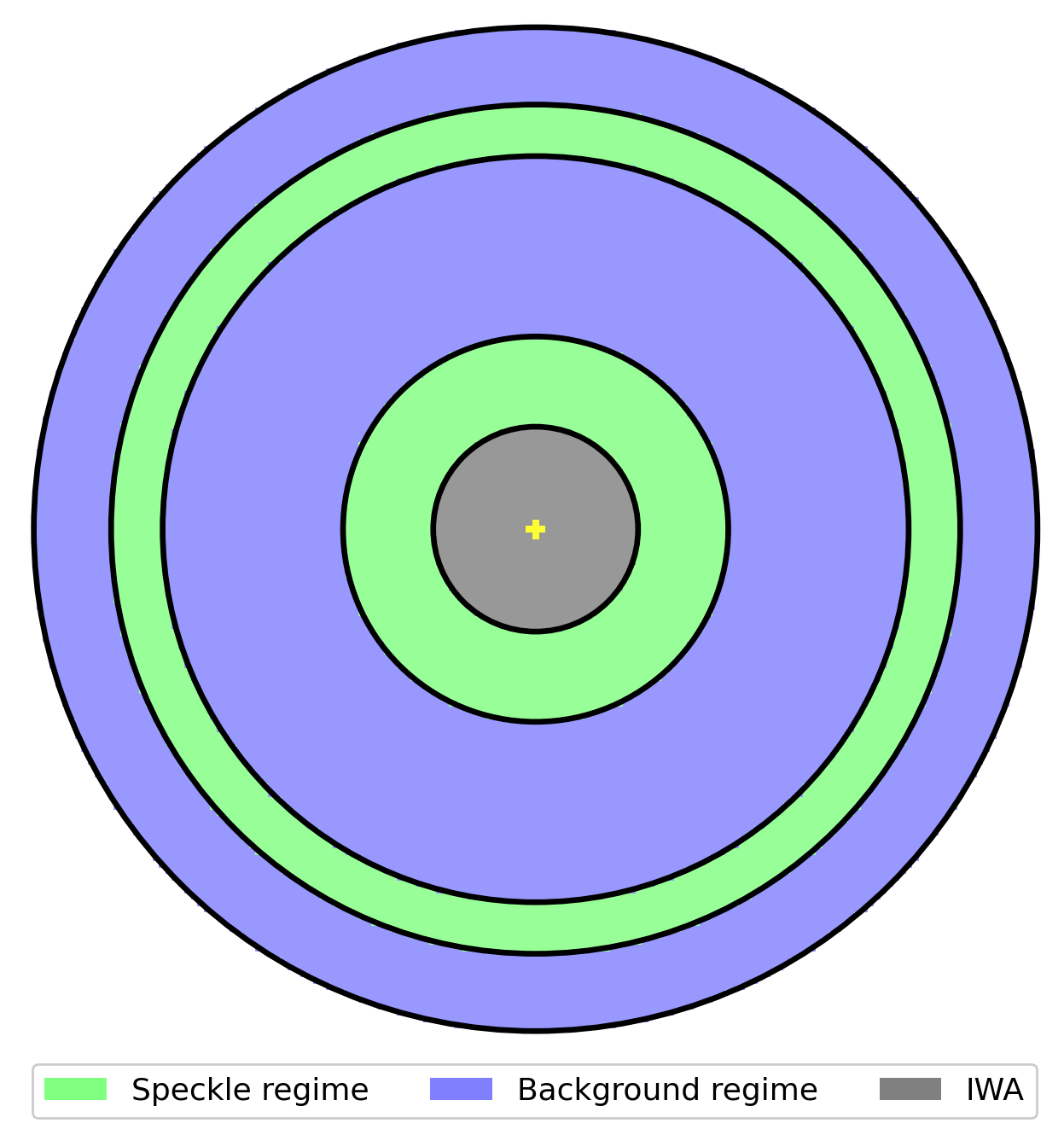}
        \caption{}
        \label{fig4:regimes}
      \end{subfigure}
      \caption{Combined normality test analysis for the \textit{sph2} \varADI{} sequence.  \textit{(a)} \varPmaps{} showing the combined p-value $\harmonicMean{p}$ both as a colour code and as values, as a function of the distance to the star through the rolling annulus ($x$ axis) and the number of principal components used in the PCA-based PSF subtraction ($y$ axis). Yellow star markers indicate when the null hypothesis \varNullHypothesis{} (Gaussian noise) is rejected. The white dashed line shows the $90\%$ CEVR at each rolling annulus. White circles in bold highlight the principal component that maximizes the S/N of fake companion recoveries. \textit{(b)}   Percentage of yellow star markers, or $H_{0}$ rejection, ($y$ axis) for each rolling annulus ($x$ axis) on the \varPmaps{}. The dashed black line highlights the selection criteria for setting the dominant noise at each rolling annulus. \textit{(c)} Final representation of estimated noise regimes along the processed frames field of view. Grey areas in each subplot highlight the inner working angle (denoted as IWA).\label{fig:pvalues_map_sph2}}
    \end{figure*}
    
    \begin{figure*}[htbp]
      \centering
      \begin{subfigure}[b]{0.6\textwidth}
        \includegraphics[width=\textwidth]{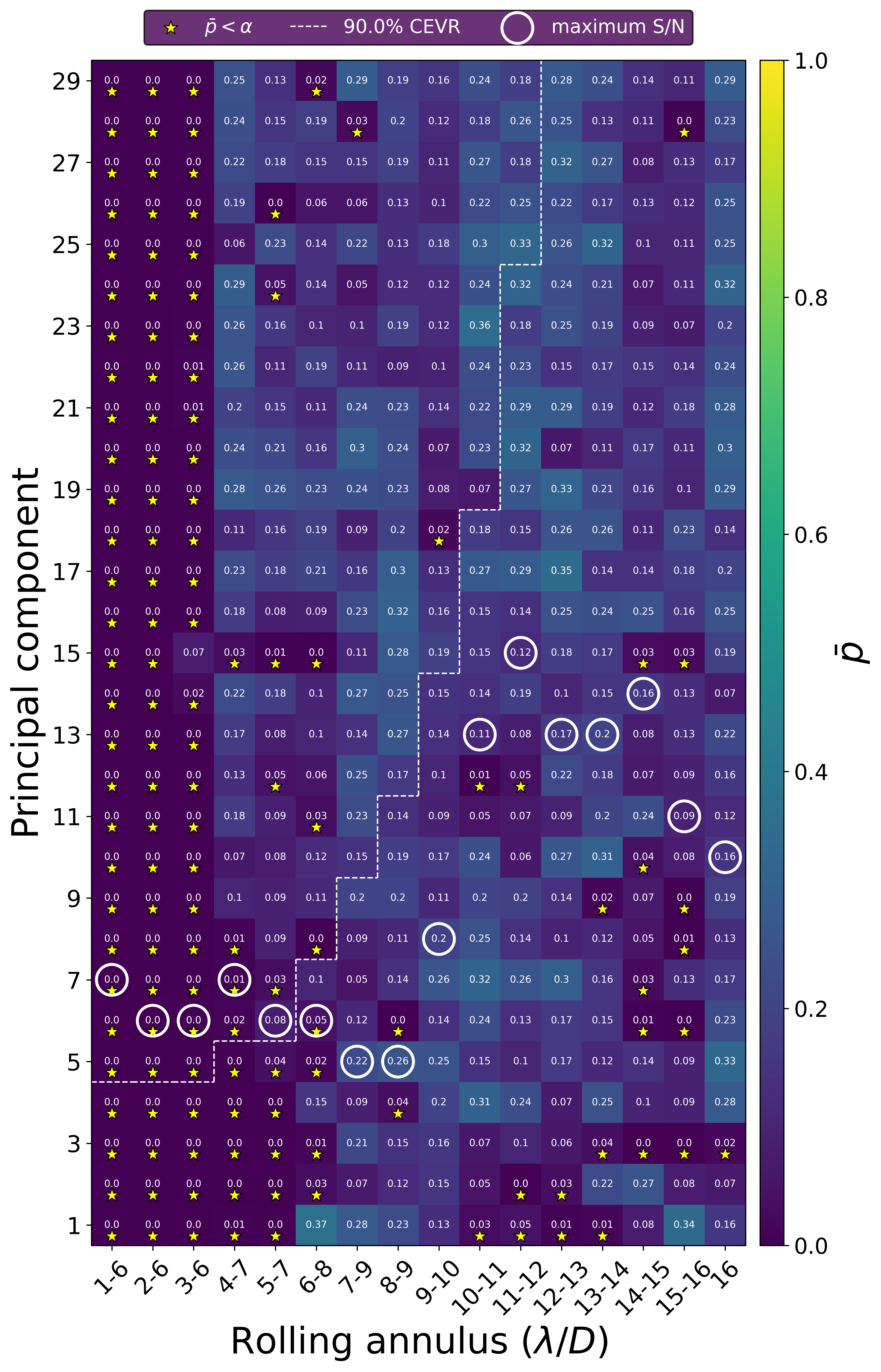}
        \caption{}
        \label{fig5:pcapmap}
      \end{subfigure}
      \hfill
      \begin{subfigure}[b]{0.39\textwidth}
        \includegraphics[width=\textwidth]{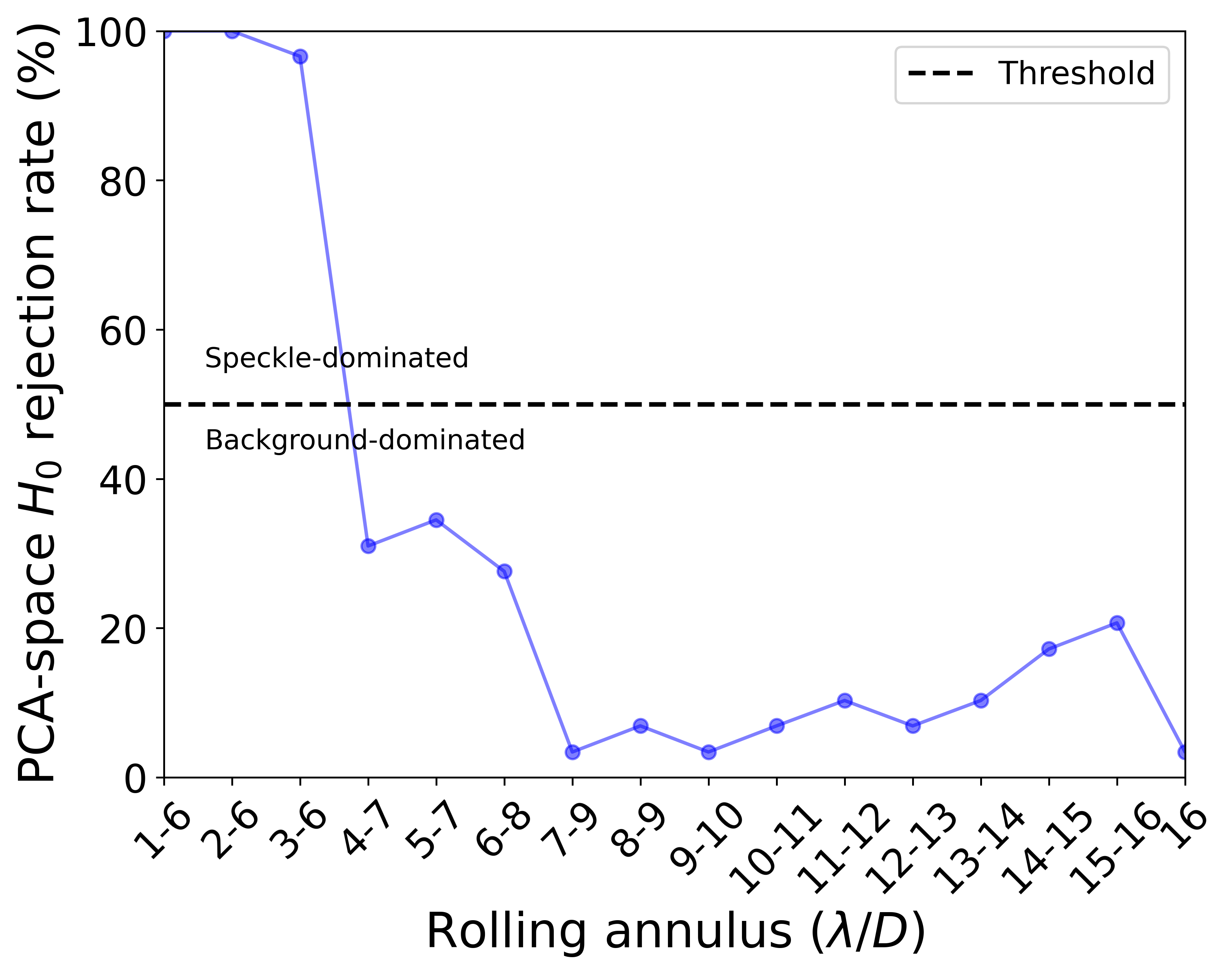}
        \caption{}
        \label{fig5:criteria}
        \vspace{2 cm}
        \includegraphics[width=\textwidth]{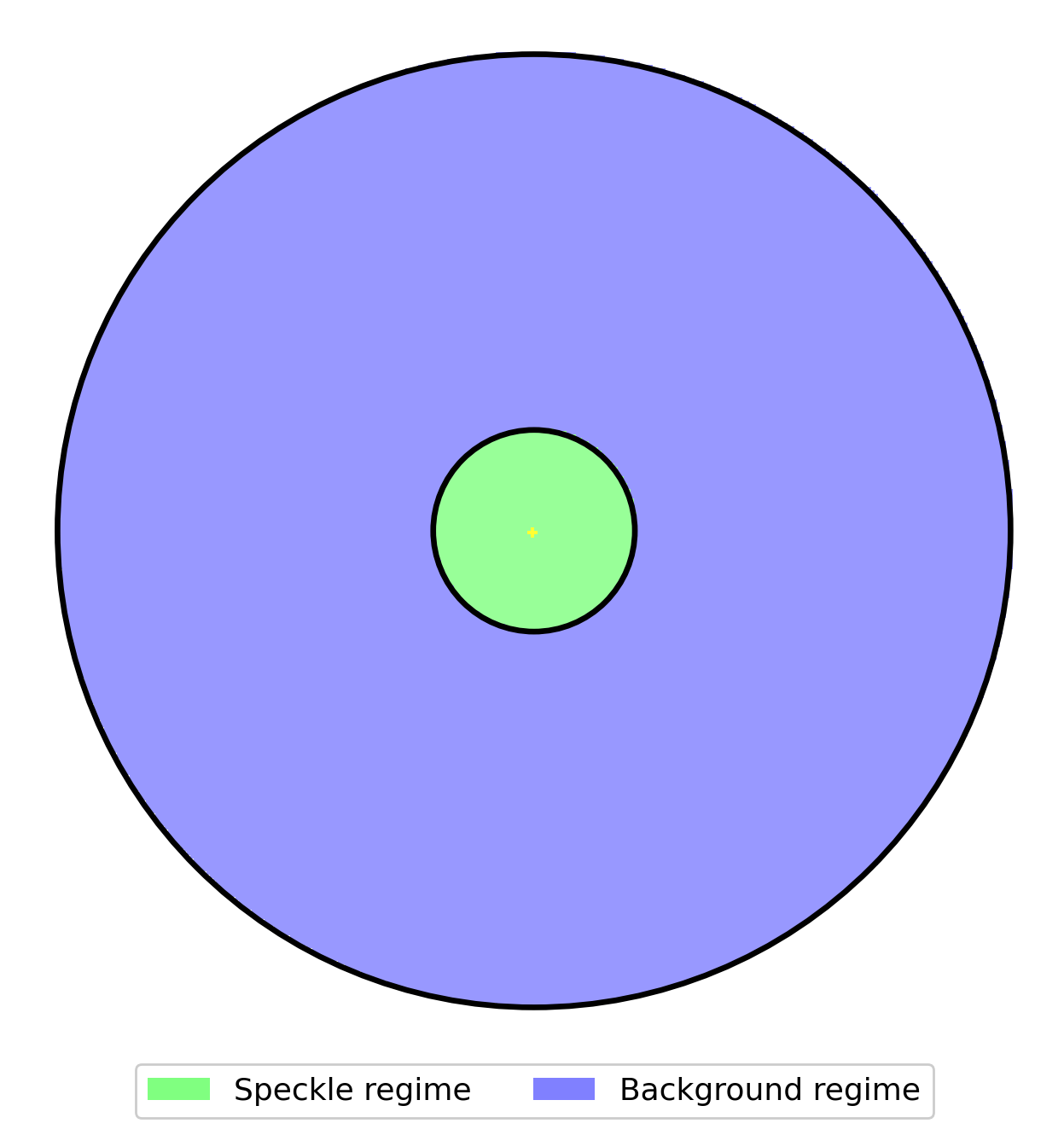}
        \caption{}
        \label{fig5:regimes}
      \end{subfigure}
      \caption{Same as Fig.~\ref{fig:pvalues_map_sph2}, but for the \textit{nrc3} \varADI{} sequence. \label{fig:pvalues_map_nirc3}}
    \end{figure*}

    Our analysis for testing the null hypothesis \varNullHypothesis{} within a specific rolling annulus of the processed frame is thus composed as follows. We begin by randomly selecting a normality test $t$ from the set $\mathcal{T}=\{sw, ad, ak, li\}$. Subsequently, we randomly choose an angular displacement $\theta$ of circular apertures for each single annulus within the rolling annulus. Assuming $N_{ann}$ single annuli, then,  $\Theta=\{\theta_{i}\}_{i=\{1, \dots ,N_{ann}\}}$, where $\Theta$ thus represents a random aperture arrangement. After defining the sample of central pixels $X(\Theta)$ for this arrangement, we use the selected statistical test $t$ to compute the p-value associated to $X(\Theta)$. We denote this p-value as $p(t,\Theta)$. This process of randomly selecting both a normality test $t$ and an aperture arrangement $X(\Theta)$ to compute $p(t,\Theta)$ is repeated $m$ times for the same rolling annulus, which produces $m$ p-values that are not statistically independent. The final step involves using the harmonic mean, as proposed by \citet{Vovk_2020}, to combine these $m$ p-values into a global p-value noted $\harmonicMean{p}$. By comparing $\harmonicMean{p}$ with a predefined significance threshold $\alpha$, we can finally  reject the null hypothesis \varNullHypothesis{} if $\harmonicMean{p}<\alpha$.
    
    By repeating this procedure for each rolling annulus in the processed frame and for various numbers of principal components in our annular-PCA post-processing algorithm, we can build what we call the PCA p-value map, or \varPmaps{} for short. Figures~\ref{fig4:pcapmap} and \ref{fig5:pcapmap} show examples of \varPmaps{}s for the \textit{sph2} and \textit{nrc3} datasets, respectively. For both, we only considered the first $29$ principal components to produce the annular-PCA space ($y$-axis in figures). Each cell in a \varPmaps{} shows, through the number in white and its background colour, the combined p-value $\harmonicMean{p}$ with $m=300$. P-values below the pre-defined threshold $\alpha$ are marked with yellow stars on the figures. In order to minimize the Type I error (false rejection of the null hypothesis), we selected the standard threshold value $\alpha=0.05$ in Figures~\ref{fig4:pcapmap} and \ref{fig5:pcapmap}. Afterward, we calculate the fraction of yellow star markers, or $H_{0}$ rejection rate, along the PCA domain for each rolling annulus in \varPmaps{}s. We then classify rolling annuli as speckle-dominated when they contain more than 50$\%$ of stars. Figures~\ref{fig4:critera} and \ref{fig5:criteria} show this selection criterion by plotting the $H_{0}$ rejection rate per rolling annulus.  In the case of \textit{sph2} (Fig.~\ref{fig:pvalues_map_sph2}), we clearly observe the presence of four noise regimes beyond the inner working angle: a first regime dominated by non-Gaussian noise due to residual speckles between $5-7 \lambda/D$ distance, a second regime where noise is more consistent with Gaussian statistics, probably dominated by background noise between $8-14 \lambda/D$, a third regime with non-Gaussian noise between $15-16 \lambda/D$, where speckles are dominating again as we approach the limit of the well-corrected area produced by the SPHERE adaptive optics \citep{Cantalloube2019Peering}, and finally, a fourth regime more consistent with Gaussian statistics again between $17-19 \lambda/D$.  The speckle-dominated regime at $15-16 \lambda/D$ would also explain the slightly leptokurtic behaviour observed at those separations in Fig.~\ref{Figure:statistical_moments}.  For the \textit{nrc3} dataset (Fig.~\ref{fig:pvalues_map_nirc3}), we only observe two noise regimes, with speckle noise dominating approximately between $1-3 \lambda/D$ distance, and background noise dominating beyond $3 \lambda/D$ (Fig.~\ref{fig5:criteria}). The white dotted line and circles overplotted on the PCA-pmaps will be explained later in Sect.~\ref{sec:nasodinn}.

    \subsection{Field-of-view splitting strategy}
    
    At this point, we can see that, for both \textit{sph2} and \textit{nrc3},  similar estimations of the noise regimes are reached using the two proposed methods: the study of statistical moments and the \varPmaps{}s. Figure~\ref{Figure:statistical_moments} provides a first insight into the spatial structure of residual noise and, thereby, brings us closer to estimating the radius split (Fig.~\ref{fig:split_assumption}) in the processed frame. Indeed, the significant increase of the variance together with the leptokurtic behaviour and the positively skewed trend at small angular separations, suggest that this regime is still dominated by residual starlight speckles. On the other hand, \varPmaps{}s contain more statistical diversity through the combination of p-values with which very similar regime estimations are reached. Thus, both analyses are complementary from a statistical perspective. Yet, from now on, we elect to use \varPmaps{}s to define the noise regime as a baseline, since they can also be used for other purposes.
    
    The noise analysis described above suggests that noise regions should be defined on a case-by-case basis. Regarding the nature of residual noise in a processed frame, our tests do not necessarily mean that residual speckle noise is non-Gaussian in the innermost, individual annuli. Instead, compound distributions could be at the origin of the non-Gaussian noise behaviour in the innermost rolling annuli. Compound distributions refer to the sampling of random variables that are not independent and identically distributed. For small angular separations (red annulus in  Fig.~\ref{fig:rolling_annulus}) where residual speckle noise dominates over background noise,  the samples are taken from distributions that might be Gaussian, but with different variances.  If they are Gaussian and their variance follows an exponential distribution, then according to \cite{Gneiting1997Normal}, the compound distribution follows a Laplacian, as observed by \citet{Pairet2019STIM}. This explanation, which is not a proof, would reconcile the belief that residual speckle noise should be locally Gaussian. Because of the small sample size, there is, however, no proper way to test this interpretation on individual annuli in the innermost regions. For all these reasons, we believe that splitting the processed frame field of view in different noise regimes is duly motivated, and, in the next sections, we detail how we have implemented this splitting to improve the detection of exoplanets.

    \section{Implementation} \label{sec:implementation}
    So far, we have focused on understanding the spatial structure of residual noise in the processed frame, which has allowed us to empirically define the regions dominated by speckle and background noise. Now, we aim to use this local noise approach in order to help post-processing algorithms enhance their detection performance. Most HCI algorithms have the potential of being applied separately to different noise regimes. Here, we are particularly interested in the case of deep learning. 
    Neural networks are good candidates to capture image noise dependencies due to their ability to recognize hidden underlying relationships in the data and make complex decisions. In order to maximize the added value of working in noise regimes and showing its benefits for the detection task, we propose to revisit \varSODINN{} \citep{Gomez2018Supervised}, the first supervised deep learning algorithm for exoplanet imaging.  In this section, we first provide a brief overview of \varSODINN{}, and then present our novel NA-\varSODINN{} algorithm, an adaptation of \varSODINN{} working on noise regimes, aided with additional handcrafted features.

    \subsection{Baseline model: The \varSODINN{} algorithm \label{sec:sodinn}}
    
\begin{figure}
        \centering
        \includegraphics[width=\columnwidth]{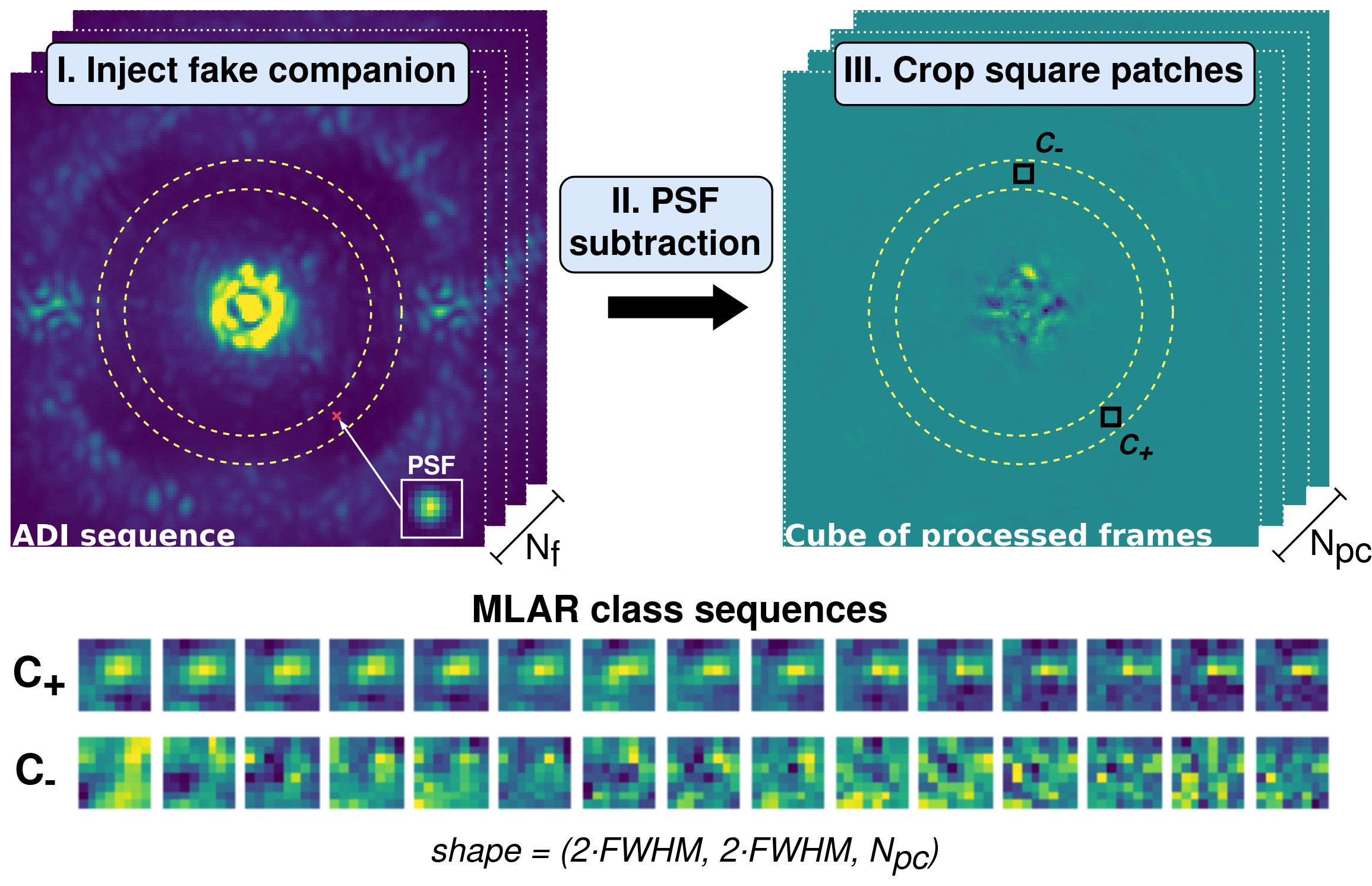}
        \caption{SODINN labelling stage. \textit{Top}: Steps for generating MLAR samples (see the text for more details). $N_{f}$ is the number of frames in the \varADI{} , and $N_{pc}$ is the number of principal components in the cube of processed frames and therefore in the final MLAR sequence. \textit{Bottom}: Example of an MLAR sequence of each class.}
        \label{fig:sodinn_label}
    \end{figure}

    \varSODINN{} stands for Supervised exOplanet detection via Direct Imaging with a deep Neural Network. It is a binary classifier that uses a convolutional neural network (CNN) to distinguish between two classes of square patch sequences: sequences that contain an exoplanet signature ($c_{+}$, the positive class), and sequences that contain only residual noise ($c_{-}$, the negative class). Figure~\ref{fig:sodinn_label} (bottom) shows an example sequence for each class, where the individual images are produced with various numbers of principal components. The first image in the sequence corresponds to the first principal component, while the last corresponds to a number of principal components with which a maximum of 90\% cumulative explained variance ratio (CEVR) is captured. \cite{Gomez2018Supervised} refer to these patch sequences as Multi-level Low-rank Approximation Residual (MLAR) samples.

    \subsubsection{Generation of the training set \label{sec:sodinn_generation}}
    
    The first step in \varSODINN{} is to build a training dataset composed of thousands of different $c_{+}$ and $c_{-}$ MLAR sequences. A $c_{+}$ sequence is formed through three consecutive steps that are summarized in Fig.~\ref{fig:sodinn_label}. (i) First, a PSF-like source is injected at a random pixel within a given annulus of the \varADI{} sequence. The flux of this injection is the result of multiplying the normalized off-axis PSF by a scale factor randomly chosen from a pre-estimated flux range that corresponds to a pre-defined range of S/N in the processed frame. The estimation of injection flux ranges is explained in Appendix~\ref{sec:appendixB}. (ii) Singular value decomposition \citep[SVD,][]{Halko2011AnAlgorithm} is then used on this synthetic \varADI{} sequence to perform PSF subtraction for different numbers of singular vectors (or principal components), thereby producing a series of processed frames. (iii) Finally, square patches are cropped around the injection coordinates for each processed frame. This forms a series of $c_{+}$ MLAR sequences, where each sequence contains the injected companion signature for different numbers of principal components. The patch size is usually defined between 1.5-2 times the FWHM of the PSF. 
    
    Likewise, we construct a $c_{-}$ sequence by extracting MLAR sequences for pixels where no fake companion injection is performed. The number and order of singular vectors are the same as those used for the $c_{+}$ sequences. For the case of $c_{-}$ sequences, \varSODINN{} must deal with the fact that, using only one ADI sequence, we obtain a single realization of the residual noise, so that the number of $c_{-}$ sequences we can grab per annuli is not enough to train the neural network without producing over-fitting. \varSODINN{} solves this problem by increasing the number of $c_{-}$ sequences in a given annulus through a dedicated data augmentation strategy that is based on four consecutive steps: (i) build a first subset by randomly grabbing $c_{-}$ sequences centred on up to ten percent of the total number of pixels; (ii) build a second subset by grabbing all the available pixels in the annulus and flip the sign of the parallactic angle when derotating the residual images, a common practice in HCI to remove possible planetary sources while preserving noise properties; (iii) randomly pick groups of three $c_{-}$ sequences from the two subsets and average them to produce new sequences; (iv) finally, perform random rotations and small shifts of the $c_{-}$ sequences obtained in the previous step to create even more diversity. The same rotation angle and shift are applied to all the slices of a given MLAR sequence. This data augmentation process ensures that we only use augmented $c_{-}$ sequences for the training. 
    
    This procedure of generating $c_{+}$ and $c_{-}$ sequences is repeated thousands of times for each annulus in the field of view. When the entire field of view is covered, MLAR sequences of the same class from all annuli are mixed, and the balanced training set (same amount of $c_{+}$ and $c_{-}$ samples) is built. 
    
    \subsubsection{Training of the network \label{sec:sodinn_training}}
    
    The training set is then used to train the \varSODINN{} neural network. This produces a detection model that is specific for the \varADI{} sequence from where MLAR sequences were generated. The \varSODINN{} network architecture is composed of two concatenated convolutional blocks. The first block contains a convolutional-LSTM \citep{Shi2015ConvLSTM} layer with 40 filters and a hyperbolic tangent activation function, and kernel and stride size of (1,1), followed by a spatial 3D dropout \citep{Srivastava2014Dropout} and a MaxPooling-3D \citep{Boureau2010ATheoretical}. 
    The second block contains the same except that it now has 80 filters, and kernel and stride size of (2,2). 
    These first two blocks extract the feature maps, capturing all spatio-temporal correlations between pixels of MLAR sequences. After that, they are flattened and sent to a fully connected dense layer of 128 hidden units. 
    Then, a rectifier linear unit \citep[ReLU,][]{Nair2010Rectified} is applied to the output of this layer followed by a dropout regularization layer. Finally, the output layer of the network consists of a sigmoid unit, which provides a normalized value between 0 and 1. This value is usually referred as a probability, however, it is known in computer vision that the output of a deep learning architecture normalized between 0 and 1 with classical activation functions (\eg the sigmoid function) tends to be more binary, and therefore, it cannot be interpreted as a real probability. 
    For this reason, from now on, we refer to this output score as the model confidence. The network weights are initialized randomly using a Xavier uniform initializer, and are learned by back-propagation with a binary cross-entropy cost function:
    \begin{equation}
        L(y_{n}, \hat{y_{n}}) = - \sum _{n} (y_{n}\ln(\hat{y}_{n})+(1-y_{n})\ln(1-\hat{y}_{n})) \, ,
    \end{equation}
    where $y_{n}$ is the true label of the $n^{th}$ MLAR sample and $\hat{y}_{n}$ is the predicted confidence that this $n^{th}$ MLAR sample  belongs to the $c_{+}$ class. \varSODINN{} uses an Adam optimizer with a step size of 0.003, and mini-batches of 64 training samples. An early stopping condition monitors the validation loss. The number of epochs is usually set to 15, with which \varSODINN{} generally reaches $\sim 99.9\%$ validation accuracy \citep{Gomez2018Supervised}.

    \subsubsection{Inference \label{sec:sodinn_inference}}
    
    Once the detection model is trained and validated, it is finally used to find real exoplanets in the same \varADI{} sequence. Because the input of the model is an MLAR structure, we first map the entire field of view by creating MLAR samples (with no injection) centred on each pixel. These MLAR samples have never been processed during the training since the $c_{-}$ class MLAR samples in the training set are built by augmentation (Sect.  \ref{sec:sodinn_generation}). The goal of the trained model is therefore to assign a confidence value between 0 (no confidence) and 1 (maximum confidence) for each of these new MLAR sequences to belong to the $c_{+}$ class. Computing a confidence score for each individual pixel leads to a confidence map, from which exoplanet detection can be performed by choosing a confidence threshold.

    \subsection{Model adaptation: The NA-\varSODINN{} algorithm \label{sec:nasodinn}}
    
    In \varSODINN{}, the training set is built by mixing all MLAR sequences from the same class, generated on every annulus in the field of view. In the presence of different noise regimes, this way of proceeding can complicate the training of the model, as the statistics of an MLAR sequence generated in the speckle-dominated regime differ from a sequence of the same class generated in the background-dominated regime instead. In order to deal with this, we train an independent \varSODINN{} detection model per noise regime instead of a unique model for the full frame field of view. Thereby, each detection model is only trained with those MLAR sequences that contain statistical properties from the same (or similar) probabilistic distribution function. Therefore,
    our region of interest in the field of view is now smaller. This means that the number of pixels available to generate MLAR sequences is reduced, and therefore, that we are losing noise diversity in comparison with a model that is trained in the full frame.  However, this loss of diversity comes with the benefit of better capturing the statistics of noise within a same noise regime, which improves the training. 
    
    \subsubsection{Adding S/N curves to the network}
    
\begin{figure}
        \centering
        \includegraphics[width=\columnwidth]{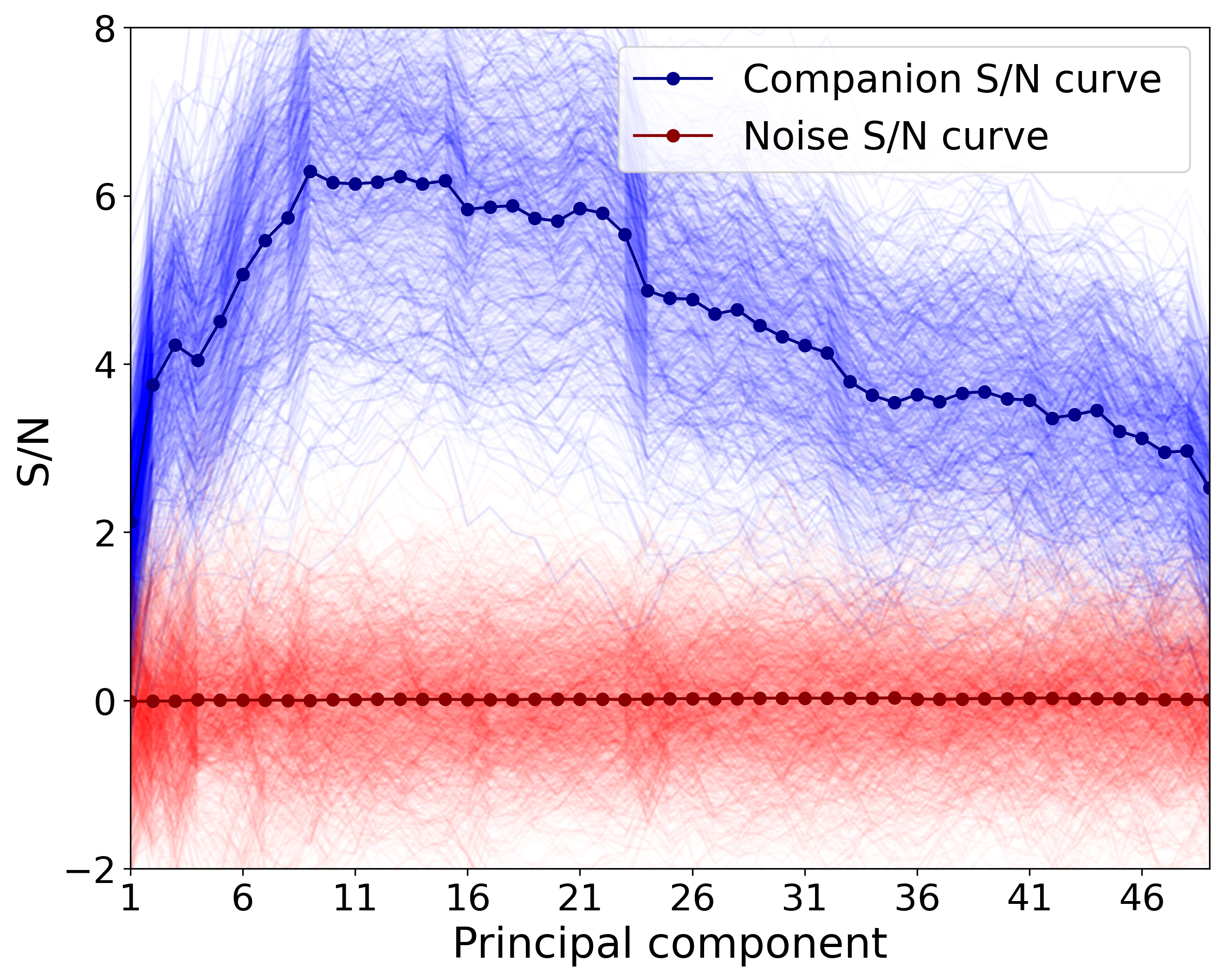}
        \caption{S/N curves generated from the $sph2$ cube of processed frames at a 8\varLambdaOverD{} distance from the star. Curves in blue contain the exoplanet signature and curves in red just residual noise. The flux of injections is randomly selected from a range that is between one and three times the level of noise. Dotted curves over populations show the mean of each class.}
        \label{fig:snr_curves}
    \end{figure}

    In order to compensate for the noise diversity loss associated with the training on individual noise regimes, we attempt to reinforce the training by means of new handcrafted features. An interesting discriminator between the $c_+$ and $c_-$ classes, which is also physically motivated, comes from their behaviour in terms of signal-to-noise ratio (S/N). The most accepted and used S/N definition in the HCI literature is from \cite{Mawet2014Fundamental}. It states that, given a $1\lambda/D$ wide annulus in a processed frame at distance $r$ (in $\lambda/D$ units) from the star, paved with $N=2\pi r$ non-overlapping circular apertures (see Fig.~\ref{fig:rolling_annulus}), the S/N for one of these apertures is defined as
    \begin{equation}
        \mathrm{S/N} = \frac{\Bar{x}_{t}-\Bar{x}_{N-1}}{\sigma_{N-1}\sqrt{1+\frac{1}{N-1}}} \, ,
        \label{eq:snr}
    \end{equation}
    where $\Bar{x}_{t}$ is the aperture flux photometry in the considered test aperture,  $\Bar{x}_{N-1}$ the average intensity over the remaining $N-1$ apertures in the annulus, and $\sigma_{N-1}$ their standard deviation. In order to maximize the S/N, image processing detection algorithms need to be tuned through finding the optimal configuration of their parameters \citep[see \eg][]{Dahlqvist2021Improving}. Here, rather than optimizing the algorithm parameters, we use the fact that we can leverage the behaviour of the S/N versus some of the algorithm parameters in our deep learning approach. This is especially the case for the number of principal components used in the PSF subtraction. We define an S/N curve as the evolution of the S/N computed for a given circular aperture as a function of the number of principal components \citep{Gomez2017VIP}. Figure~\ref{fig:snr_curves} shows an example of $1000$ S/N curves generated from the $sph2$ \varADI{} sequence. We clearly see in Fig.~\ref{fig:snr_curves} that, in the presence of an exoplanet signature (blue curves), the S/N curve first increases and then decreases, which leads to the appearance of a peak at a given number of principal components. This behaviour, capturing the competition between noise subtraction and signal self-subtraction, was already documented elsewhere \citep[\eg][]{Gomez2017VIP}. The peak in the S/N curve indicates the number of principal components for which the contrast between the companion and the residual noise in the annulus is maximum. Hereafter, we denote as $k$ the principal component at which this S/N peak is located.
    
    \begin{figure*}
        \centering
        \includegraphics[width=\textwidth]{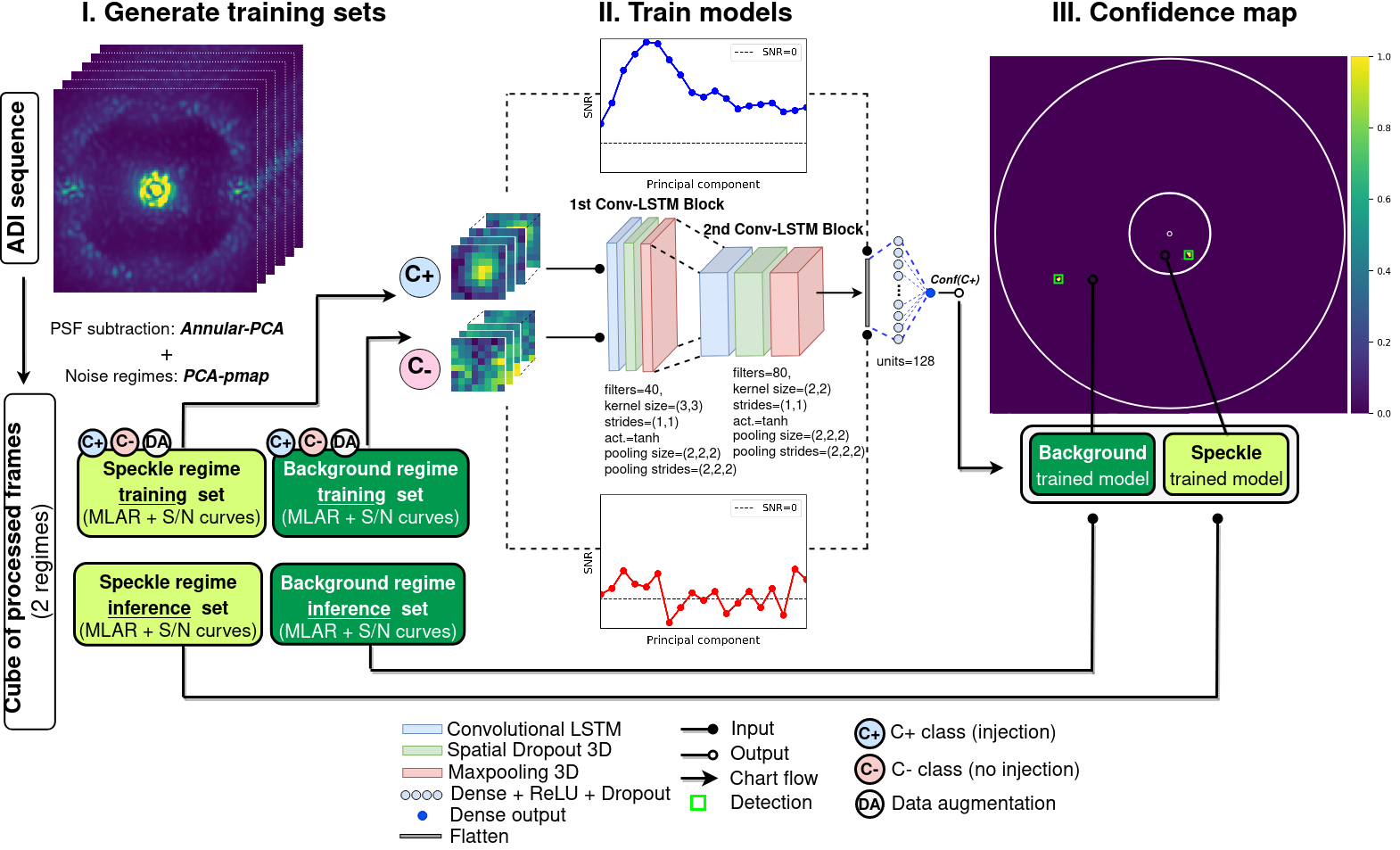}
        \caption{Illustration of the three steps within the NA-\varSODINN{} algorithm working flow. \textit{Left:} Generation of the training set. NA-\varSODINN{} uses the annular-PCA algorithm to perform PSF subtraction and produce the cube of processed frames. Then, it detects residual noise regimes by applying the \varPmaps{} technique to this cube, and builds both the training and inference datasets at each regime, which are composed of both MLAR samples and S/N curves. \textit{Middle:} Model training. NA-\varSODINN{} trains as many detection models as detected noise regimes using their respective training datasets (for the sake of simplicity, we have not duplicated the central deep neural network). This case contains two noise regimes, the speckle- and background-dominated noise regimes, so that two models are trained. \textit{Right:} Detection map. Finally, NA-\varSODINN{} uses each trained model to assign a confidence value to belong to the $c_{+}$ class to each pixel of the corresponding noise regime field of view. \label{fig:nasodinn}}
    \end{figure*}
    
    For a given 1-FWHM circular aperture, the MLAR sequence (no matter the class) and the S/N curve are linked from a physical point of view. Actually, the evolution of the S/N as a function of the number of principal components can be readily extracted from intermediate products used in the production of the training dataset. Therefore, the information conveyed through the S/N curve is already partly contained in the MLAR patches. But while the MLAR sequence contains localized information on the signal and noise behaviour, the S/N curve conveys annulus-wise information, obtained through aperture photometry. Indeed, each aperture's S/N estimation depends on the noise in the rest of the annulus (Eq.~\ref{eq:snr}), so it also contains information that connects with other circular apertures at the same angular separation from the star. This dependency is not captured in MLAR sequences. S/N curves make this rich summary statistics directly available to the neural network to improve the neural network training. 
    One complication in using S/N curves in the training relates to data augmentation, which is mandatory to build up a sufficiently large training dataset for \varSODINN{}.
    Because these augmentation operations modify the intensity and distribution of pixels in the MLAR sequence, there is no direct way to compute the associated S/N curve of an augmented MLAR sequence through Eq.~\ref{eq:snr}. To deal with this, we make simplifying assumptions for each augmentation operation in \varSODINN{}: (i) image rotations do not affect the S/N curve as the same pixels are kept in the final sequence, (ii) averaging two sequences can be approximated as averaging their S/N curves, and (iii) image shifts do not affect the S/N curve as long as the shift is sufficiently small.

    By adding the noise regime approach and the S/N curves to \varSODINN{}, we are building a new detection algorithm. We refer to this novel framework, depicted in Fig.~\ref{fig:nasodinn}, as Noise-Adaptive \varSODINN{}, or NA-\varSODINN{} for short.  As its predecessor, NA-\varSODINN{} is composed of the same three steps: producing the training set from an \varADI{} sequence (Section~\ref{sec:nasodinn_generation}), training a detection model with this training set, and applying the model to find companions in the same \varADI{} sequence (Section~\ref{sec:nasodinn_training_inference}).

    \subsubsection{Generation of the training set \label{sec:nasodinn_generation}}
    
    NA-\varSODINN{} generates as many training sets as detected residual noise regimes. Each of these sets is composed of MLAR sequences and their corresponding S/N curves generated from the corresponding noise regime, including data augmentation. 
    
    Unlike SODINN, which makes use of the CEVR to define the appropriate range of principal components to generate the MLAR sequences \citep{Gomez2018Supervised}, the selection of the principal components for producing both MLAR sequences and S/N curves in NA-\varSODINN{} is instead determined through a novel metric derived from the PCA-pmap. For each rolling annulus, the PCA-pmap can be used to estimate the principal component $k$ that maximizes the S/N for any planetary injection at any position within the annulus (see the peak on the blue curves of Fig.~\ref{fig:snr_curves}). The underlying motivation behind the identification of $k$ is that MLAR sequences and their S/N curves can then be defined around this principal component, thus maximizing the gap between planetary and noise signals in the training set.
    
    To identify $k$ at a given angular separation and for a pre-defined S/N interval of injections, the PCA-pmap relies on two steps: (i) through the data-driven procedure of Appendix~\ref{sec:appendixB}, it pre-estimates the injection flux range that corresponds to the selected S/N range; (ii) once this flux range is estimated, it is used to randomly select fluxes within the range to inject many fake companions, within the annulus at random coordinates, and retrieve their S/N curves (\eg Fig.~\ref{fig:snr_curves}). The $k$ can finally be estimated by averaging all these S/N curves. Here, we select the injected companion fluxes to produce an S/N ranging between 1 and 3 in the final PCA-processed map obtained with one single principal component, which was experimentally found to be appropriate for the NA-SODINN training as it generally produces companions close to the detection limit for a larger number of PCs. We indicate the $k$ obtained for this S/N range as white circles in Figs.~\ref{fig4:pcapmap} and \ref{fig5:pcapmap}. By comparing $k$ with the principal components where the 90\% CEVR is reached in \varPmaps{}s for both $sph2$ and $nrc3$ \varADI{} sequences (Figs.~\ref{fig4:pcapmap} and \ref{fig5:pcapmap}), we observe that at some angular separations, $k$ is not well captured by the CEVR metric. This suggests that the use of CEVR as a figure of merit for choosing the principal components is not always optimal. Therefore, while more training data is generally beneficial, the range of PCs around the value of $k$ can be chosen differently each time NA-SODINN is employed. A range between 15-30 PCs is generally optimal.

    \subsubsection{Training and inference \label{sec:nasodinn_training_inference}}
    
    NA-\varSODINN{} trains an independent detection model for each regime by using its corresponding training set. For each MLAR sequence in the training set, the feature maps created through convolutional blocks are now concatenated with their respective S/N curves after the flattened layer (Fig.~\ref{fig:nasodinn}). NA-\varSODINN{} generally reaches a $\sim 99.9\%$ validation accuracy with 5-8 epochs. In the last step, NA-\varSODINN{} makes inferences for each individual noise regime. It applies the trained model of each regime to infer its corresponding confidence map of the same regime (Fig.~\ref{fig:nasodinn}). 
    Finally, NA-\varSODINN{} builds the final confidence detection map by joining all confidence regime maps inferred with each detection model. Thus, our  NA-\varSODINN{} algorithm is conceived to keep the main characteristics of the pioneering \varSODINN{} algorithm \citep{Gomez2018Supervised}, such as its architecture,  and adapt its optimization process to our local noise approach.

    \section{Model evaluation} \label{sec:evaluation}
    Now that NA-\varSODINN{} has been introduced, we aim to thoroughly evaluate its detection ability. In the first part of this section, we explain the evaluation strategy and benchmark NA-\varSODINN{} with respect to its predecessor \varSODINN{} using the same \textit{sph2} and \textit{nrc3} \varADI{} sequences. Then, in the second part, we apply NA-\varSODINN{} to the first phase of the EIDC \citep{Cantalloube2020Exoplanet}, providing confidence maps for each \varADI{} sequence in the data challenge and running the same statistical analysis to compare the NA-\varSODINN{} performance with the rest of the HCI algorithms.

    \subsection{Performance assessment \label{sec:roc}}
    The evaluation of HCI detection algorithms consists of minimizing the false positive rate (FPR) while maximizing the true positive rate (TPR) at different detection thresholds applied in the final detection map. This information is summarized by a curve in the receiver operating characteristics (ROC) space, where each point in the curve captures both metrics at a given threshold value \citep{Gomez2018Supervised, Dahlqvist2020Regime}. In order to produce ROC curves for various versions of \varSODINN{} applied to a given \varADI{} sequence $D$, we first build the evaluation set $\mathcal{D}_{eval}=\{D_{1}, D_{2}, D_{3}, \dots, D_{s}\}$ containing $s$ synthetic datasets $D_i$, where each synthetic dataset is a copy of $D$ with one fake companion injection per noise regime. Here, we limit the number of injected companions per noise regime to one at a time to avoid any risk of cross-talk between companions in the detection algorithms themselves (\eg because multiple companions can affect the PCA), or in their evaluation (\eg if they get too close and merge in terms of confidence patch). The coordinates of these injections are randomly selected within the considered noise regime boundaries, and their fluxes are randomly set within a pre-defined range of fluxes that correspond to an S/N range between 0.5 and 2 in the processed frame. This pre-defined range of fluxes is estimated through the same data-driven strategy explained in Appendix~\ref{sec:appendixB} and illustrated in Fig.~\ref{fig:fluxes_estimation}. Hence, each algorithm provides $s$ final detection maps, from which true positives (TPs), false positives (FPs), true negatives (TNs) and false negatives (FNs) indicators are computed across the whole noise regime field of view at different detection thresholds. Then, all these indicators are averaged, and the corresponding ROC curve for the considered noise regime is produced. Instead of using the FPR metric as in standard ROC curves, here we used the mean number FPs within the whole field of view, which is more representative of the HCI detection task and facilitates the interpretation of our performance simulations.

    We perform the proposed ROC curve analysis on both $sph2$ and $nrc3$ \varADI{} sequences, with $s=100$ for each. For this assessment, a detection is defined as a blob in the final detection map with at least one pixel above the threshold inside a circular aperture of diameter equal to the FWHM centred at the position of each injection. With the aim of benchmarking NA-\varSODINN{}, we include in this evaluation the annular-PCA algorithm \citep{Absil2013Searching} as implemented in the \texttt{VIP} Python package \citep{Gomez2017VIP,Christiaens2023VIP}, the \varSODINN{} framework by \cite{Gomez2018Supervised}, and two hybrid detection models. These hybrid models are modifications of \varSODINN{} to include only one of the two additional features introduced in NA-\varSODINN{}: the adaptation to noise regimes, or the addition of S/N curves in the training. Hereafter, we refer to them respectively as \varSODINN{}+Split and \varSODINN{}+S/N. In the same spirit as an ablation study, these two hybrid models are included in our evaluation in order to provide information about the added value of each approach separately for the task of detection. It is worth mentioning that instead of retraining all considered \varSODINN{}-based models every time a different fake companion is injected into each evaluation set, we train them once per ADI sequence. While retraining would be more accurate, as the presence of an injected fake companion could slightly perturb the $c_{-}$ class, we assume that our augmentation strategy (Sect.~\ref{sec:sodinn_generation}) mitigates this perturbation and does not significantly impact the training process and the model's performance. Moreover, using the same model to detect all fake companion injections in a single ADI sequence saves computation time.

    An important aspect to consider when comparing algorithms in ROC space is to optimally choose their model parameters. In the case of annular-PCA, we use one, five, and ten principal components for each annulus as a good compromise to analyse its performance.
    For the various versions of \varSODINN{}, we need to define two main parameters: the list of principal components $\mathcal{PC} = (pc_1, pc_2, \ldots, pc_m)$ that are used to produce each sample in both the MLAR sequence and S/N curve, and the level of injected fluxes used for making $c_+$ class samples (see Sect. \ref{sec:sodinn}).
    For \varSODINN{}, we used the criterion based on the CEVR, as proposed by \citet{Gomez2018Supervised}, to define the $\mathcal{PC}$ list. For NA-\varSODINN{} and the hybrid models, we instead rely on the novel \varPmaps{}s technique, and we choose a list of $m=15$ principal components centred around $k$ (Sect.~\ref{sec:nasodinn_generation}). Regarding the injected fake companion fluxes, we choose for all SODINN-based models a range of fluxes that correspond to an S/N between one and three in the PCA-processed frame with one principal component (Appendix~\ref{sec:appendixB}). This range of fluxes does not generally lead to class overlapping, where $c_{+}$ and $c_{-}$ class samples would look too similar. However, in order to avoid FPs in the final detection map, the user may consider higher flux ranges. Finally, to build the ROC curve, we consider a list of S/N thresholds ranging from 0.1 to 4.5 in steps of 0.5 for annular PCA, while for the SODINN-based models, we use a list of confidence thresholds from 0.09 to 0.99 in steps of 0.1.  All SODINN-based models are trained on balanced training sets containing around $10^{5}$ samples for each class using an NVIDIA GeForce RTX 3070 graphics processing unit (GPU).

    \begin{figure}
        \centering
        \includegraphics[width=0.92\columnwidth]{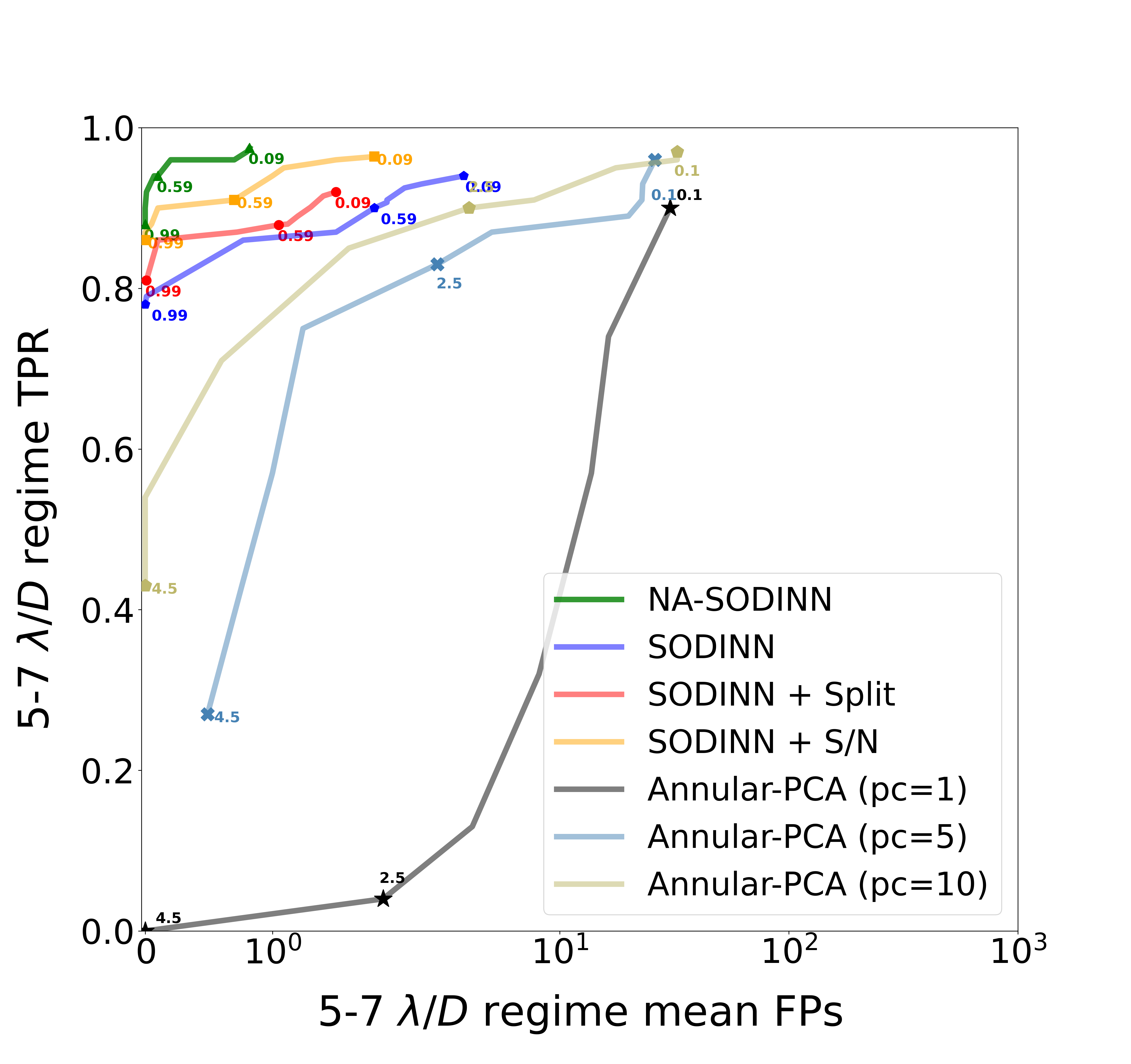}
        \includegraphics[width=0.92\columnwidth]{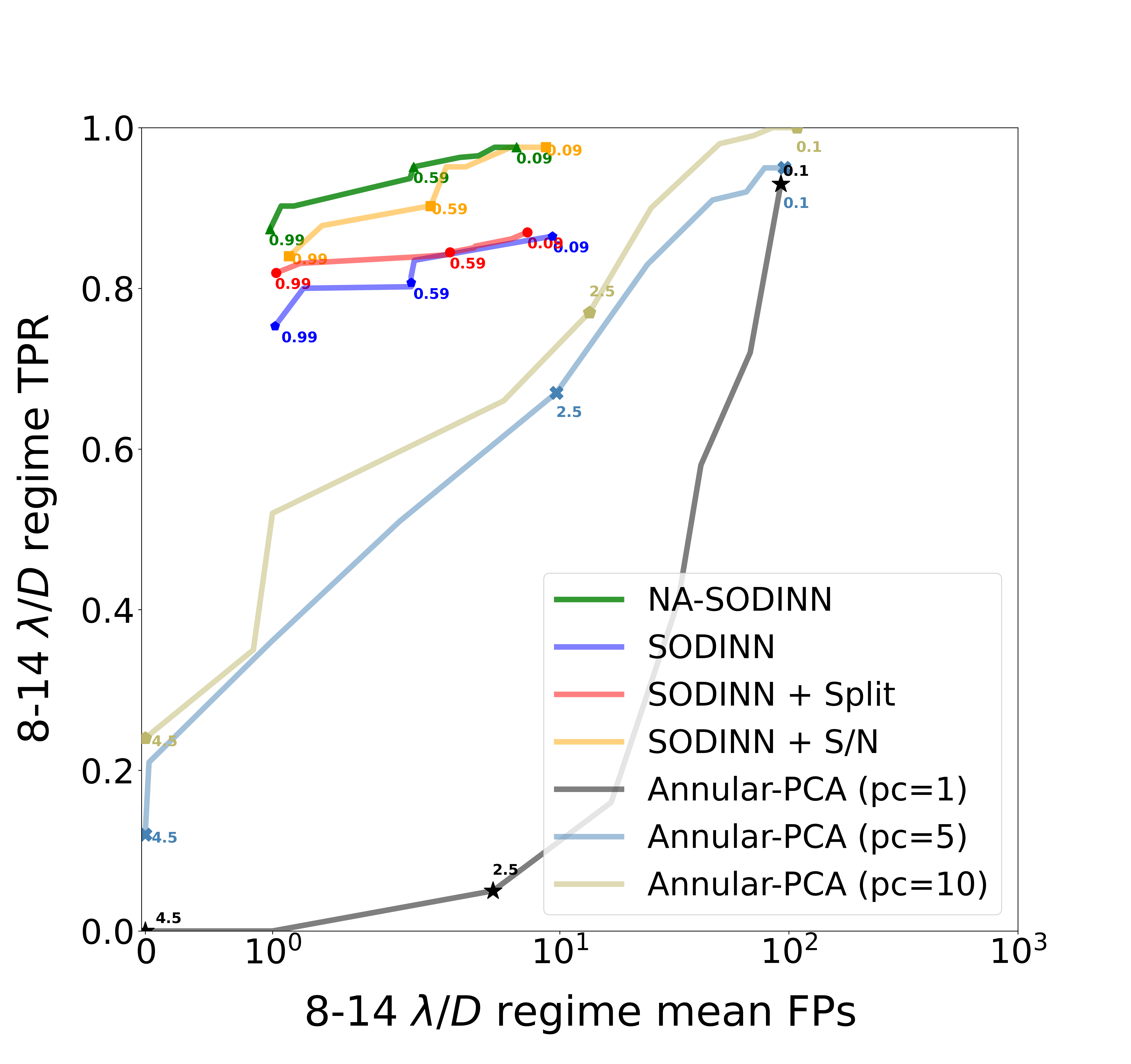}
        \includegraphics[width=0.92\columnwidth]{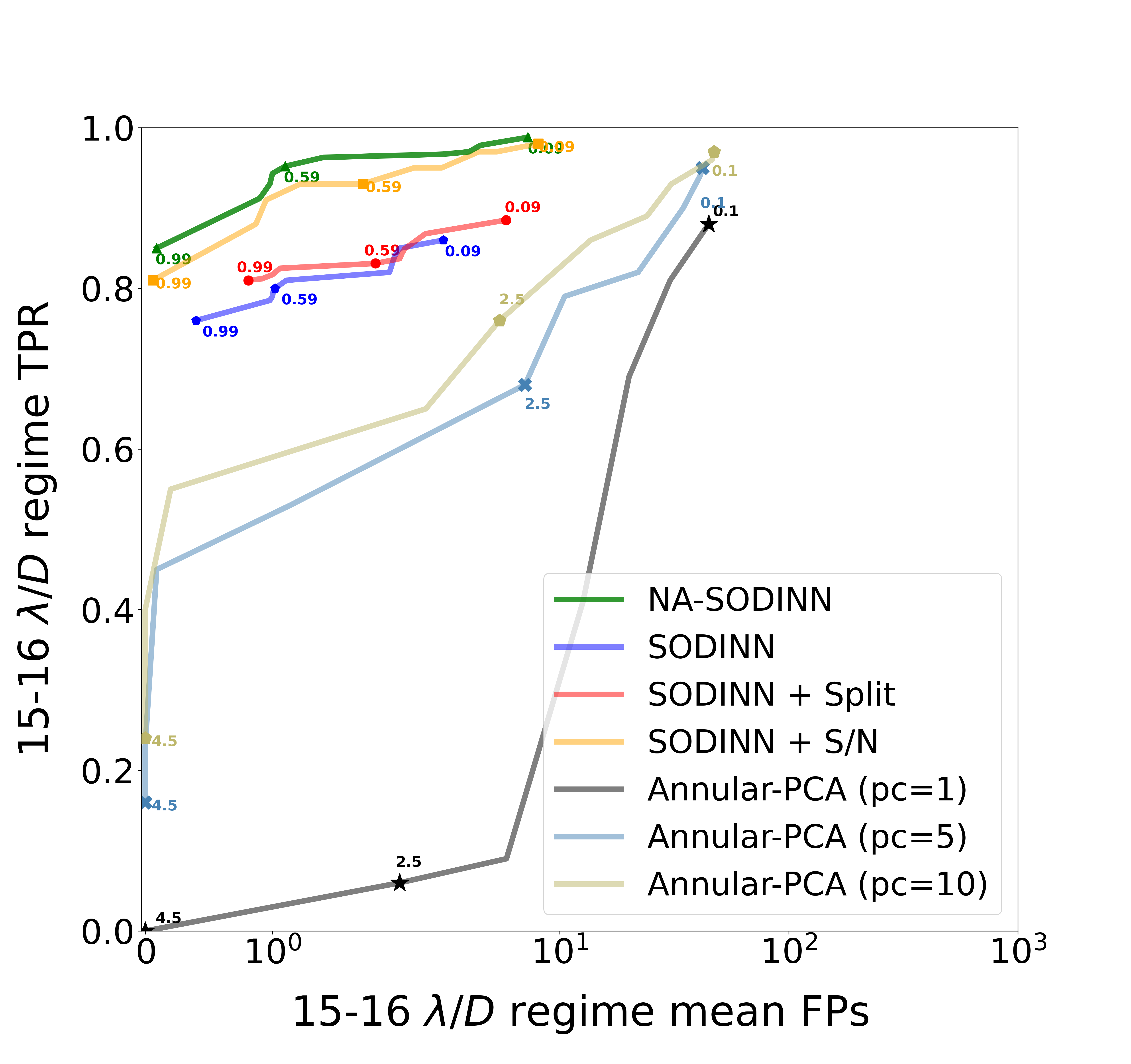}
        \caption{ROC analysis per noise regime for the \textit{sph2} dataset showing the performance of SODINN, NA-\varSODINN{}, annular-PCA, and hybrid SODINN models. The values plotted alongside each curve highlight some of the selected thresholds.}
        \label{fig:roc_curves_sph2}
    \end{figure}
    
    \begin{figure}
        \centering
        \includegraphics[width=0.92\columnwidth]{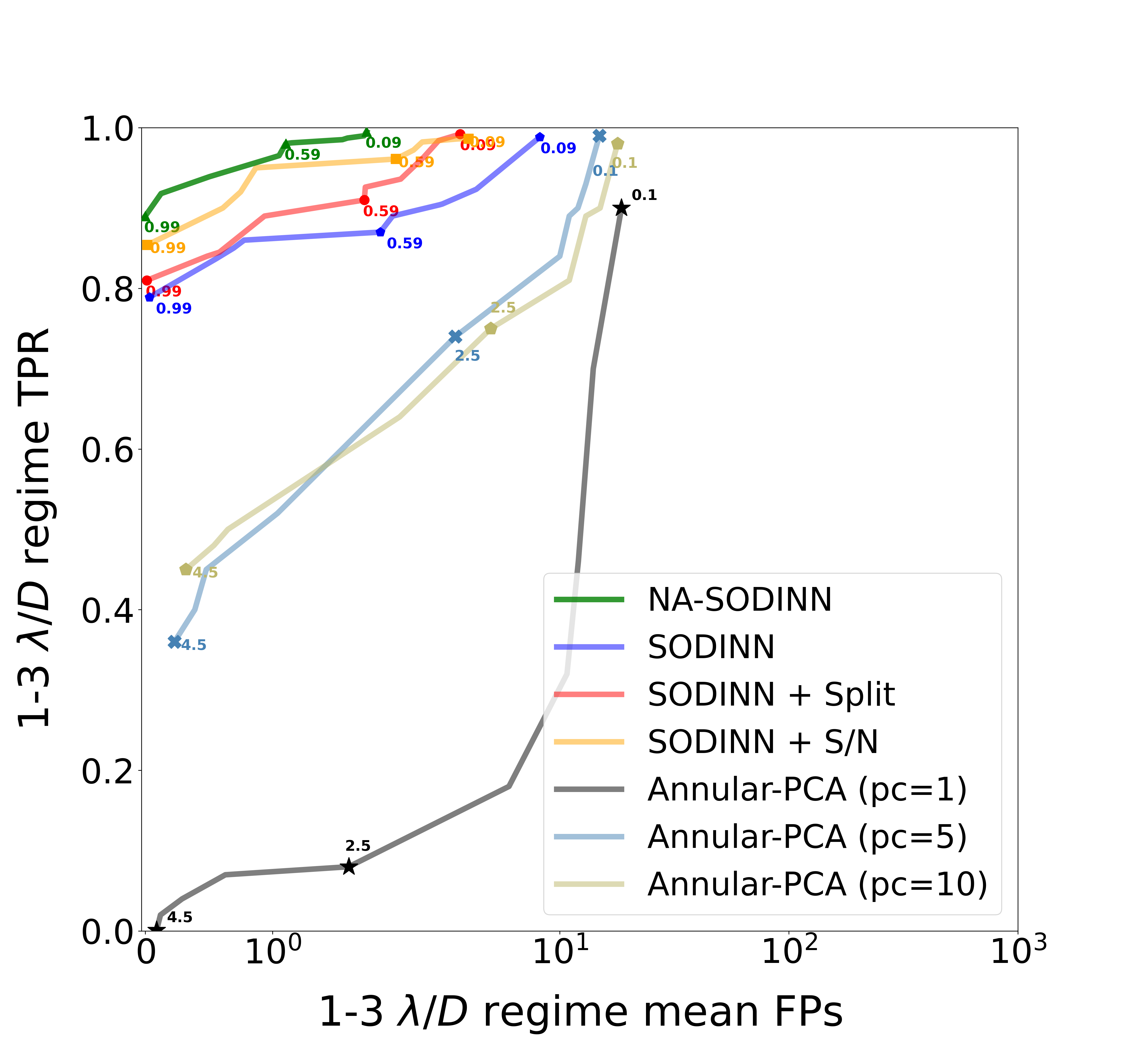}
        \includegraphics[width=0.92\columnwidth]{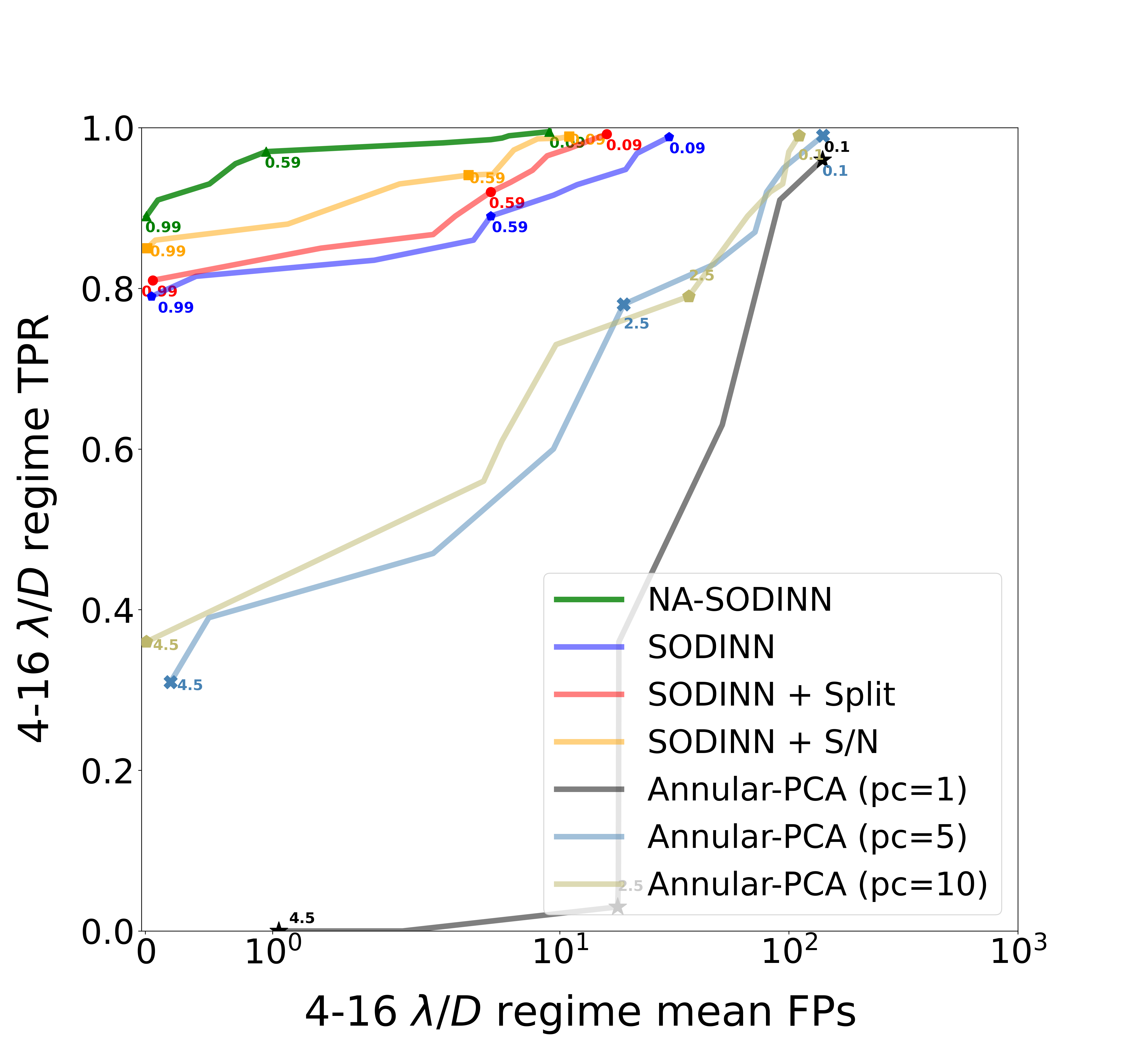}
        \caption{Same as Fig.~\ref{fig:roc_curves_sph2}, but for the \textit{nrc3} dataset.}
        \label{fig:roc_curves_nrc3}
    \end{figure}

    \begin{figure*}
    \begin{center}
    \begin{tabular}{ccc}
      \includegraphics[width=0.31\textwidth]{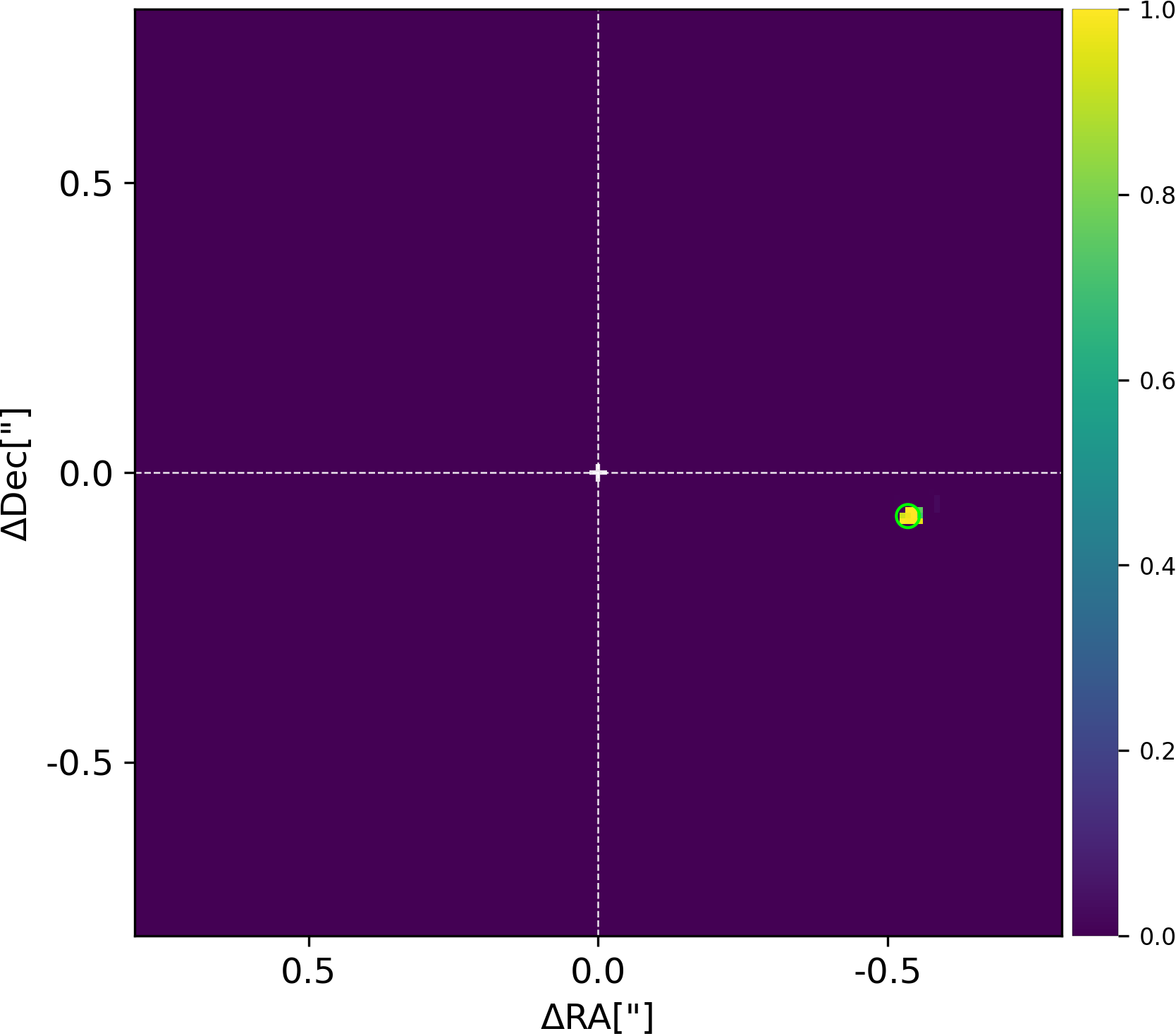} & 
      \includegraphics[width=0.31\textwidth]{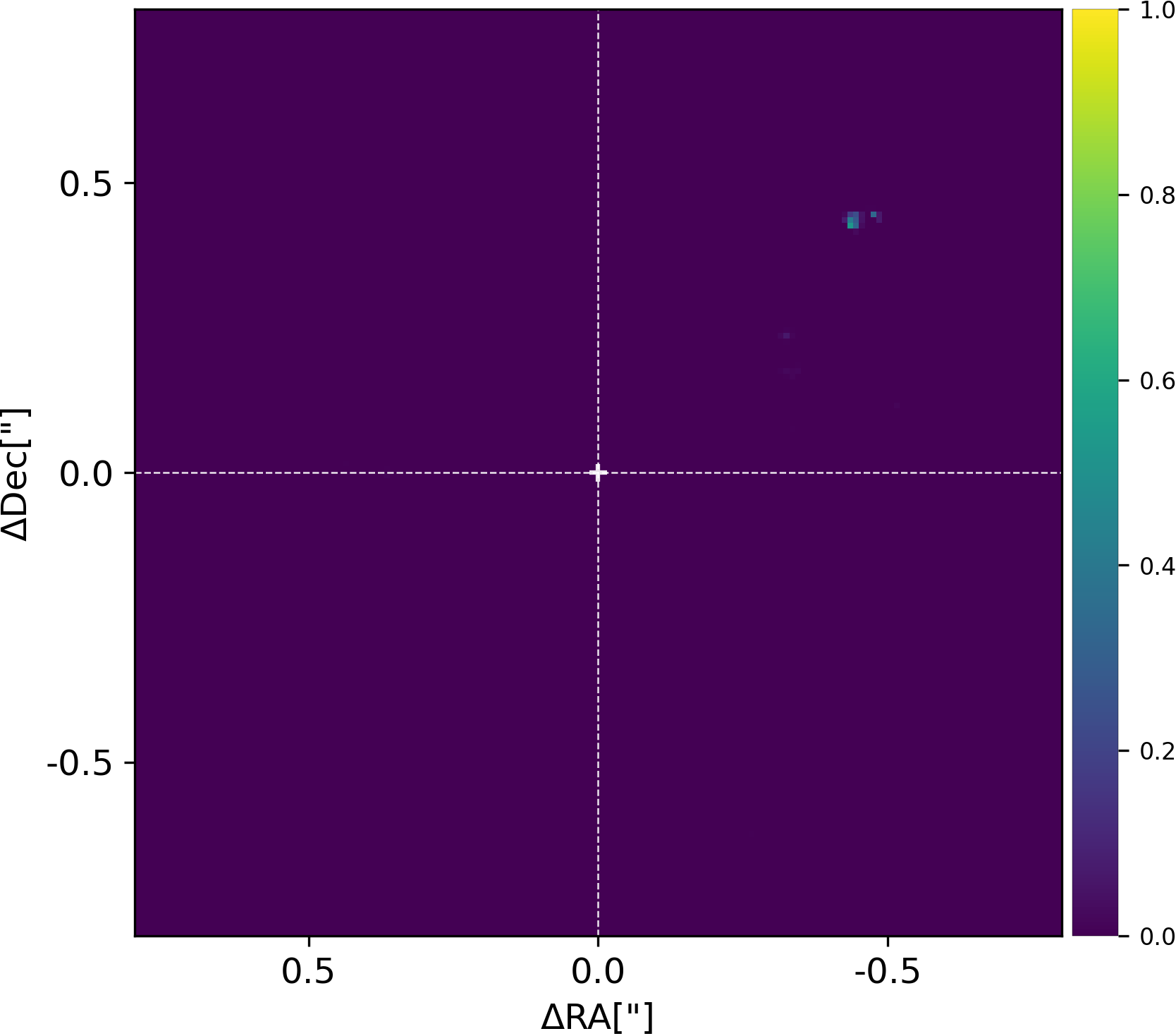} &
      \includegraphics[width=0.31\textwidth]{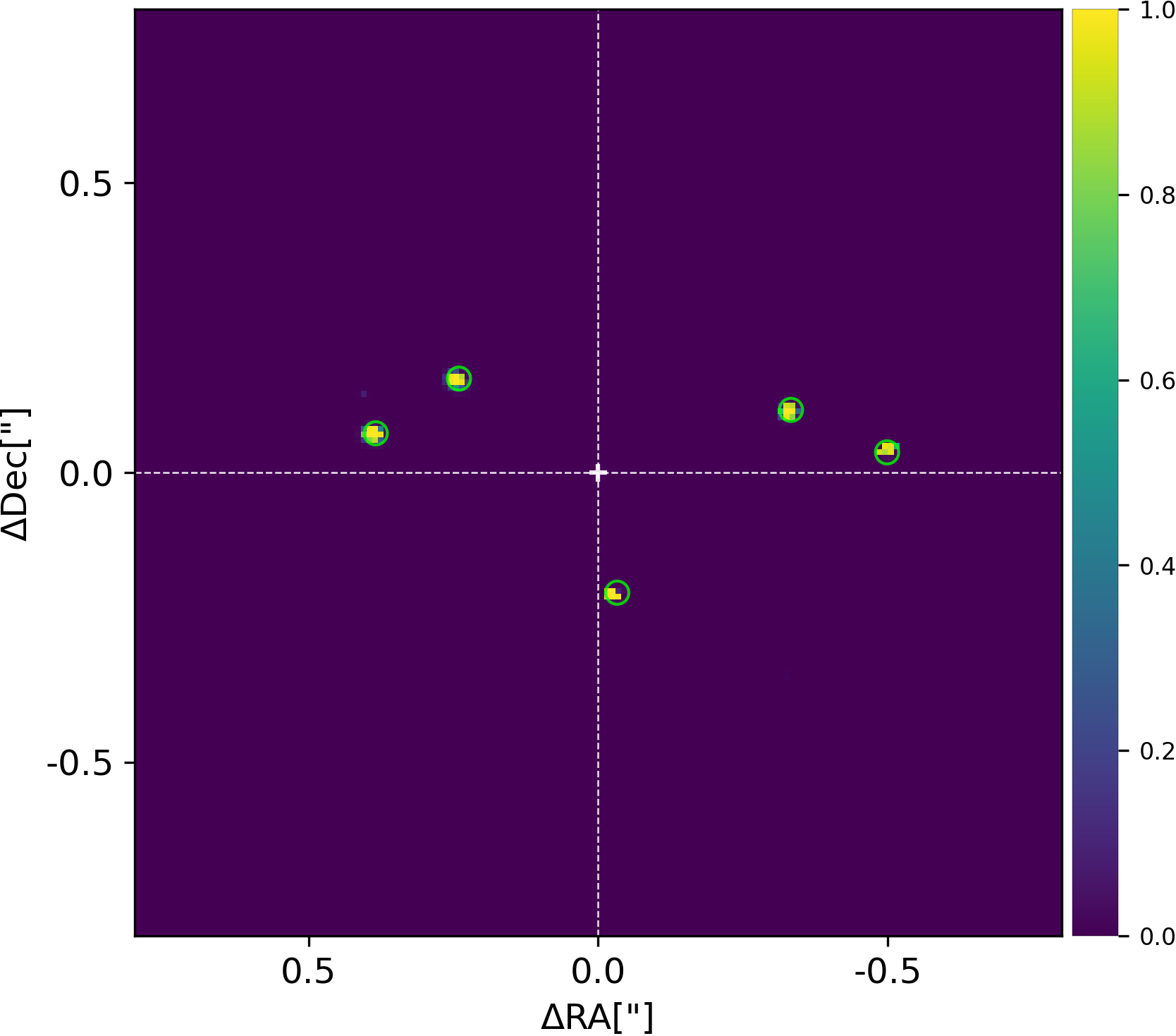} \\
      \textit{sph1} - 1/1 detections & \textit{sph2} - 0/0 detections & \textit{sph3} - 5/5 detections\\[6pt]\\
     \includegraphics[width=0.31\textwidth]{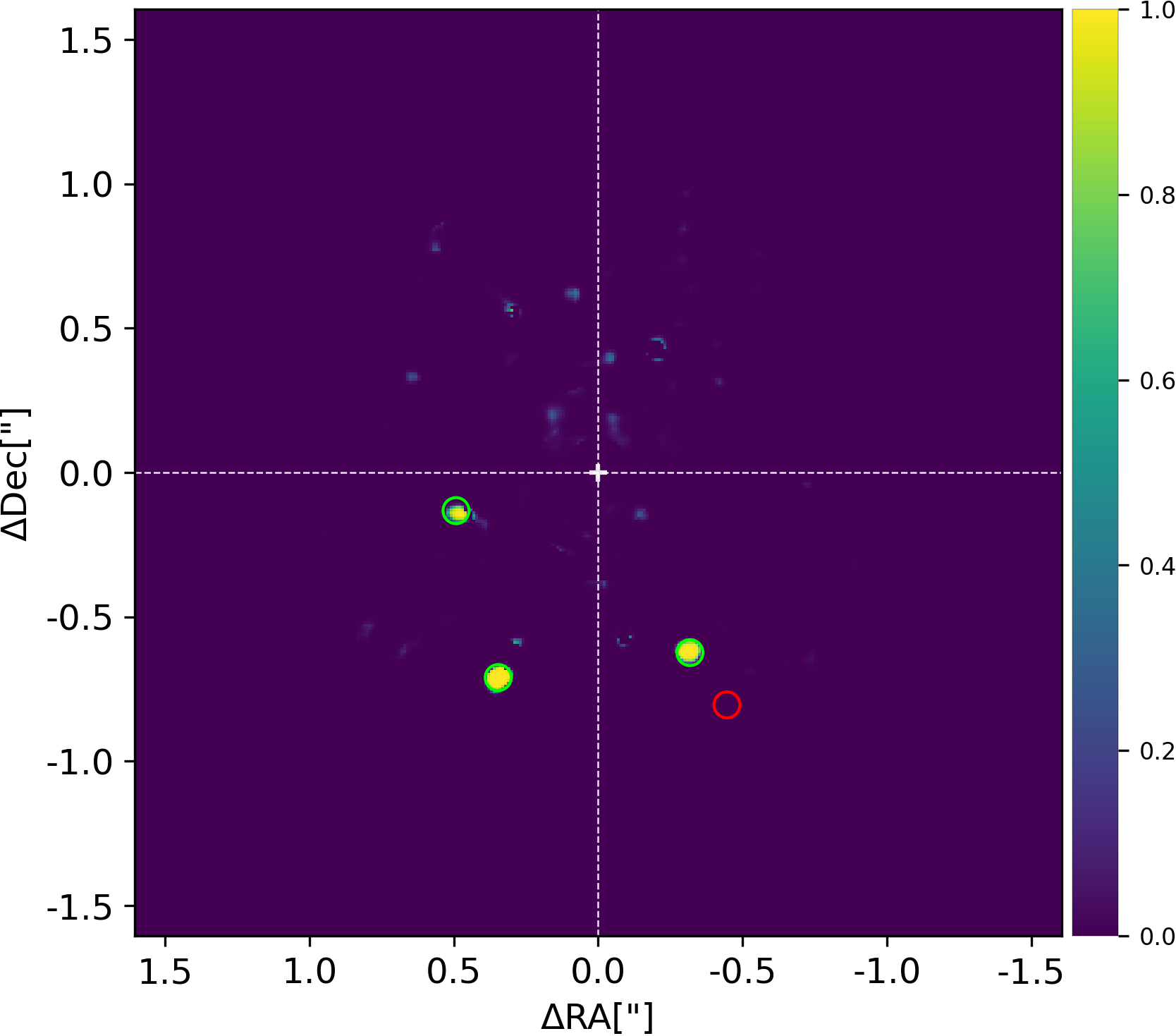} & 
     \includegraphics[width=0.31\textwidth]{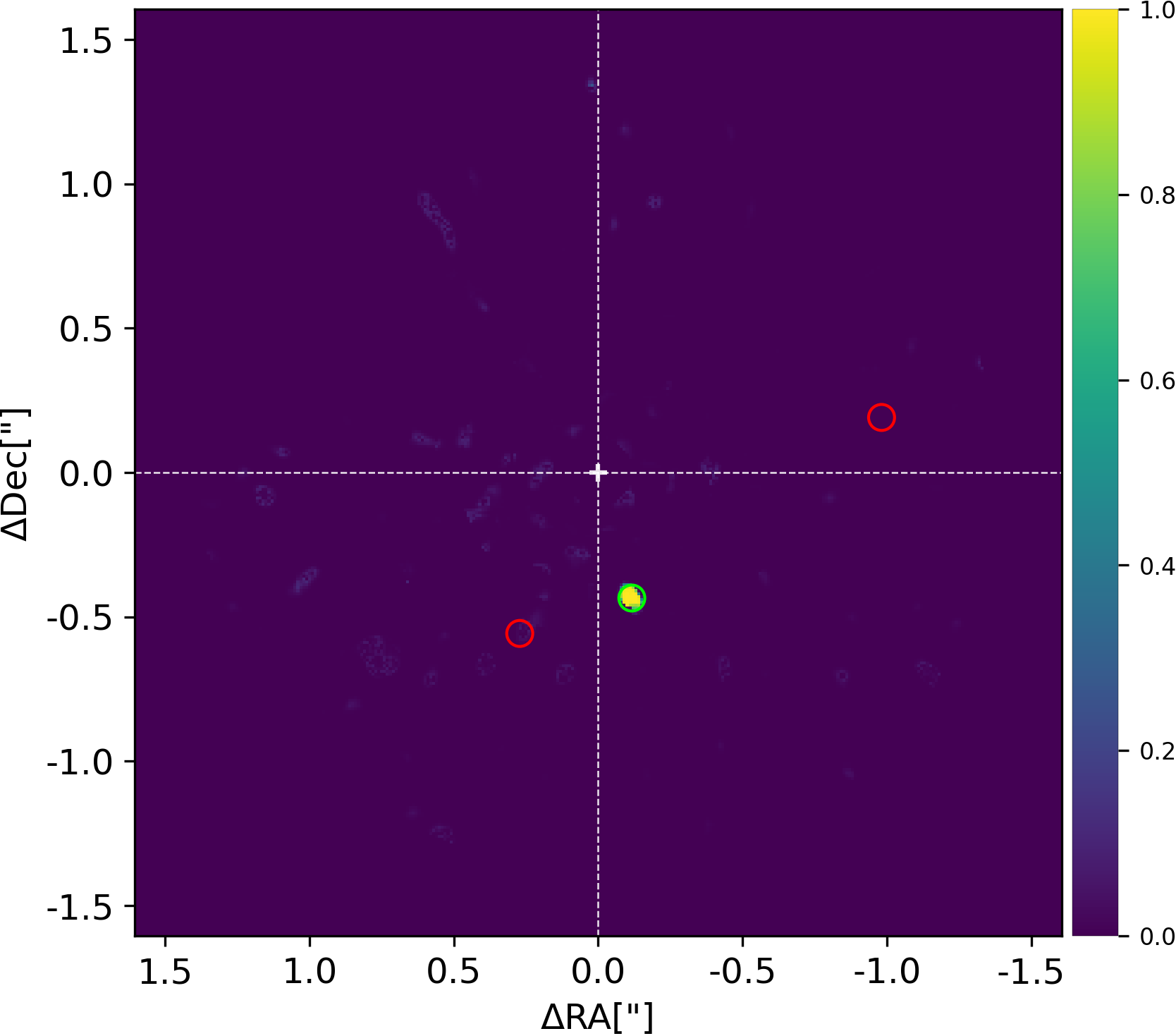} &
     \includegraphics[width=0.31\textwidth]{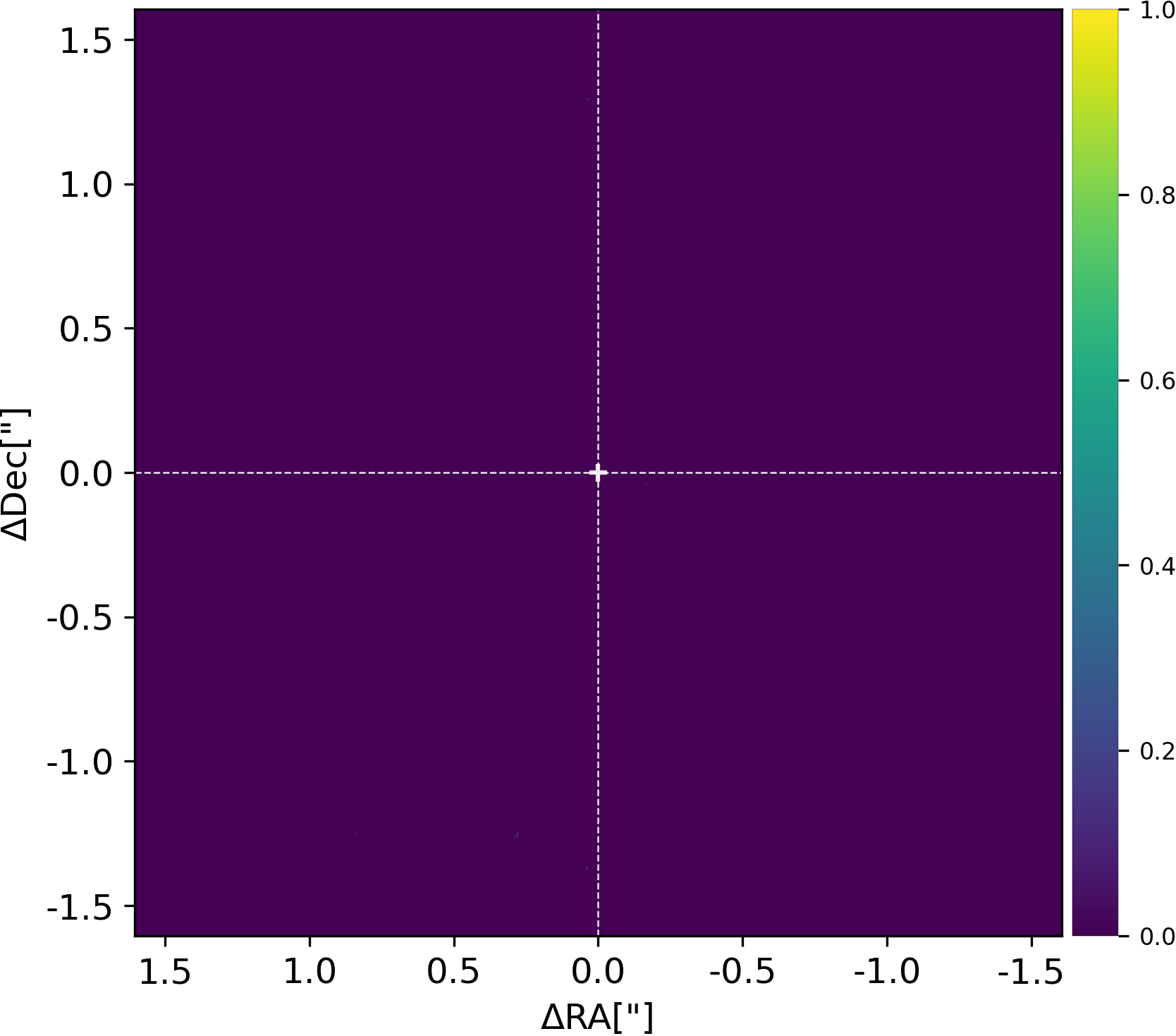} \\
    \textit{nrc1} - 3/4 detections & \textit{nrc2} - 1/3 detections & \textit{nrc3} - 0/0 detections \\[6pt]\\
     \includegraphics[width=0.31\textwidth]{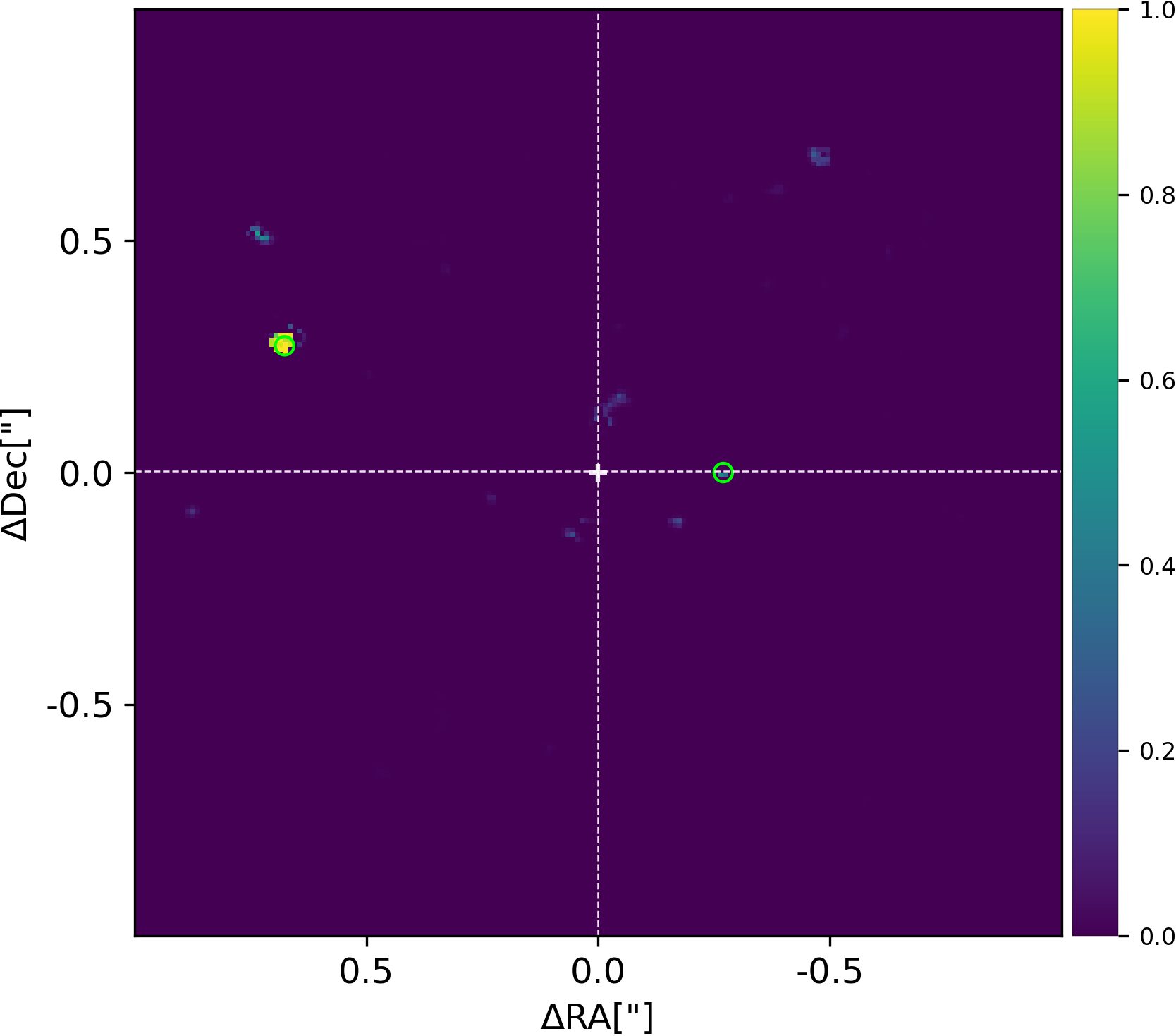} & 
     \includegraphics[width=0.31\textwidth]{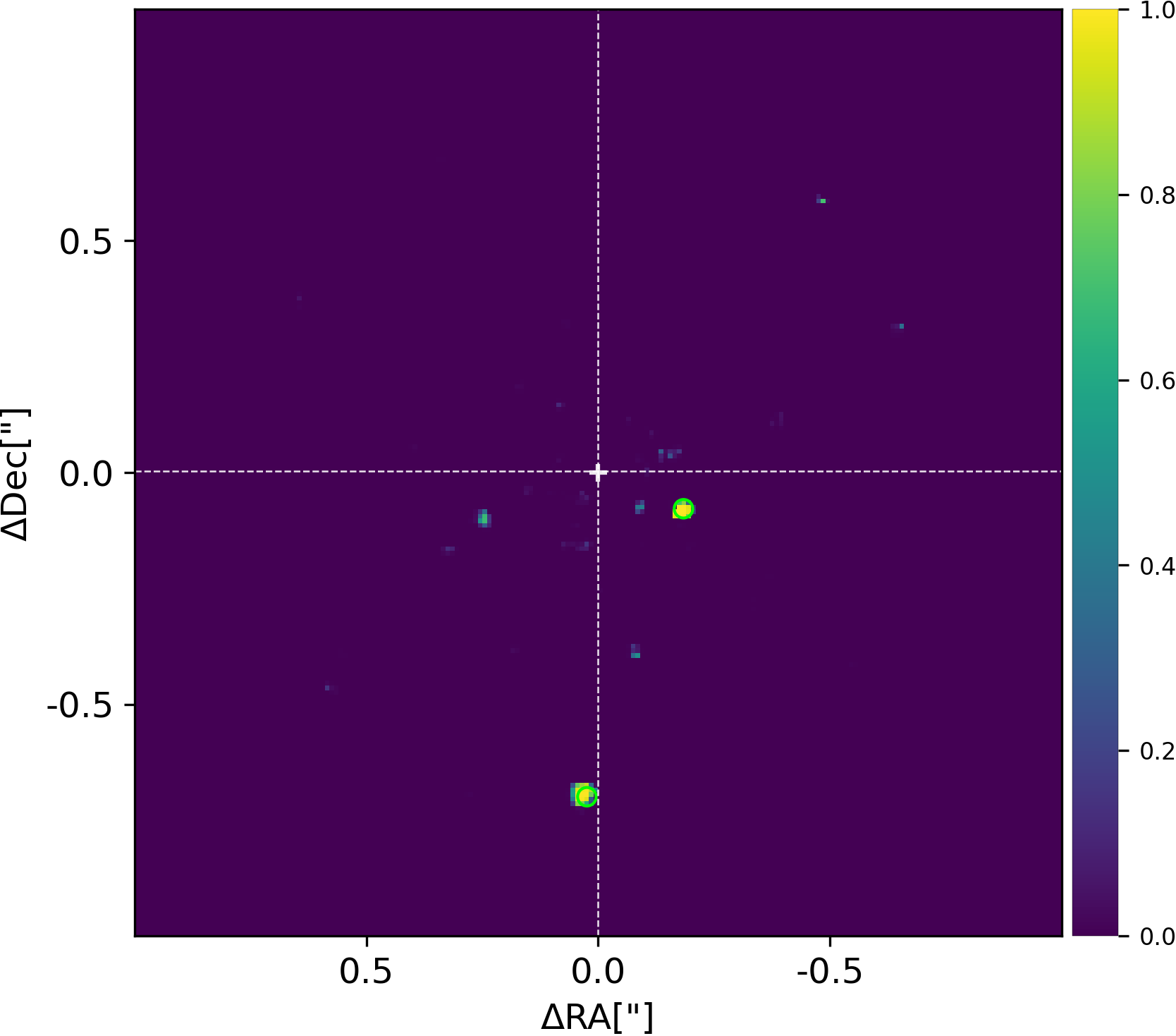} &
     \includegraphics[width=0.31\textwidth]{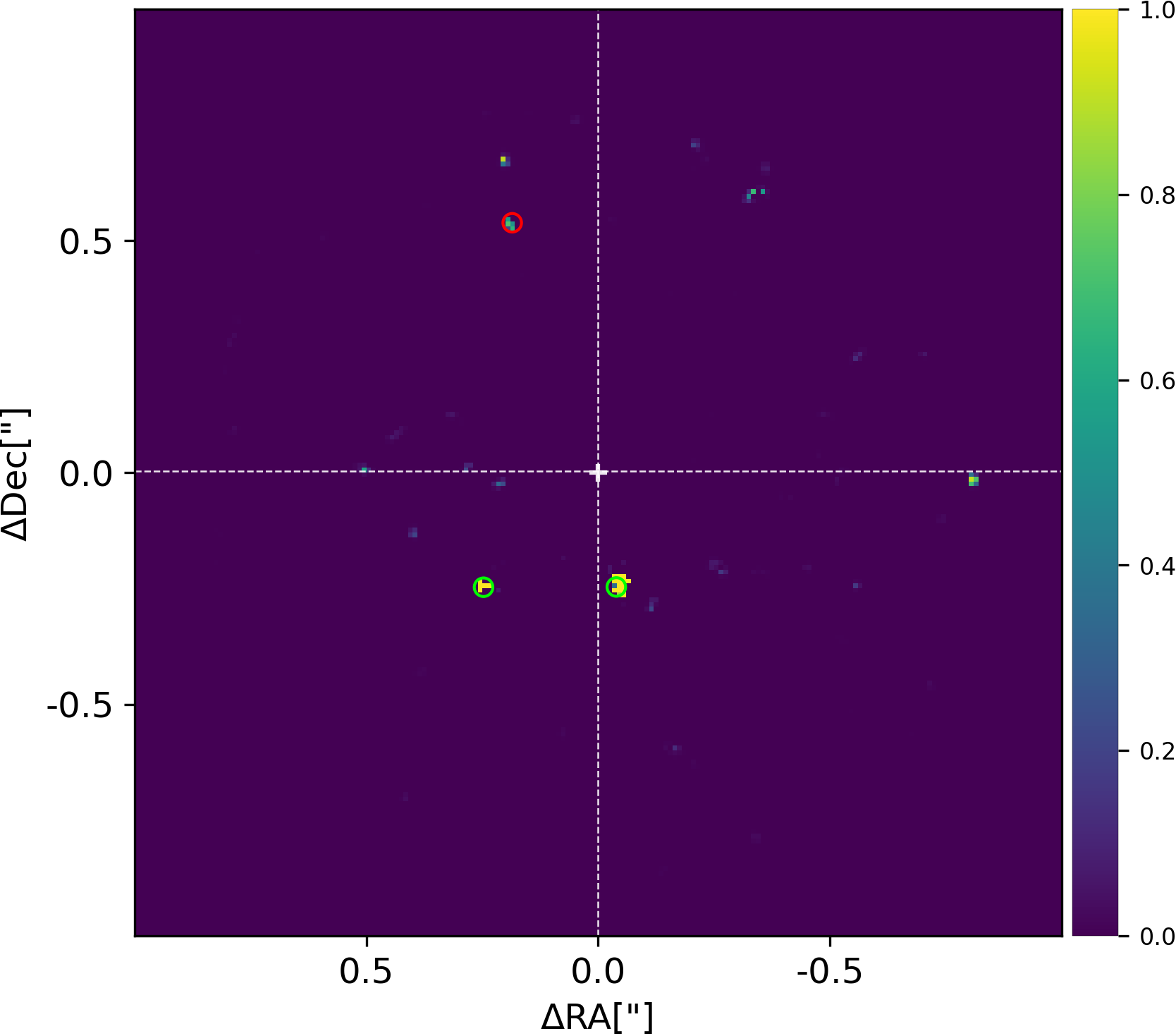} \\
    \textit{lmi1} - 1/2 detections & \textit{lmi2} - 2/2 detections & \textit{lmi3} - 2/3 detections \\[6pt]
    \end{tabular}
    \end{center}
    \caption{NA-\varSODINN{} confidence maps obtained on the whole set of EIDC \varADI{} sequences (Table~\ref{Table:datasets}). For the submitted confidence threshold $\tau=0.90$, we highlight with green circles the correct detection of injected companions (true positives), and with red circles the non-detection of injected companions (false negatives). The circles have a FWHM diameter. No false positive is reported in our maps, as all the remaining non-circled peaks in the confidence maps are below the threshold. \label{fig:eidc_grid}}
    \end{figure*}
    
    Figures~\ref{fig:roc_curves_sph2} and \ref{fig:roc_curves_nrc3} display a series of ROC spaces -- one for each detected noise regime --, respectively for the \textit{sph2} and \textit{nrc3} \varADI{} sequences. For the sake of simplicity, we do not consider the detected regime comprised between 17-19 \varLambdaOverD{} in \textit{sph2} (Fig.~\ref{fig:pvalues_map_sph2}) for this analysis. Each of these ROC spaces displays one ROC curve per algorithm, which informs about its detection performance on that specific noise regime for different thresholds. We observe from both figures that NA-\varSODINN{} outperforms its predecessor, the hybrid models, and the annular-PCA technique for each noise regime. This behaviour is further illustrated in Appendix~\ref{sec:appendixC}, with Figs.~\ref{fig:maps_sph2_5to7},\ref{fig:maps_sph2_8to14} and \ref{fig:maps_sph2_15to16} for the case of \textit{sph2}, and Figs.~\ref{fig:maps_nrc3_1to3},\ref{fig:maps_nrc3_4to16} for \textit{nrc3}, where the confidence maps from each algorithm are compared at different threshold levels. Regarding hybrid models, we generally observe that they land between the \varSODINN{} and NA-\varSODINN{} detection performance, with \varSODINN{}+S/N generally being the best hybrid model. It can also be observed that annular-PCA with PC=5 and PC=10 perform better than with PC=1 for all regimes. We associate this behaviour to the fact that for PC=5 and PC=10, we are closer to the principal component $k$  where the S/N is maximized, and therefore, the star-planet contrast is improved. 
    
    Based on these results and our experiments, we observe a general trend for both approaches separately. While splitting the field of view in noise regimes tends to reduce the number of false positives, especially when residual speckle noise is significant, adding an S/N curve for each MLAR sequence tends to enhance the algorithm's sensitivity to detect signals. These findings imply that both techniques, when combined in the neural network, considerably improve the \varSODINN{} detection performance.

    \subsection{NA-\varSODINN{} in EIDC}
    
    By design, the Exoplanet Imaging Data Challenge  \citep[EIDC,][]{Cantalloube2020Exoplanet}  can be used as a laboratory to compare and evaluate new detection algorithms against other state-of-the-art HCI detection algorithms. For instance, \cite{Dahlqvist2021Auto-RSM} used the EIDC to highlight the improvement of the automated version of their RSM algorithm. Here, we use the first sub-challenge of the EIDC to generalize the ROC analysis presented above, and evaluate how NA-\varSODINN{} performs with respect to the state-of-the-art HCI algorithms that entered the data challenge. Besides the $sph2$ and $nrc3$ datasets used so far, the first EIDC sub-challenge includes seven additional \varADI{} sequences in which a total of 20 planetary signals with different contrasts and position coordinates were injected. Two of these seven \varADI{} sequences are from the SPHERE instrument \citep{Beuzit2019SPHERE}, identified as \textit{sph1} and \textit{sph3}, two more from the NIRC-2 instrument \citep{Serabyn2017TheWW}, identified as \textit{nrc1} and \textit{nrc3}, and the remaining three from the LMIRCam instrument \citep{Skrutskie2010TheLarge}, with \textit{lmr1}, \textit{lmr2} and \textit{lmr3} ID names. For each of these nine datasets, EIDC provides a pre-processed temporal cube of images, the parallactic angles variation corrected from true north, a non-coronagraphic PSF of the instrument, and the pixel-scale of the detector. Each algorithm entering the EIDC had to provide a detection map for each ADI sequence. The following standard metrics are then used to assess the detection performance of each submitted detection map:
    \begin{itemize}
        \item True Positive Rate: $\mbox{TPR} = \frac{TP}{TP+FN}$,
        \item False Positive Rate: $\mbox{FPR} = \frac{FP}{FP+TN}$,
        \item False Discovery Rate: $\mbox{FDR} = \frac{FP}{FP+TP}$,
        \item F1-score: $\mbox{F1} = \frac{2 \cdot TP}{2 \cdot TP+FP+FN}$.
    \end{itemize}

    \begin{figure}
        \centering
        \includegraphics[width=\columnwidth]{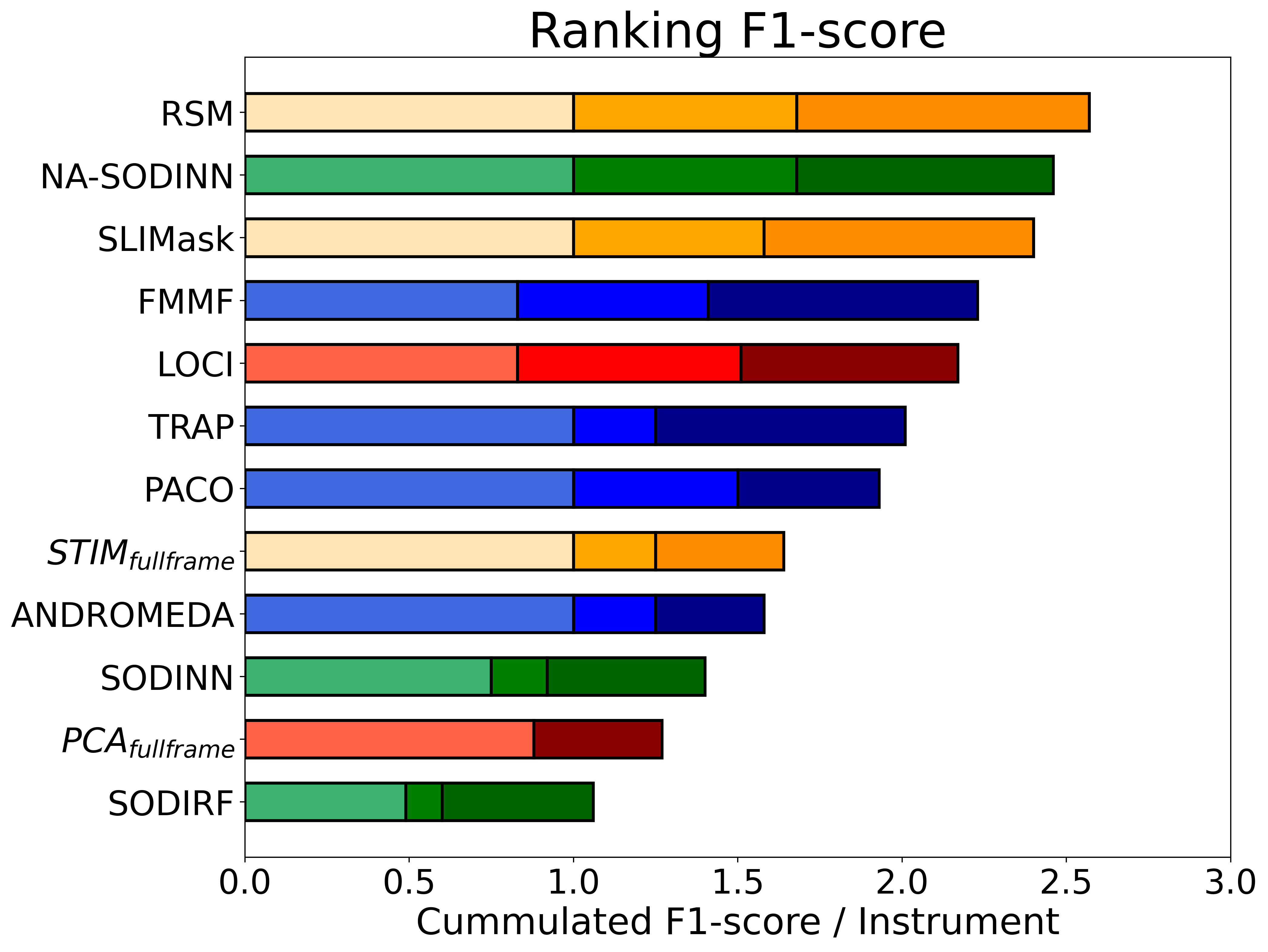}\vspace{0.4cm}
        \includegraphics[width=\columnwidth]{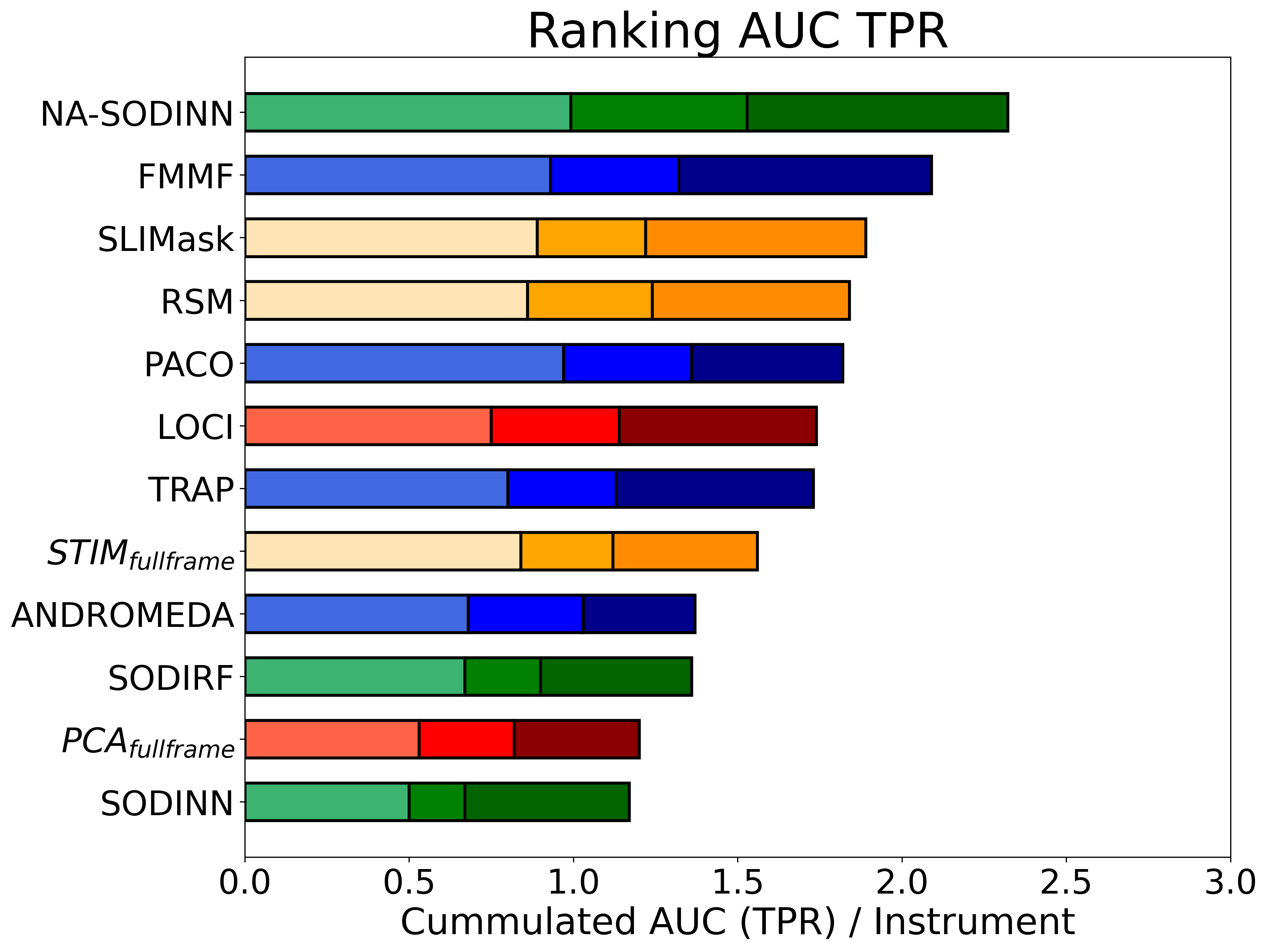}\vspace{0.4cm}
        \includegraphics[width=\columnwidth]{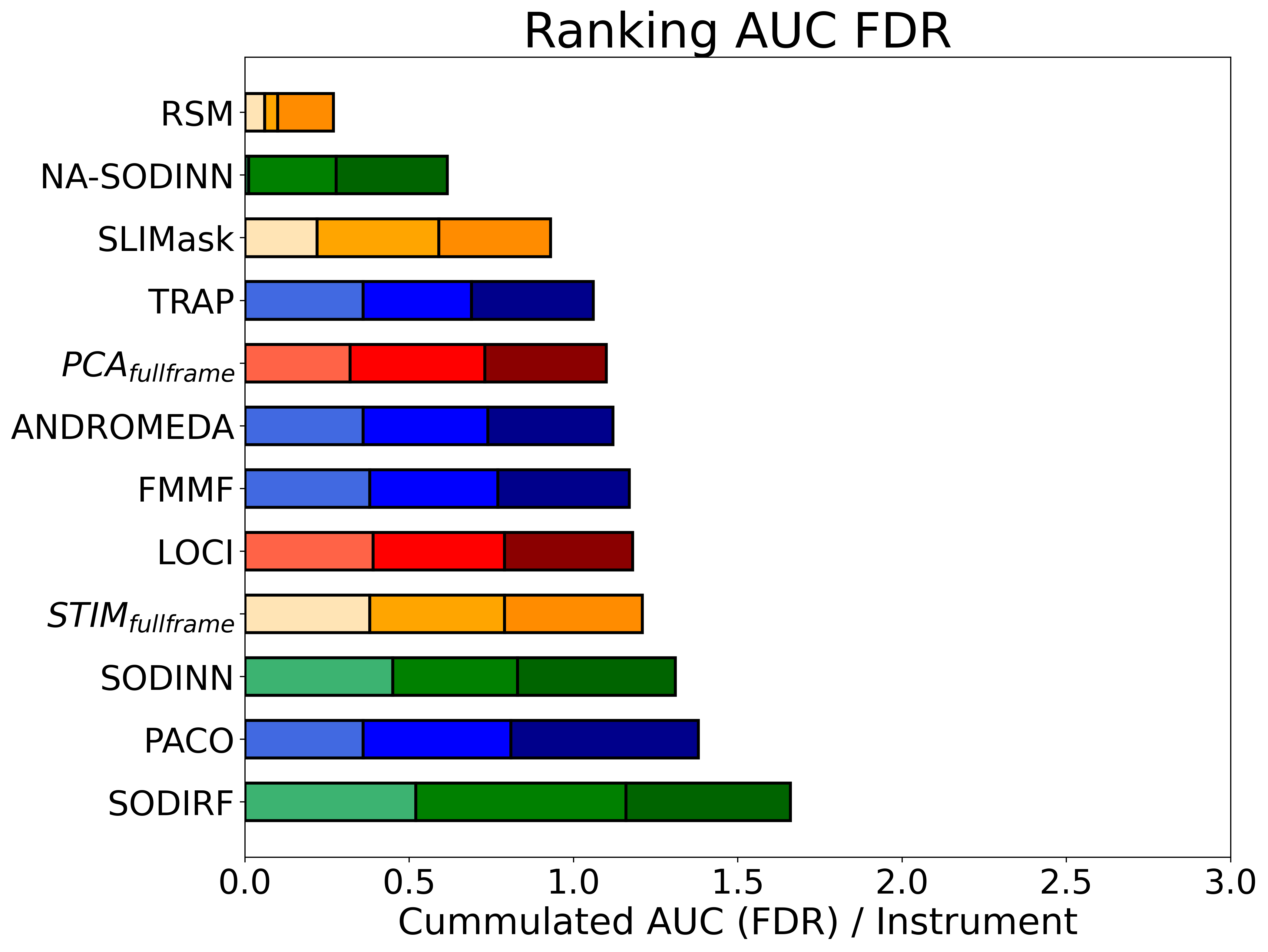}
        \caption{Updated EIDC leaderboard after the NA-\varSODINN{} submision. Ranking based on the F1-score (on top), the AUC of the TPR (in the middle) and the AUC of the FDR (on bottom). Colours refer to HCI detection algorithm families: PSF-based subtraction techniques providing residual maps (red) or detection maps (orange), inverse problems (blue) and supervised machine learning (green). The light, medium and dark tonalities correspond to SPHERE, NIRC-2, and LMIRCam datasets, respectively.}
        \label{fig:ranking}
    \end{figure}
    
    We apply our NA-\varSODINN{} framework to the EIDC, and as in the ROC analysis, we use \varPmaps{}s as a tool for both estimating residual noise regimes and choosing the list of principal components $\mathcal{PC}$ at each angular separation. For the injection flux ranges, we use an S/N range between one and four times the level of noise in the processed frame. Each model is trained with balanced training sets that contain around $10^{5}$ samples per class. Because all three LMIRCam cubes contain more than $3{,}000$ frames (Table~\ref{Table:datasets}), we decided to reduce this number to around 250-300 frames to limit the computational time. To do that, we average a certain number of consecutive frames along the time axis in the sequence. Figure~\ref{fig:eidc_grid} shows a grid of all the  resulting NA-\varSODINN{} confidence maps for the EIDC ADI sequences where we observe, by visual inspection, that NA-\varSODINN{} finds most of the injected fake companions, while producing only faint false positives that all fall below our default detection threshold $\tau=0.9$. In order to quantify this information, we follow the same approach as in \cite{Cantalloube2020Exoplanet} by considering the area under the curve (AUC) for the TPR, FPR, and FDR as a function of the threshold, which allows mitigating the arbitrariness of the threshold selection by considering their evolution for a pre-defined range. The $\rm AUC_{TPR}$ should be as close as possible to one, and the $\rm AUC_{FPR}$ and $\rm AUC_{FDR}$ as close as possible to zero. The F1-score ranges between zero and one, where one corresponds to a perfect algorithm, and is computed only on a single threshold $\tau_{sub}$ that is chosen by the participant. 
    
    Figure~\ref{fig:eidc_grid_metrics} shows the result of this analysis for all NA-\varSODINN{} confidence maps of Fig.~\ref{fig:eidc_grid}, in which all TPR, FPR, and FDR metrics (and their respective AUCs) are computed for different confidence threshold values ranging from zero to one. Here, we mainly see that the $\rm AUC_{FDR}$ is generally higher along the range of thresholds for NIRC-2 and LMIRCam than for SPHERE datasets, the $\rm AUC_{FPR}$ is close to zero for all datasets, and the $\rm AUC_{TPR}$ is almost perfect for SPHERE datasets. To compute the F1-score, we choose a $\tau_{sub}=0.9$ confidence threshold. From our test with NA-\varSODINN{}, we consider this value as the minimum confidence threshold for which one can rely on the significance of detections, maximizing TPs while minimizing FPs. Thus, any pixel signal above this $\tau_{sub}$ on each confidence map of Fig.~\ref{fig:eidc_grid} is considered as a detection for the computation of the F1-score. Finally, through the $\rm AUC_{TPR}$, $\rm AUC_{FDR}$ and F1-score metrics obtained with the NA-\varSODINN{} algorithm, we are able to update the general EIDC leaderboard \citep{Cantalloube2020Exoplanet}. Figure~\ref{fig:ranking} shows how NA-\varSODINN{} ranks compared to the algorithms originally submitted to the EIDC, for each considered metric. We clearly observe that NA-\varSODINN{} ranks at the top, or close to the top, for each of the EIDC metrics, with results generally on par with the RSM algorithm by \cite{Dahlqvist2020Regime}. In particular, NA-SODINN provides the highest area under the true positive curve, while preserving a low false discovery rate.

    \section{Conclusions} \label{sec:conclusions}
    In this paper, we explore the possibility of enhancing exoplanet detection in the field of HCI by training a supervised classification model that takes into account the noise structure in the PCA-processed frame.  \varSODINN{} \citep{Gomez2018Supervised}, a pioneering deep-learning detection algorithm in HCI, has been adapted to learn from different noise regimes in the processed frame and local discriminators between the exoplanet and noise, such as S/N curves. With these two approaches working in synergy, we built a new detection algorithm, NA-\varSODINN. Although our findings related to the spatial structure of noise distributions are showcased by adapting the \varSODINN{} detection framework, we believe that other algorithms dealing with processed frames could be adapted similarly.

    The NA-\varSODINN{} detection capabilities were tested through two distinct analyses. First, we performed a performance assessment based on ROC curves using two \varADI{} sequences provided by the VLT/SPHERE and Keck/NIRC-2 instruments. Here, NA-\varSODINN{} is evaluated with respect to annular-PCA, the original SODINN, and two SODINN-based hybrid models that use only one of the two proposed approaches, that is, the noise regime splitting or the S/N curves' addition. We found that hybrid models improve the detection performance of SODINN in all noise regimes, which demonstrates the interest of the local noise approaches considered in this paper. Moreover, we found that NA-\varSODINN{} reaches an even higher detection performance, especially in the speckle noise regime, by combining both approaches in the same framework. Then, in order to benchmark NA-\varSODINN{} against other state-of-the-art HCI algorithms, we applied NA-\varSODINN{} to the first phase of EIDC \citep{Cantalloube2020Exoplanet}, a community-wide effort meant to offer a platform for a fair and common comparison of exoplanetary detection algorithms. In this analysis, we observed that NA-\varSODINN{} is ranked at the top (first or second position) of the challenge leaderboard for all considered evaluation metrics, providing in particular the highest true positive rate among all entries, while still keeping a low false discovery rate.   
    
    We identified some limitations that could be addressed in future work to improve the effectiveness and practicality of our NA-\varSODINN{} method. While the algorithm currently performs well in noise regimes over PCA-processed frames, it relies on previous noise analyses to define these regime boundaries, limiting its independence. Future avenues would include modifying the network architecture to enable the identification of noise regimes during training, which could enhance the detection performance. Another limitation of our approach is the challenge of setting an appropriate detection threshold in the final detection map. This is typically based on the presence of obvious false positives, which may affect the application of NA-\varSODINN{} in certain contexts. However, this limitation can be mitigated by using dedicated metrics such as ROC space to assess detection performance. We also note that NA-\varSODINN{} and its predecessor rely on data augmentation techniques to generate a diverse training set. To supplement these techniques, we suggest exploring generative neural networks to train more robust supervised models that can generalize better. Lastly, extending the application of NA-\varSODINN{} to work on other observing strategies and detect extended sources such as protoplanetary disks would be a valuable avenue to increase the flexibility of the algorithm.
    
    The NA-\varSODINN{} framework represents a significant step forward in the search for new and unconfirmed worlds in individual datasets and large surveys using ADI-based techniques. This framework offers greater accuracy in identifying exoplanets across all angular separations, making it particularly well suited for improving our understanding of the demographics of directly imaged exoplanets.

    \begin{acknowledgements}
        The authors would like to thank the python open-source scientific community, and in particular the developers of the Keras deep learning library \citep{tensorflow2015-whitepaper} and the VIP high-contrast imaging package \citep{Gomez2017VIP, Christiaens2023VIP}. The authors acknowledge stimulating discussions with Faustine Cantalloube, Rakesh Nath, Markus Bonse, and Emily O.~Garvin, as well as the whole Exoplanet Imaging Data Challenge team. This project has received funding from the European Research Council (ERC) under the European Union's Horizon 2020 research and innovation programme (grant agreement No 819155), and from the Wallonia-Brussels Federation (grant for Concerted Research Actions).

    \end{acknowledgements}

    \bibliographystyle{aa}
    \bibliography{references}

    \begin{appendix}
        \section{EIDC datasets} \label{sec:appendixA}
            
            \begin{table}[h]
            \centering
                    \caption{Features of the nine \varADI{} sequences from EIDC: The number of frames in the sequence ($N_{t}$), the frame size ($N_{\rm img}$),  the wavelength ($\lambda_{obs}$), and the field rotation ($\Delta_{rot}$).  \label{Table:datasets}}
            \setlength{\tabcolsep}{3pt}
            \resizebox{\columnwidth}{!}{
                    \begin{tabular}{cccccccc} 
                        \hline \hline
                        ID & Telescope/Instr. & FWHM & $N_{t}$ & $N_{\rm img}$  & $\lambda_{obs}$  & $\Delta_{rot}$  & Inj. \\
                        & & [px] & & [px$\times$px] & [$\mu$m] & [º] \\
                        \hline
                        sph1 & VLT/SPHERE     & 4    & 252     & 160$\times$160 & 1.625 ± 0.29 & 40.3 & 1        \\ 
                        sph2 & VLT/SPHERE     & 4    & 80      & 160$\times$160 & 1.593 ± 0.052 & 31.5 & 0         \\
                        sph3 & VLT/SPHERE     & 4    & 228     & 160$\times$160 & 1.593 ± 0.052 & 80.5 & 5         \\
                        nrc1 & Keck/NIRC-2     & 9    & 29      & 321$\times$321 & 3.776 ± 0.70 & 53.0 & 3         \\ 
                        nrc2 & Keck/NIRC-2     & 9    & 40      & 321$\times$321 & 3.776 ± 0.70 & 37.3 & 4        \\ 
                        nrc3 & Keck/NIRC-2     & 9    & 50      & 321$\times$321 & 3.776 ± 0.70 & 166.9 & 0        \\ 
                        lmr1 & LBT/LMIRCAM    & 5    & 4838    & 200$\times$200 & 3.780 ± 0.10 & 153.4 & 2         \\ 
                        lmr2 & LBT/LMIRCAM    & 4    & 3219    & 200$\times$200 & 3.780 ± 0.10 & 60.6 & 2          \\ 
                        lmr3 & LBT/LMIRCAM    & 4    & 4620    & 200$\times$200 & 3.780 ± 0.10 & 91.0 & 3         \\ \hline
                    \end{tabular}
            }
            \end{table}

            \FloatBarrier
        
        \section{Injection fluxes estimation \label{sec:appendixB}}
        
            In HCI, a planetary injection is defined as the process of pasting the AO-corrected instrumental PSF (centred, cropped, and normalized) to every frame in the image sequence at specific coordinates $(r, \theta)$ following field rotation. To control the flux of this injection, the standard procedure is to multiply the normalized PSF by a flux scale factor $\alpha$. Estimating an injection flux range that corresponds to a given S/N range in the post-processed frame implies estimating its respective flux scale factor range $\alpha_{R}=[\alpha_{min}, \alpha_{max}]$. Given a desired S/N range and an angular separation, $\alpha_{min}$ and $\alpha_{max}$ are estimated through the following data-driven procedure: 
            \begin{enumerate}
                \item inject a companion in the raw image sequence at random coordinates $(r, \theta)$ within the annuli and with a random scale factor $\alpha$;
            
                \item compute the ADI-PCA processed frame for this synthetic image sequence using one single principal component in the PCA approximation of the speckle field;
            
                \item apply Eq.~\ref{eq:snr} on the processed frame at the injection coordinates $(r, \theta)$, retrieving the companion S/N value;
            
                \item repeat 1-3 steps $N_{\mathrm{inj}}$ times;
            
                \item plot all S/N values retrieved from all $N_{inj}$ injections of step 4 as a function of their corresponding scale factor;
            
                \item linearly fit the data plotted in step 5, and define $\alpha_{min}$ and $\alpha_{max}$ as the intersection between the linear fit and the corresponding S/N range boundaries.
            \end{enumerate}
            
\begin{figure*}
                \centering
                \includegraphics[width=0.48\textwidth]{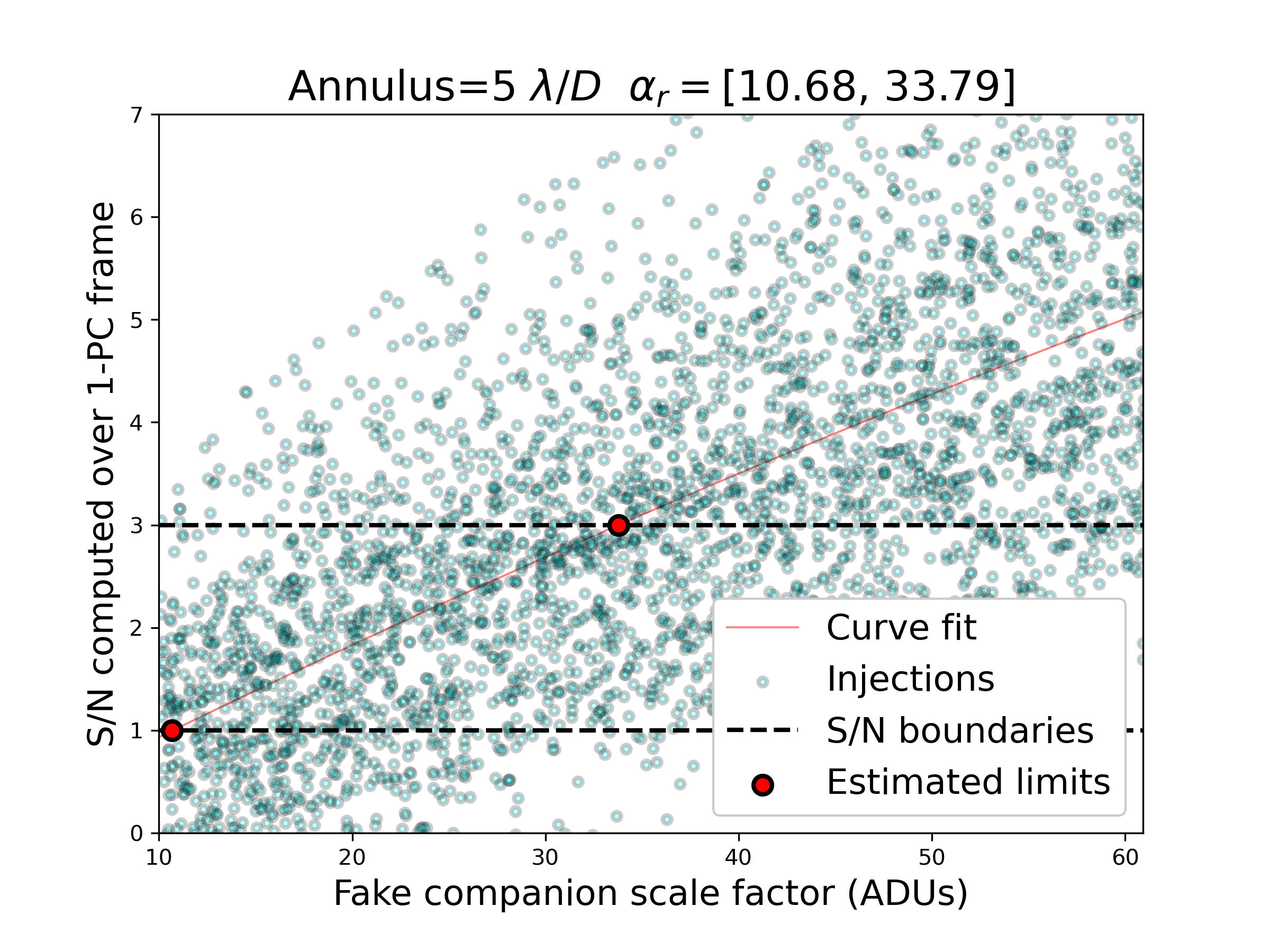}
                \includegraphics[width=0.48\textwidth]{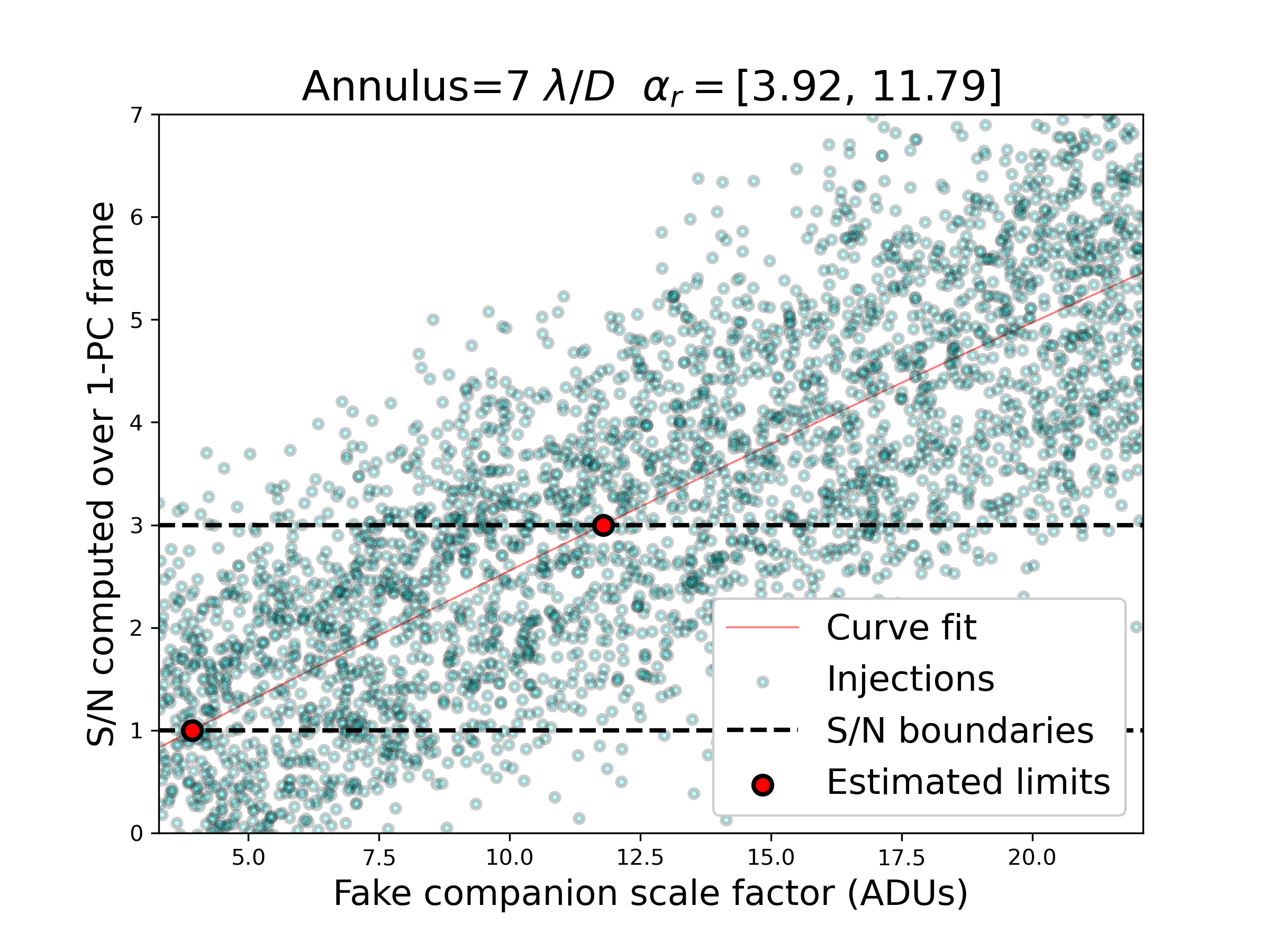}
                \includegraphics[width=0.48\textwidth]{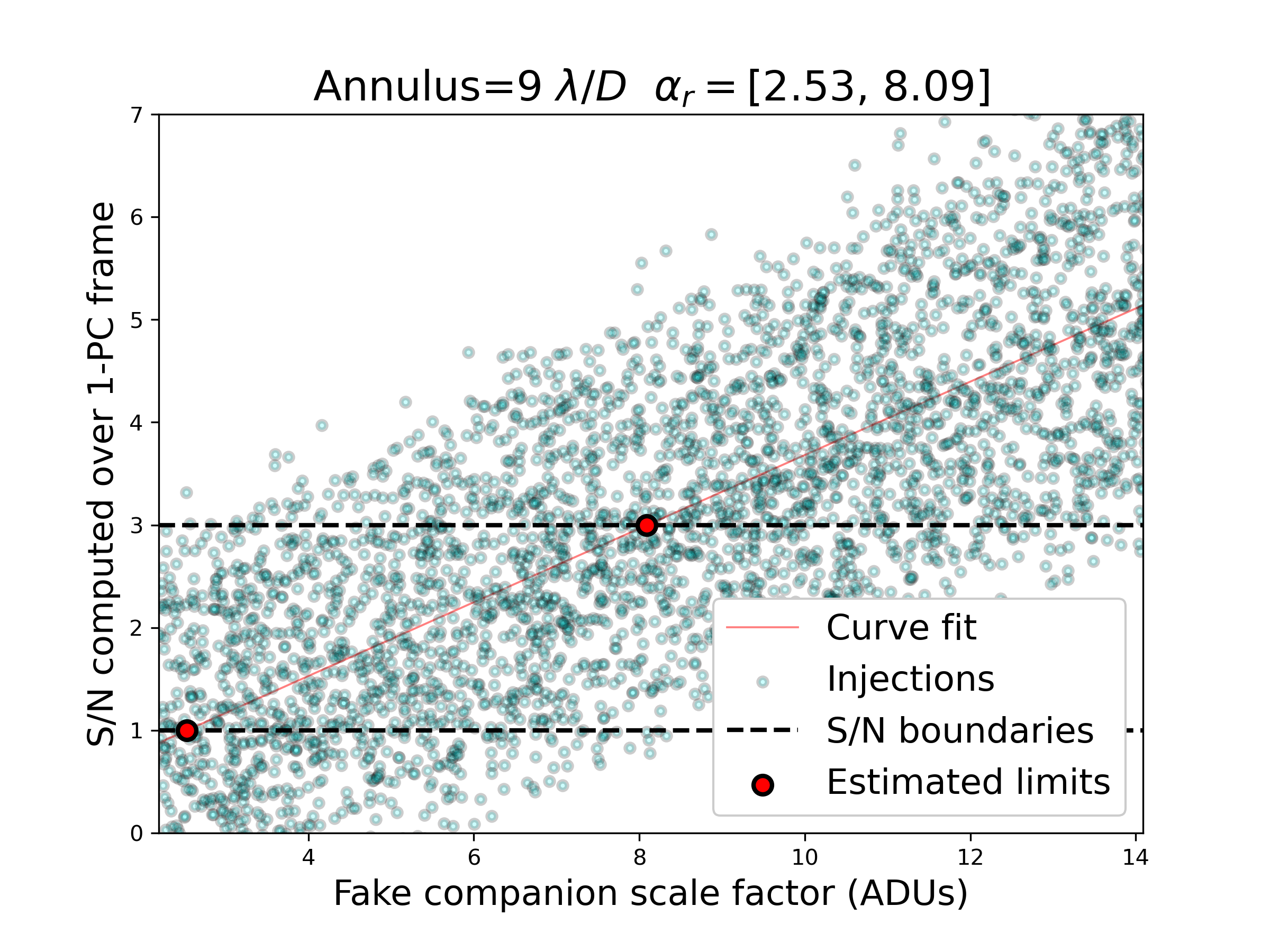}
                \includegraphics[width=0.48\textwidth]{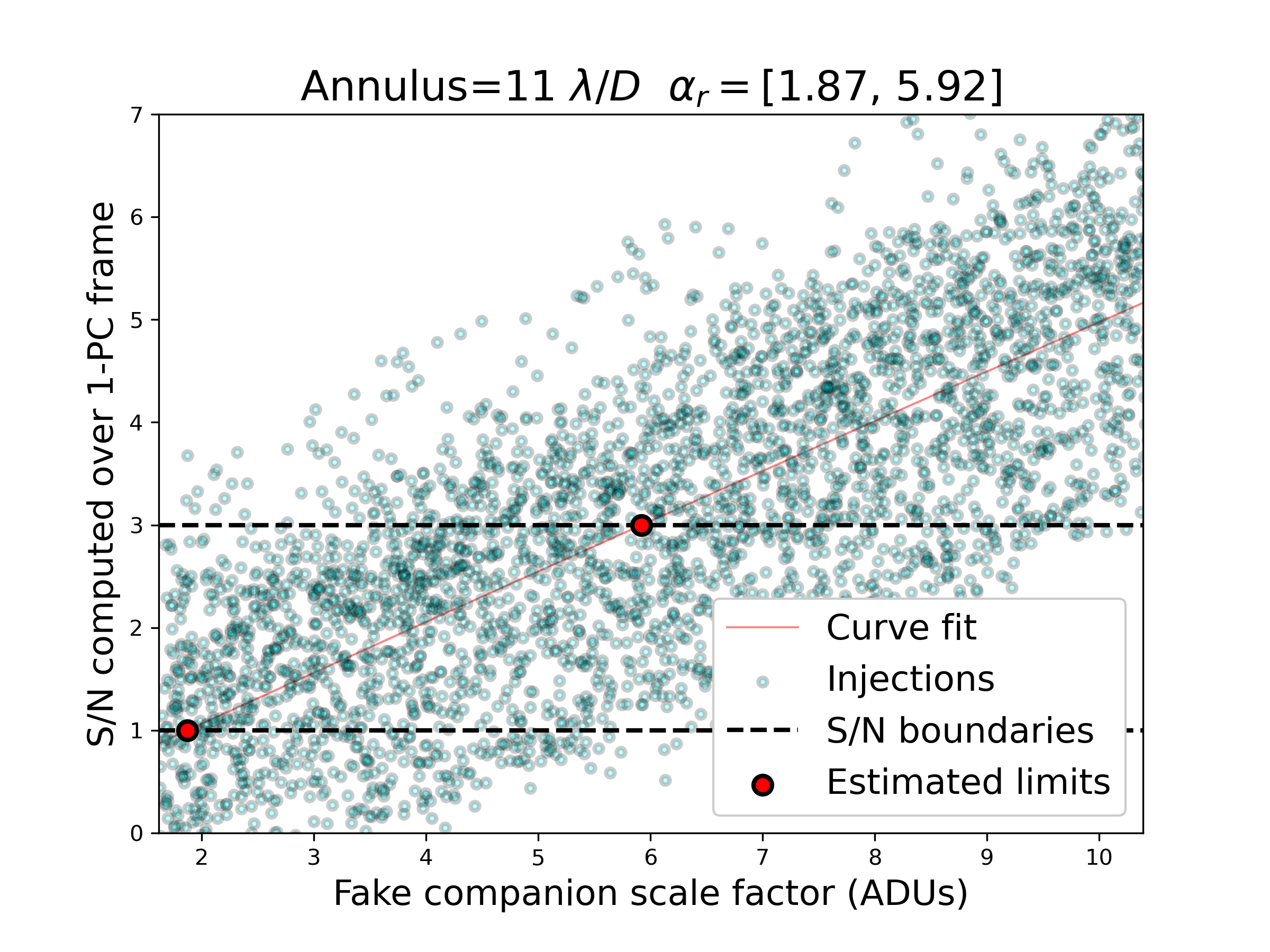}
                \includegraphics[width=0.48\textwidth]{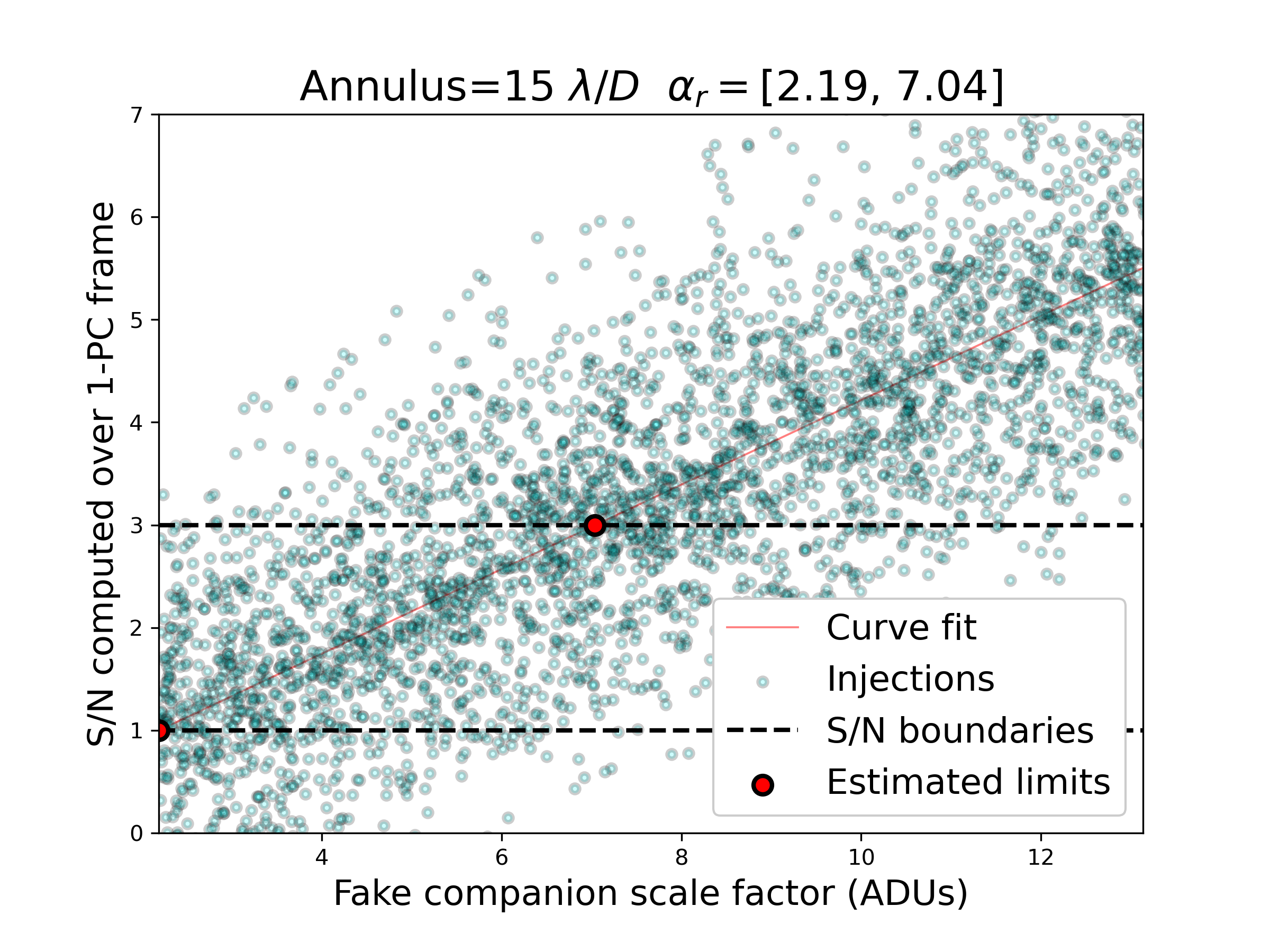}
                \includegraphics[width=0.48\textwidth]{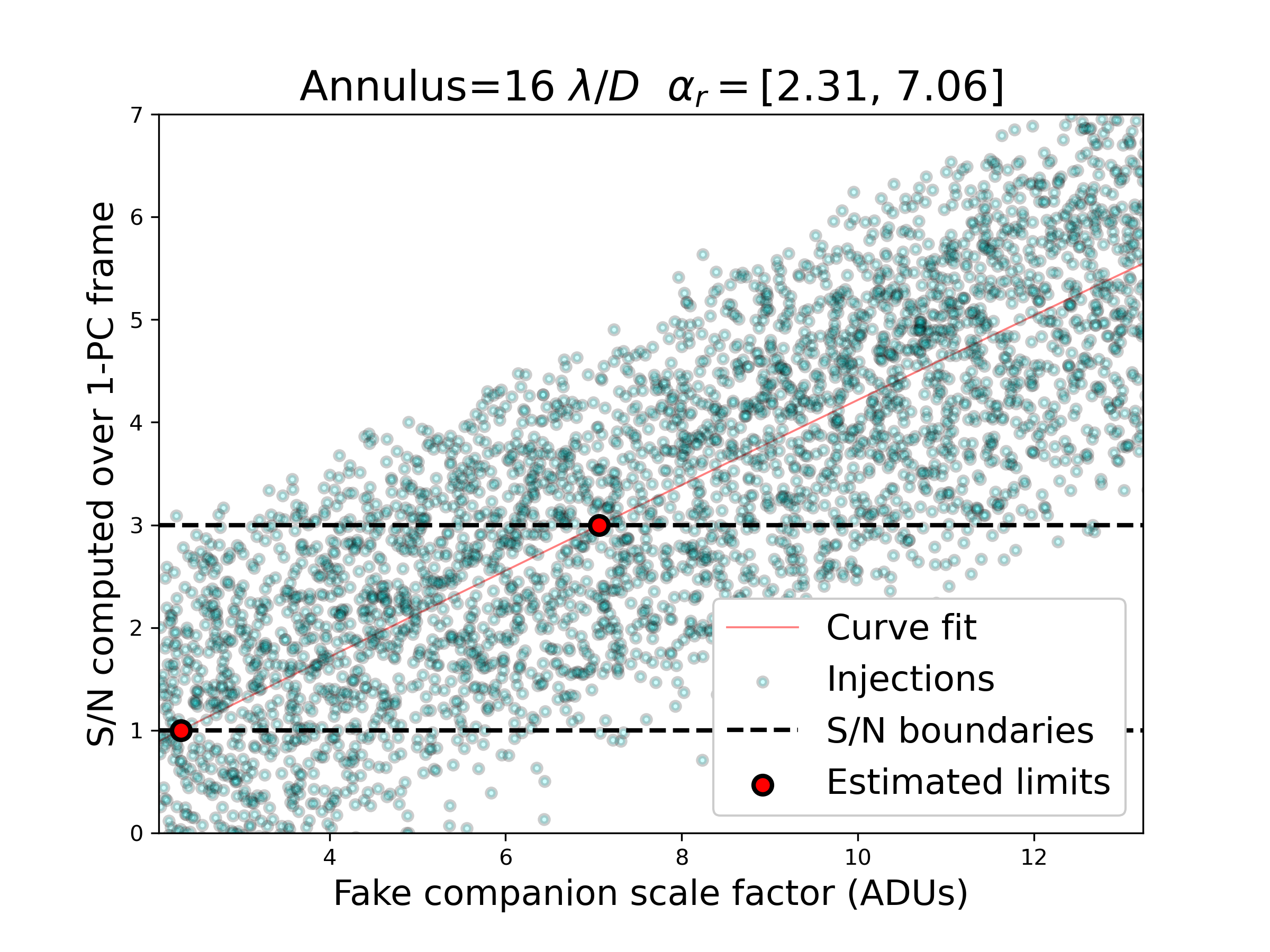}
                \caption{Example of the injection flux estimation method for the case of the \textit{sph2} sequence. Each subplot refers to a different angular separation, and shows the S/N of an injection ($y$ axis), retrieved from the PCA post-processed frame with one principal component, as a function of its scale factor ($x$ axis). Each point in cyan on subplots thus represents a fake companion, which has been injected in the ADI sequence at random coordinates within the corresponding annulus and with a random scale factor. The thin red line is the curve fit of all injections, and  dashed horizontal curves in black delimit the chosen S/N range, which is from one to three in this case. The two red dots show the intersection between the curve fit and the S/N limits (step 6), which therefore define the range of the scale factor corresponding to the chosen S/N limits.}
                \label{fig:fluxes_estimation}
            \end{figure*}

            This process is repeated for each angular separation in the field of view in such a way that a different flux scale factor range $\alpha_{R}$ is estimated for each annulus. Figure~\ref{fig:fluxes_estimation} illustrates this data-driven procedure for the case of the \textit{sph2} dataset, showing the plots of step 5 for different annuli, each with $N_{\mathrm{inj}}=3000$ injections (step 4). From Fig.~\ref{fig:fluxes_estimation}, we observe a general trend: the estimated scale factor range decreases as the angular distance increases. This observation aligns with our expectations, given that the spatial component of speckle noise intensity displays a radial dependency in the raw HCI data, a characteristic that persists even after the ADI processing. However, for this \textit{sph2} dataset, a notable departure from this trend occurs between $15-16 \lambda/D$ separations. In this specific interval, we observe in Fig.~\ref{fig:fluxes_estimation} an anomalous increase in the estimated scale factors instead. We associate this behaviour with the fact that, at these angular separations for  \textit{sph2}, speckle dominates over background noise, as concluded in Sect.~\ref{sec:regimes}.
            
            \section{Detection maps for \textit{sph2} and \textit{nrc3} datasets} \label{sec:appendixC}
               
            \begin{figure*}
                \centering
                \hspace{-0.02cm}\includegraphics[width=0.918\textwidth]{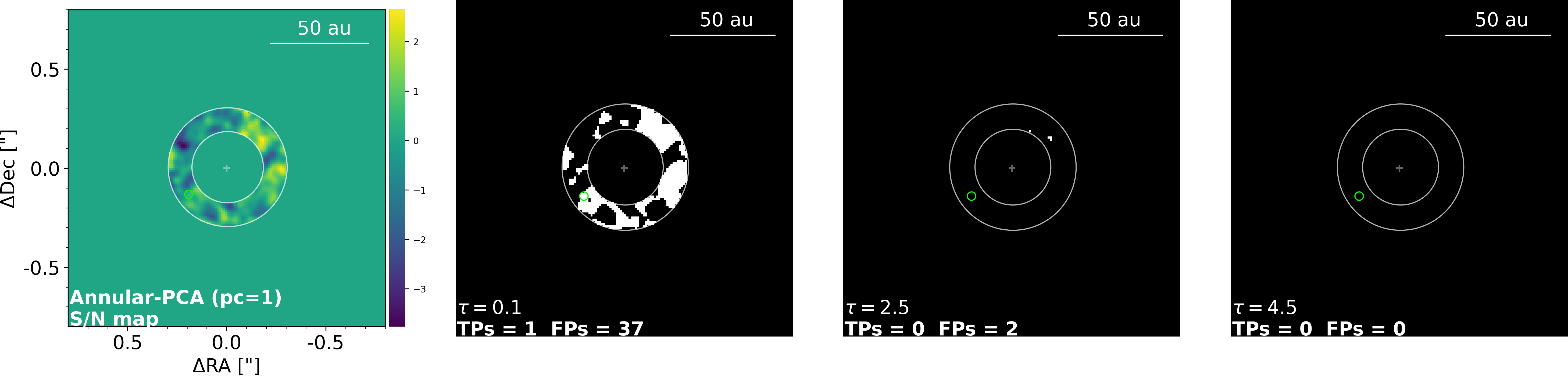}

                \vspace{-0.33cm}
                \hspace{-0.02cm}\includegraphics[width=0.917\textwidth]{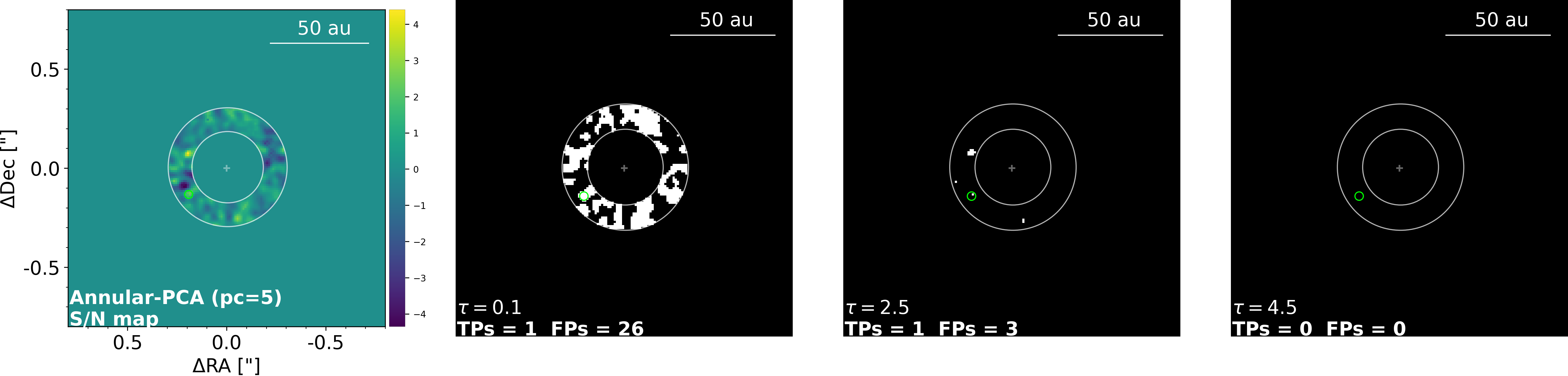}

                \vspace{-0.33cm}
                \includegraphics[width=0.92\textwidth]{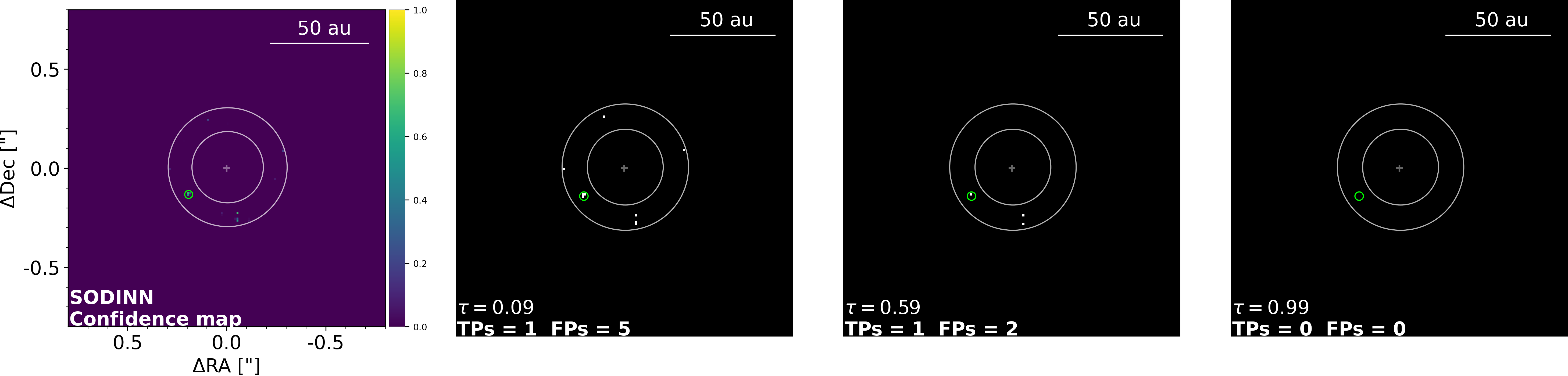}

                \vspace{-0.33cm}
                \includegraphics[width=0.92\textwidth]{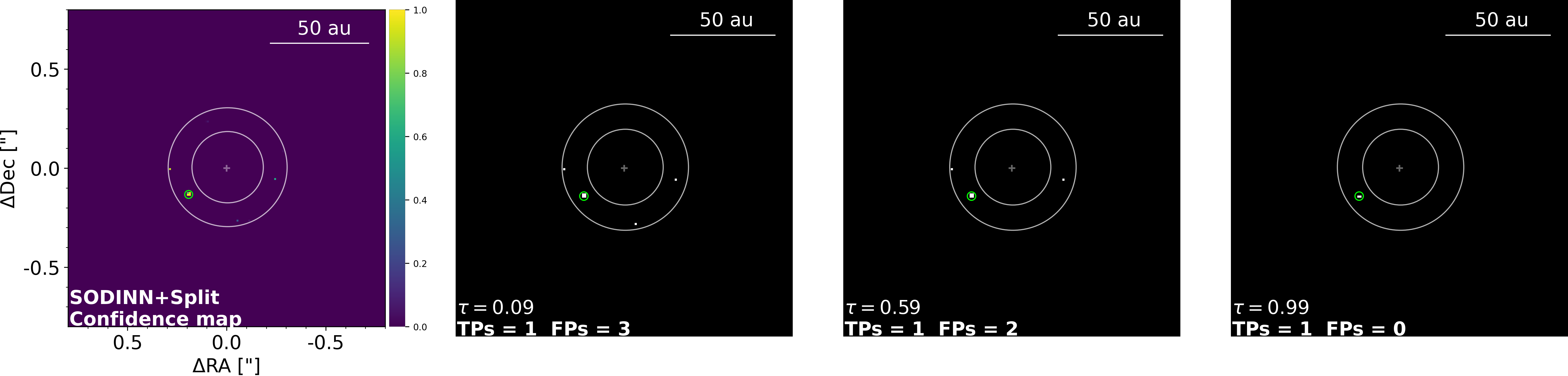}

                \vspace{-0.33cm}
                \includegraphics[width=0.92\textwidth]{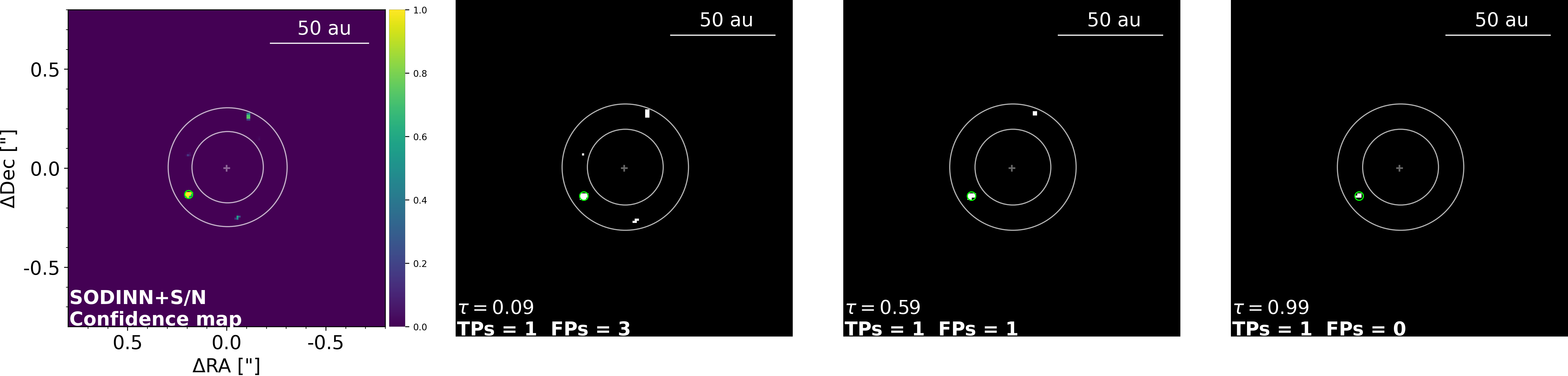}

                \vspace{-0.33cm}
                \includegraphics[width=0.92\textwidth]{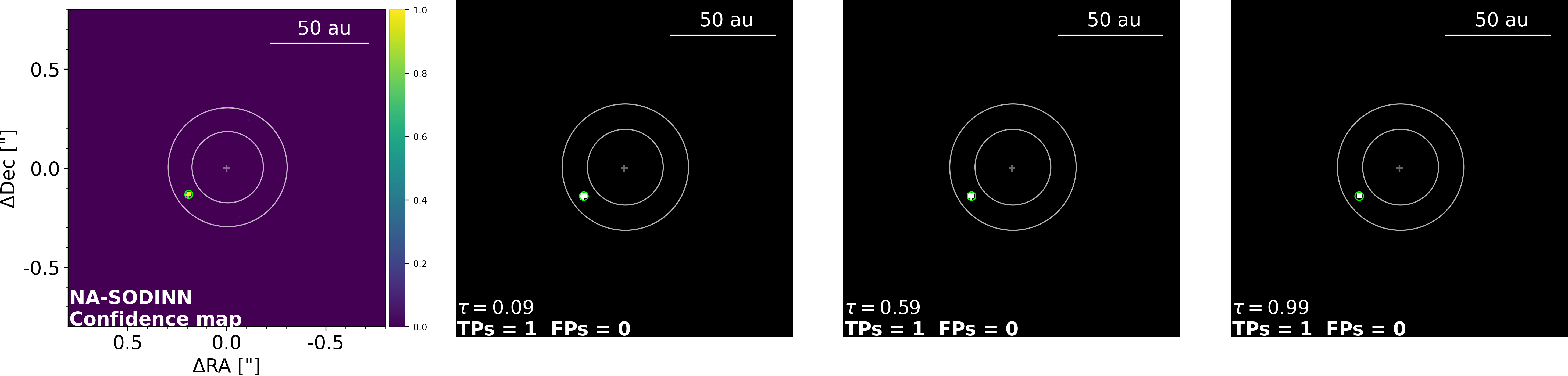}
            
                \caption{Evaluation example of Annular-PCA and all SODINN-based algorithms over the \textbf{5-7} $\lambda/D$ regime of \textbf{sph2}, where a fake companion has been injected with \textbf{S/N=0.75} (computed in the PCA-processed frame using the first principal component). Each row corresponds to a different algorithm, where its detection map is on left column, and its three thresholds (binary maps) are on the right. The threshold $\tau$, TPs and FPs are highlighted over each binary map. White concentric circles indicate the regime's boundaries. Other noise regimes are masked. Small green circles indicate the position of the injection.}
                \label{fig:maps_sph2_5to7}
            \end{figure*}
            
            \begin{figure*}
                \centering
                \hspace{-0.02cm}\includegraphics[width=0.918\textwidth]{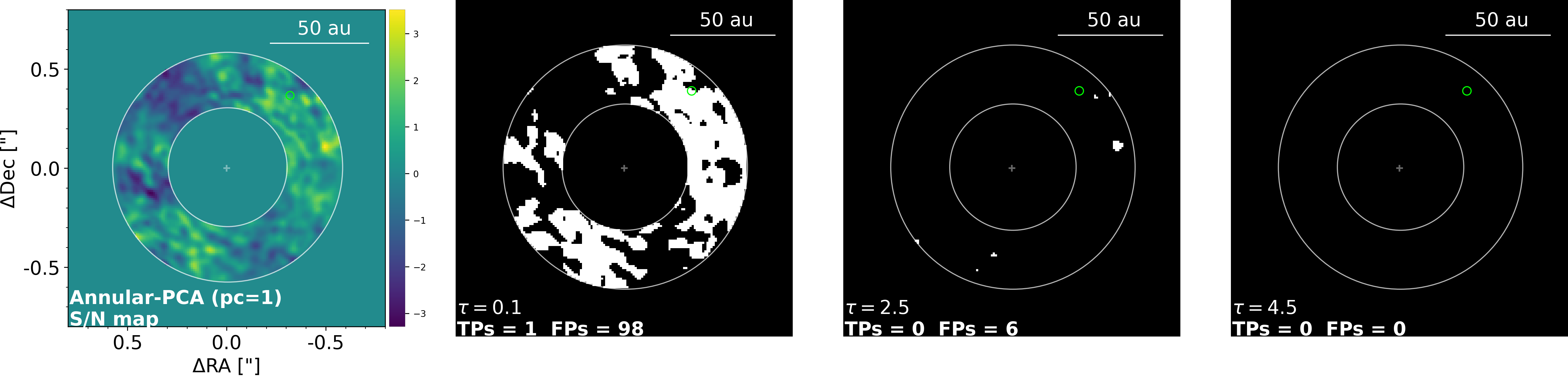}

                \vspace{-0.33cm}
                \hspace{-0.02cm}\includegraphics[width=0.917\textwidth]{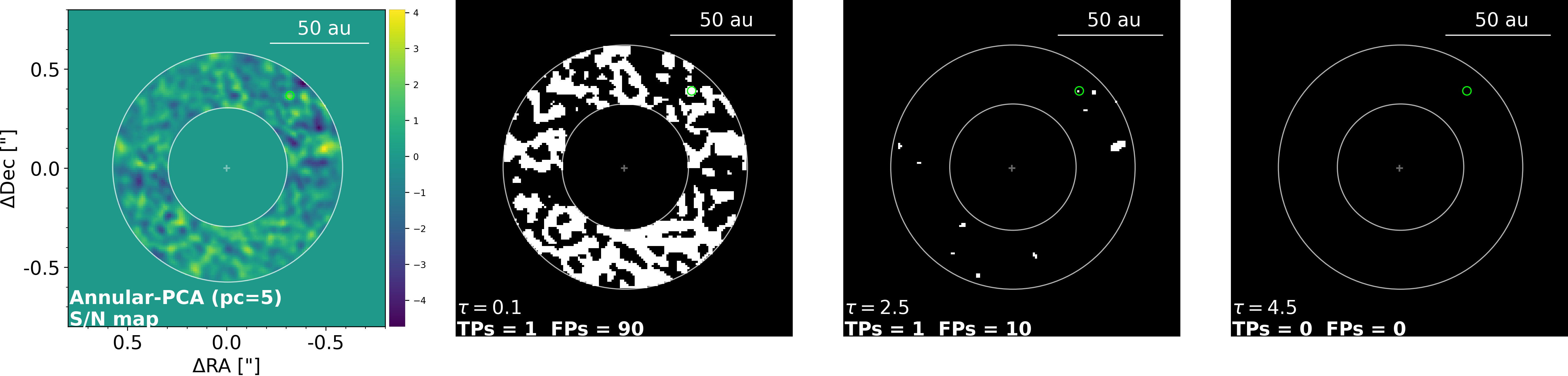}

                \vspace{-0.33cm}
                \includegraphics[width=0.92\textwidth]{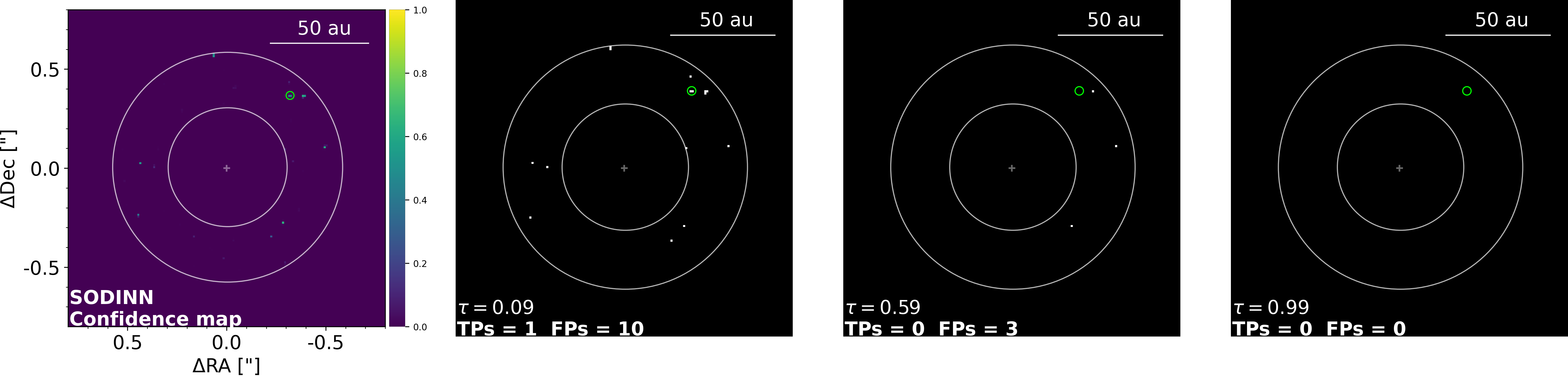}

                \vspace{-0.33cm}
                \includegraphics[width=0.92\textwidth]{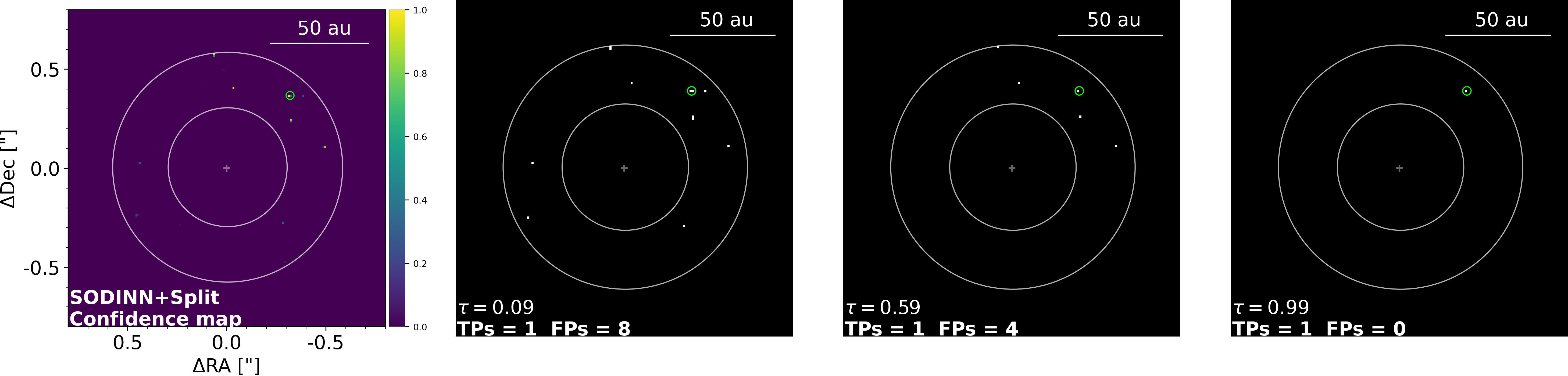}

                \vspace{-0.33cm}
                \includegraphics[width=0.92\textwidth]{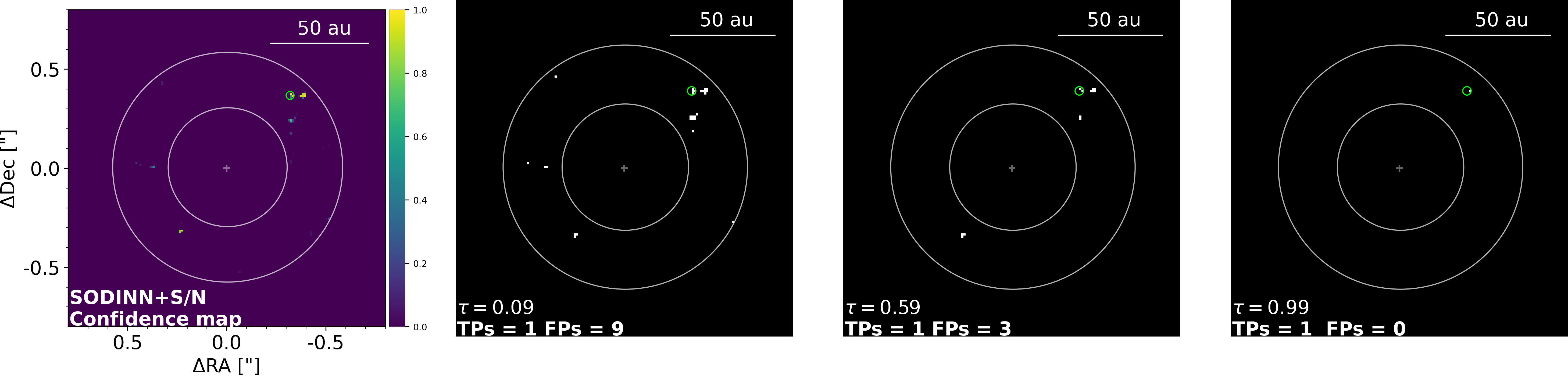}

                \vspace{-0.33cm}
                \includegraphics[width=0.92\textwidth]{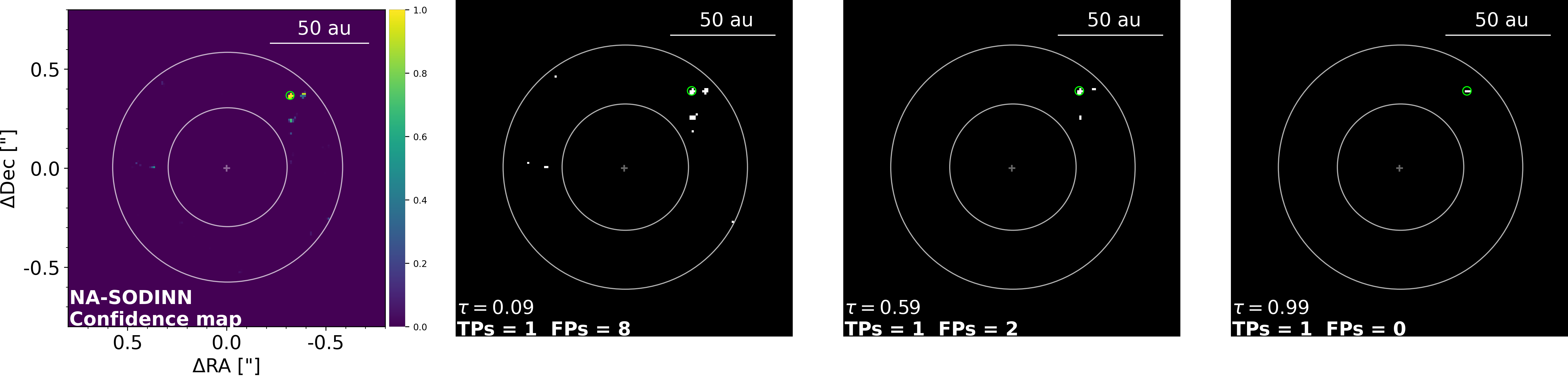}
            
                \caption{Same of Fig.~\ref{fig:maps_sph2_5to7} for the regime \textbf{8-14} $\lambda/D$ on \textbf{sph2}, where a fake companion has been injected with \textbf{S/N=0.89}.}
                \label{fig:maps_sph2_8to14}
            \end{figure*}

            \begin{figure*}
                \centering
                \hspace{-0.02cm}\includegraphics[width=0.918\textwidth]{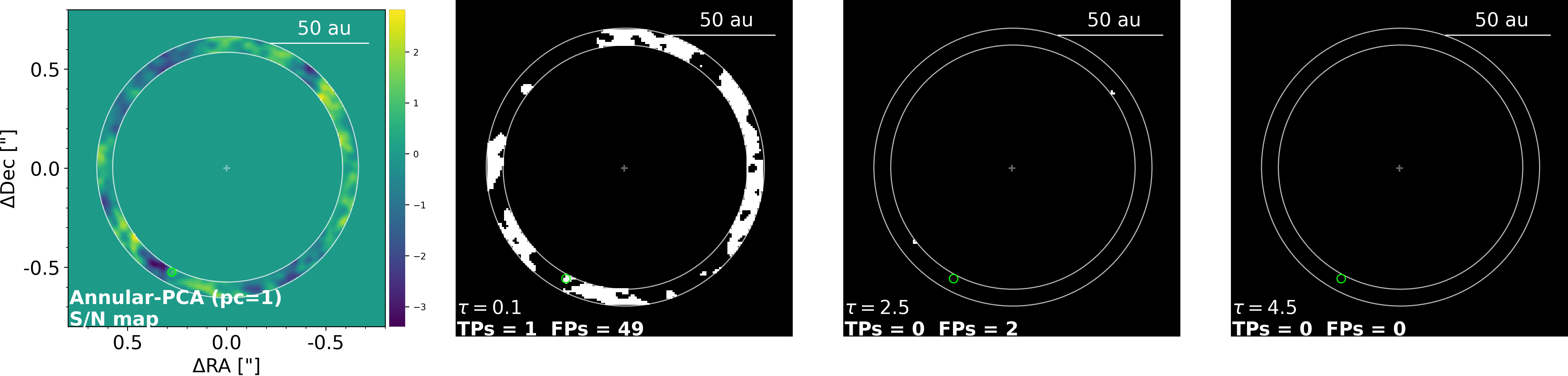}

                \vspace{-0.33cm}
                \hspace{-0.02cm}\includegraphics[width=0.917\textwidth]{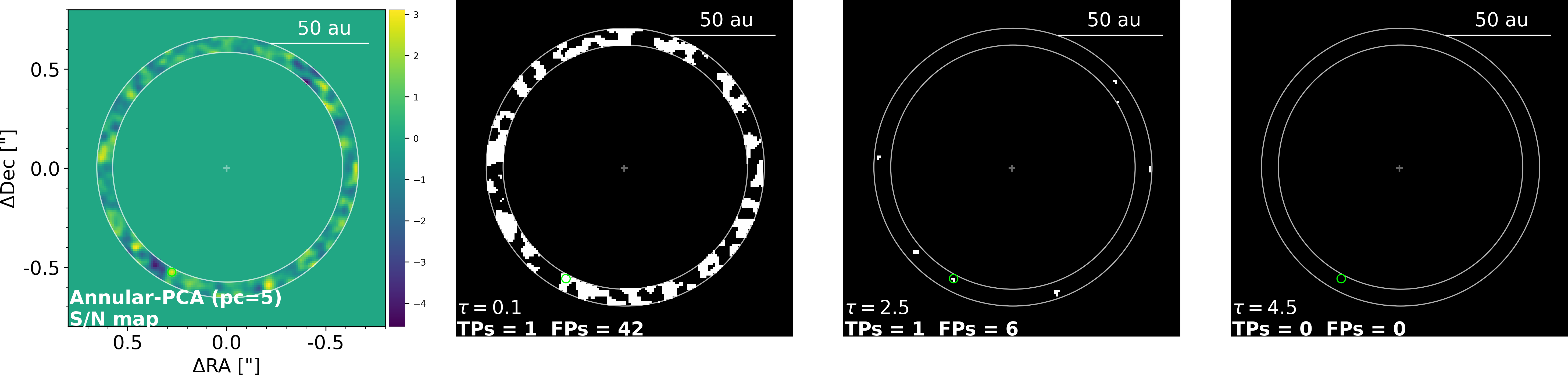}

                \vspace{-0.33cm}
                \includegraphics[width=0.92\textwidth]{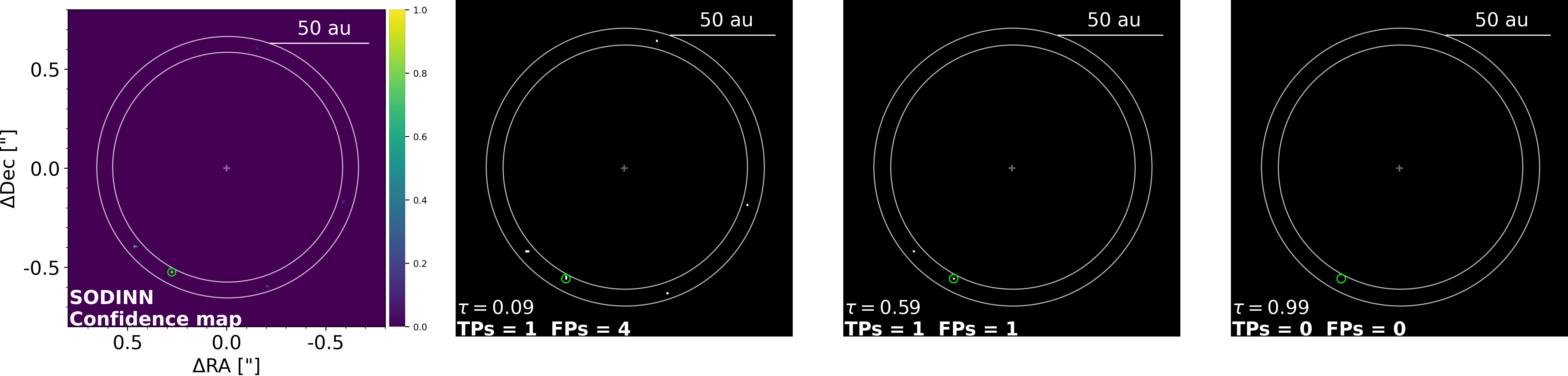}

                \vspace{-0.33cm}
                \includegraphics[width=0.92\textwidth]{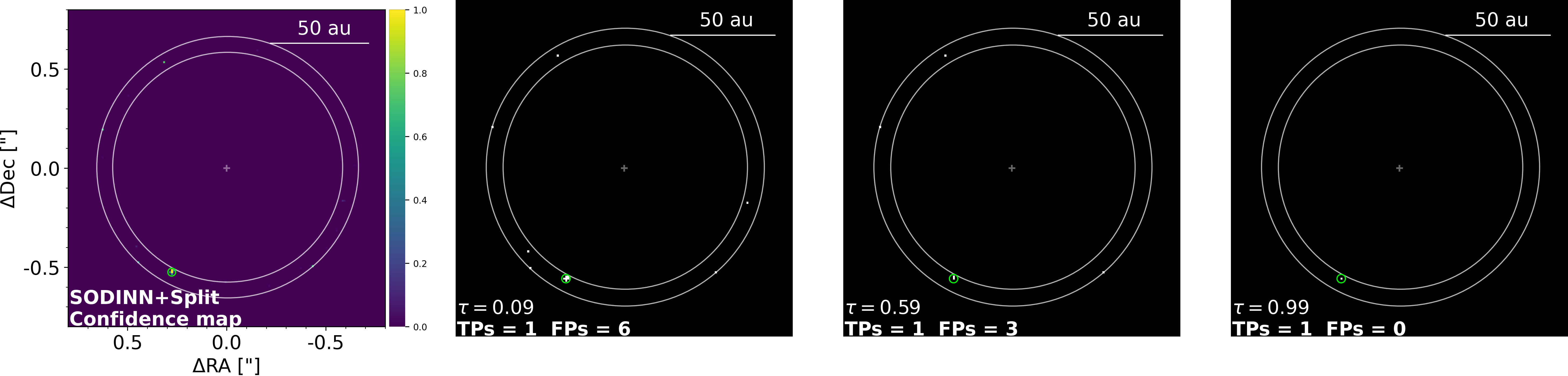}

                \vspace{-0.33cm}
                \includegraphics[width=0.92\textwidth]{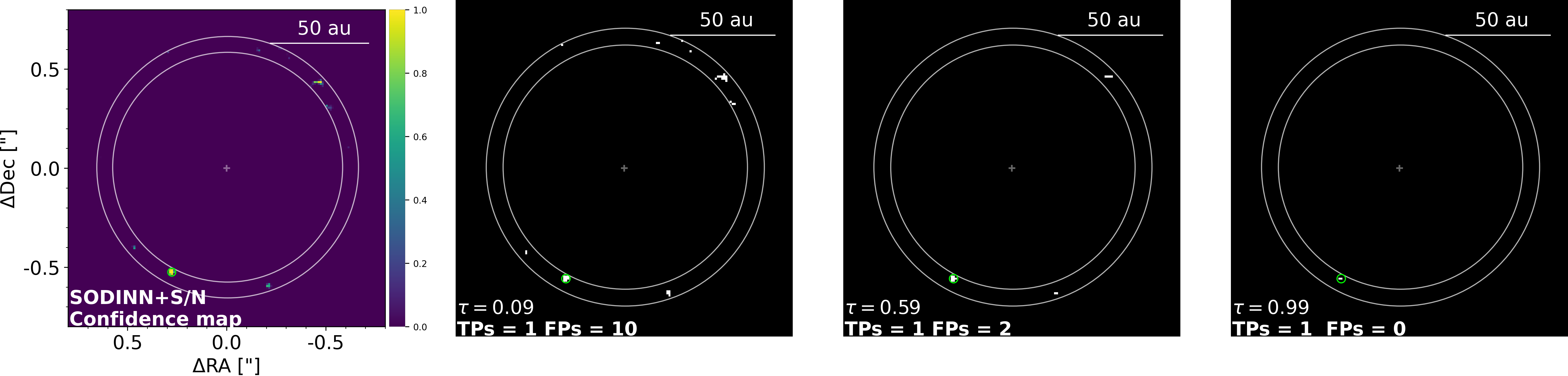}

                \vspace{-0.33cm}
                \includegraphics[width=0.92\textwidth]{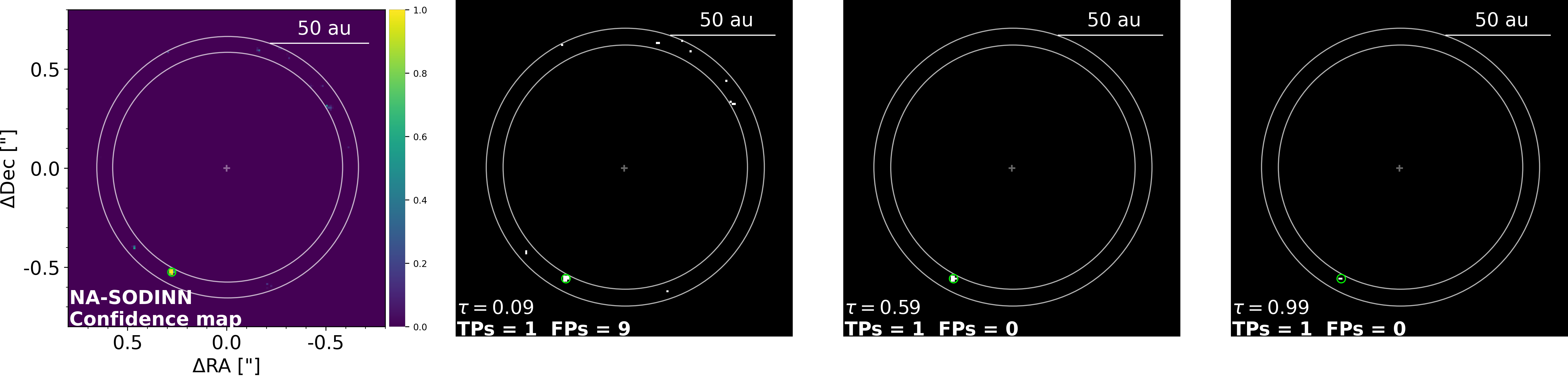}
            
                \caption{Same of Fig.~\ref{fig:maps_sph2_5to7} for the regime \textbf{15-16} $\lambda/D$ on \textbf{sph2}, where a fake companion has been injected with \textbf{S/N=0.78}.}
                \label{fig:maps_sph2_15to16}
            \end{figure*}

            \begin{figure*}
                \centering
                \hspace{-0.02cm}\includegraphics[width=0.918\textwidth]{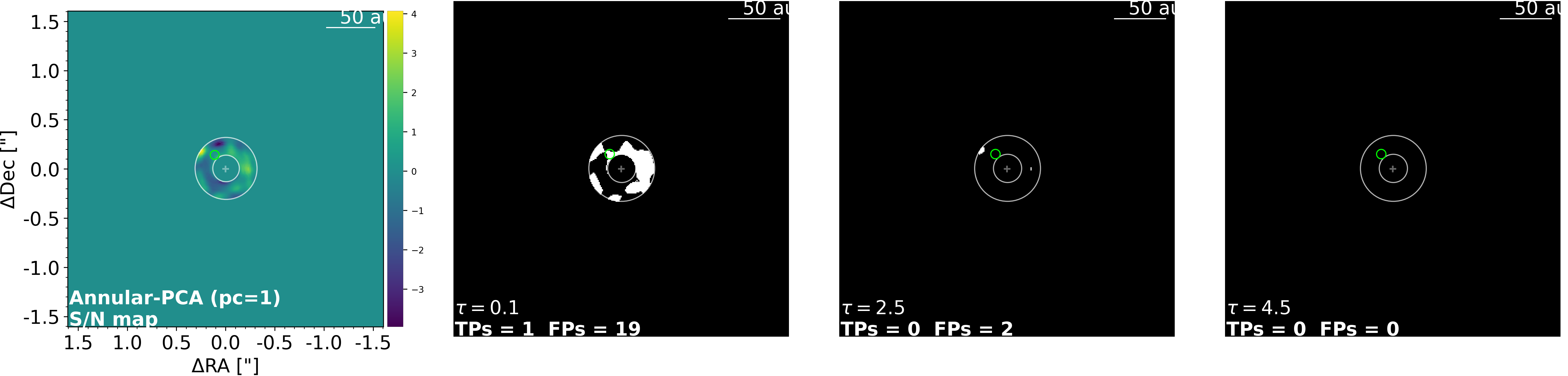}

                \vspace{-0.34cm}
                \hspace{-0.02cm}\includegraphics[width=0.917\textwidth]{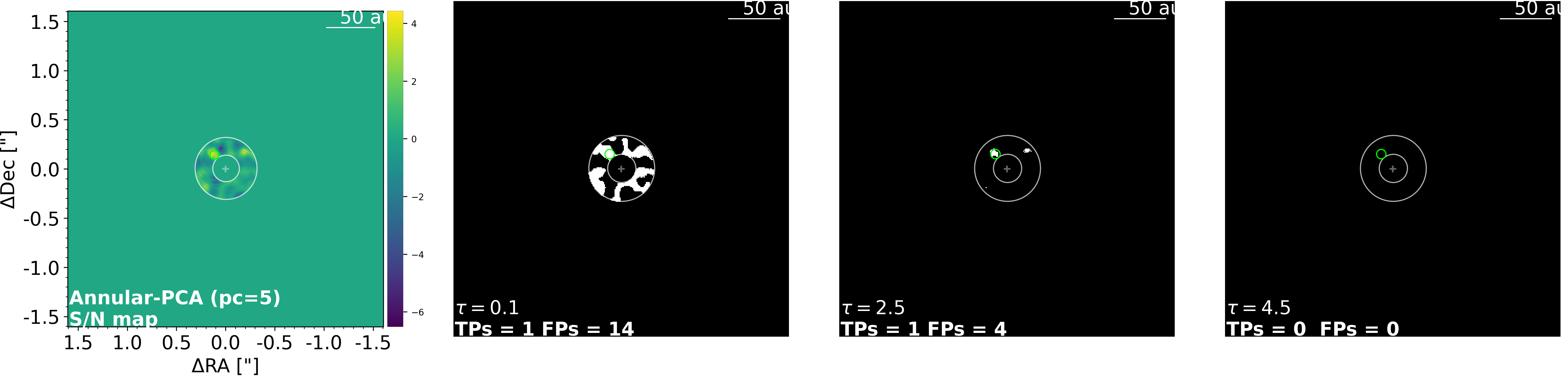}

                \vspace{-0.34cm}
                \includegraphics[width=0.92\textwidth]{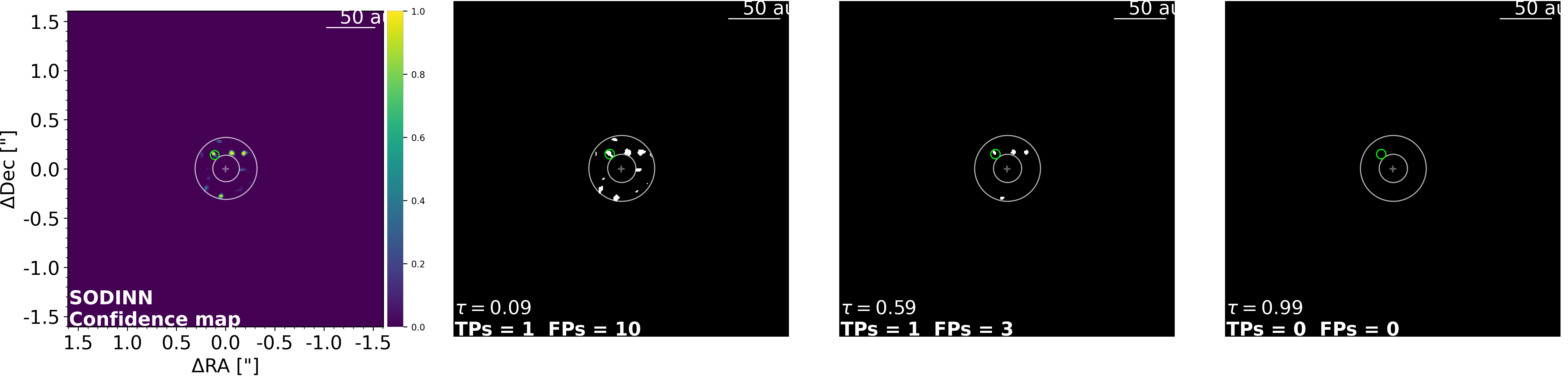}

                \vspace{-0.34cm}
                \includegraphics[width=0.92\textwidth]{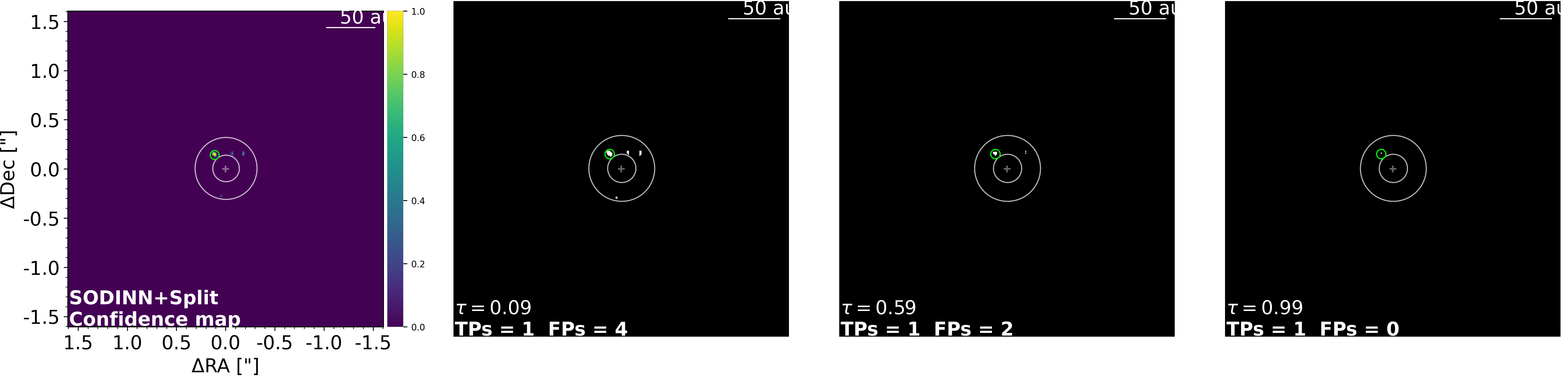}

                \vspace{-0.34cm}
                \includegraphics[width=0.92\textwidth]{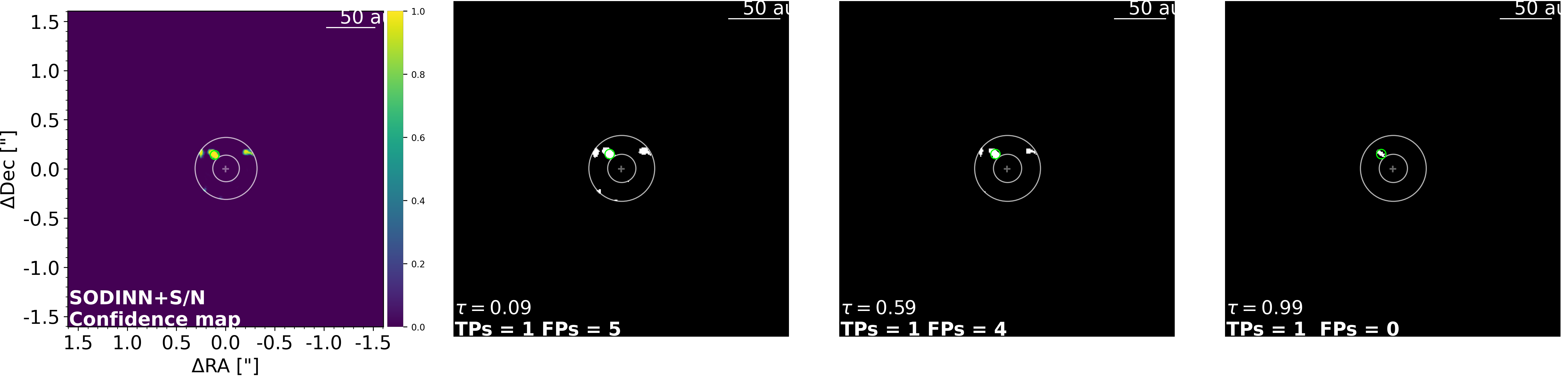}

                \vspace{-0.34cm}
                \includegraphics[width=0.92\textwidth]{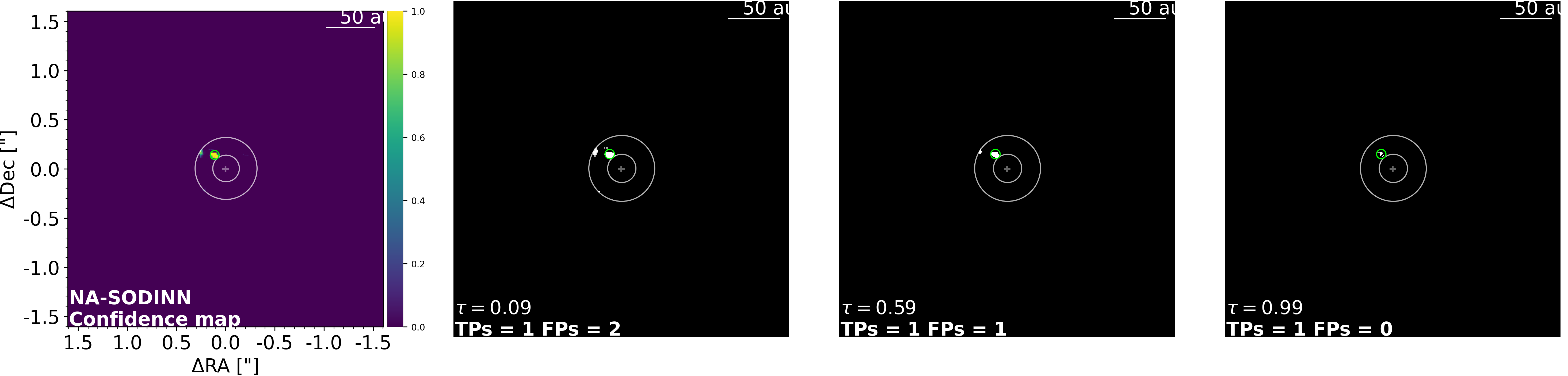}
            
                \caption{Same of Fig.~\ref{fig:maps_sph2_5to7} for the regime \textbf{1-3} $\lambda/D$ on \textbf{nrc3}, where a fake companion has been injected with \textbf{S/N=0.78}.}
                \label{fig:maps_nrc3_1to3}
            \end{figure*}

            \begin{figure*}
                \centering
                \hspace{-0.02cm}\includegraphics[width=0.918\textwidth]{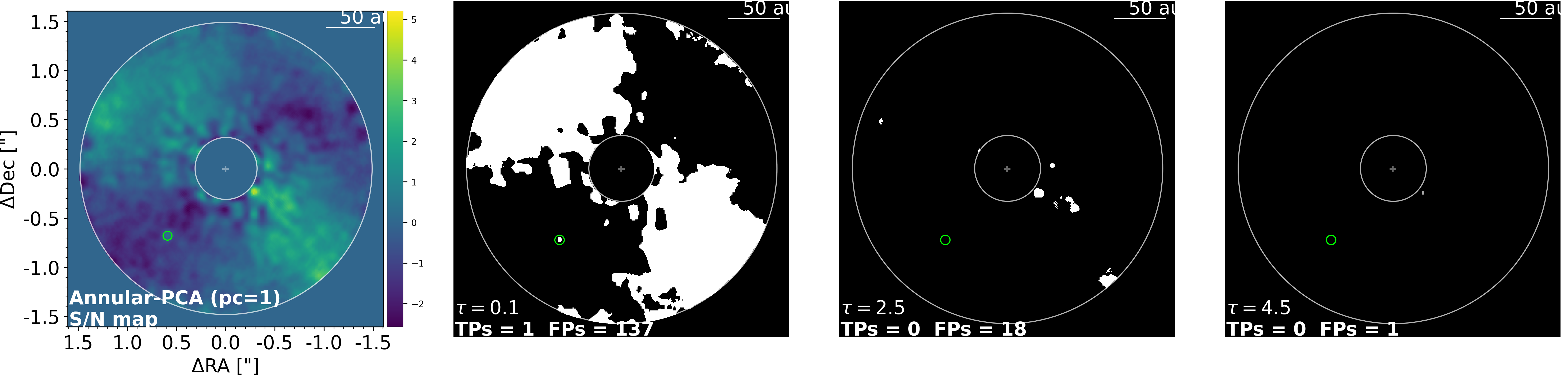}

                \vspace{-0.34cm}
                \hspace{-0.02cm}\includegraphics[width=0.917\textwidth]{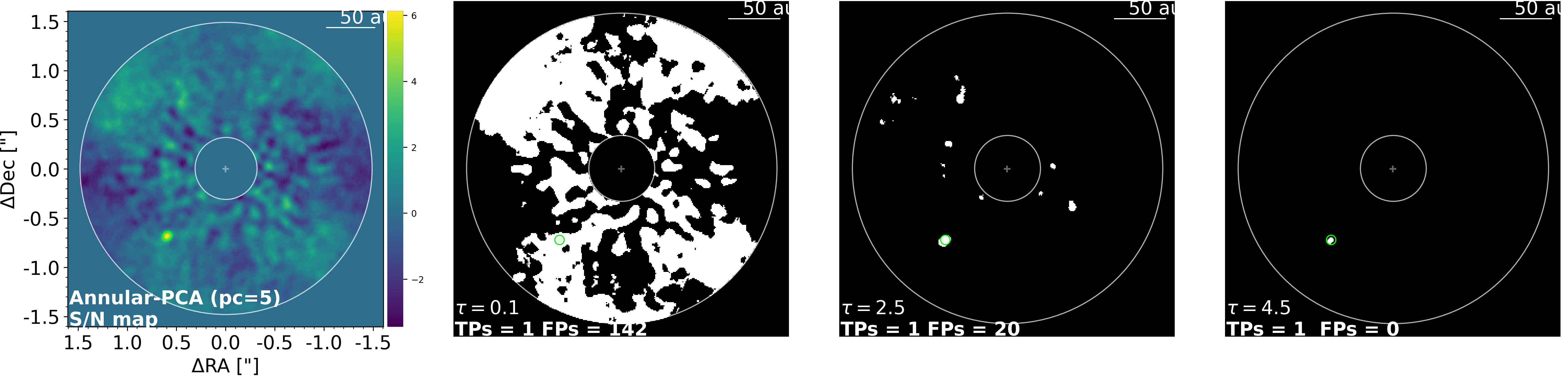}

                \vspace{-0.34cm}
                \includegraphics[width=0.92\textwidth]{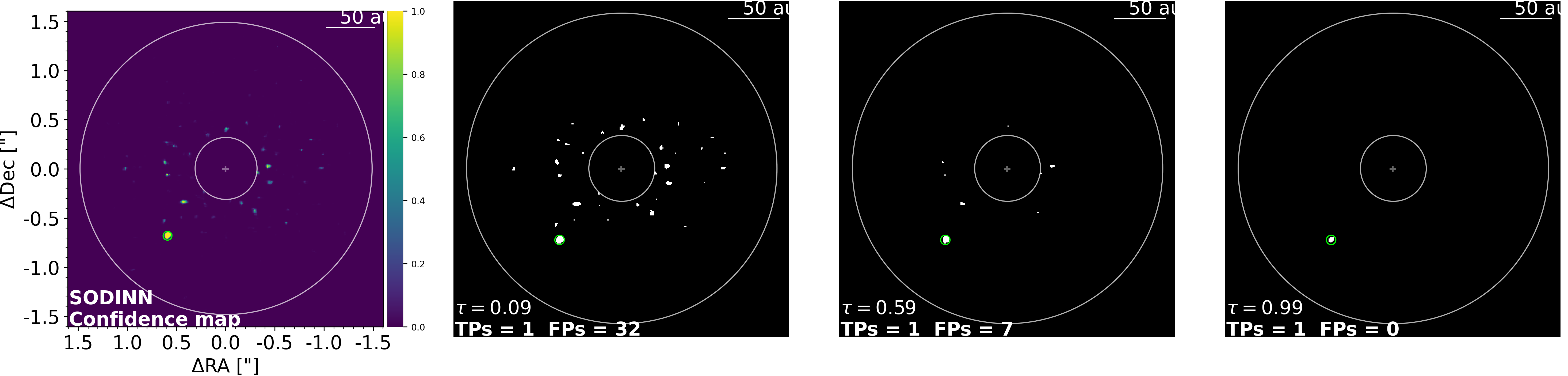}

                \vspace{-0.34cm}
                \includegraphics[width=0.92\textwidth]{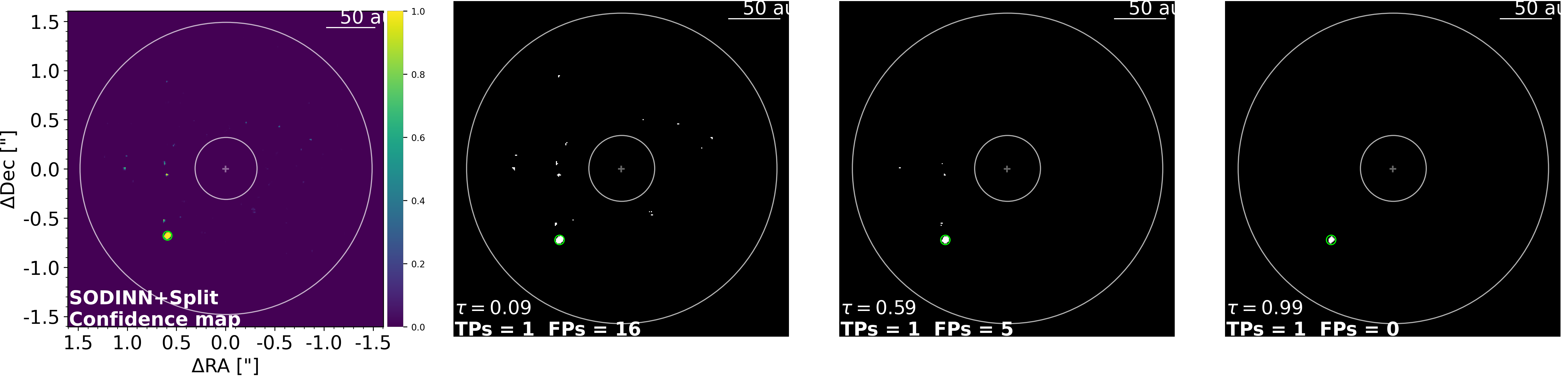}

                \vspace{-0.34cm}
                \includegraphics[width=0.92\textwidth]{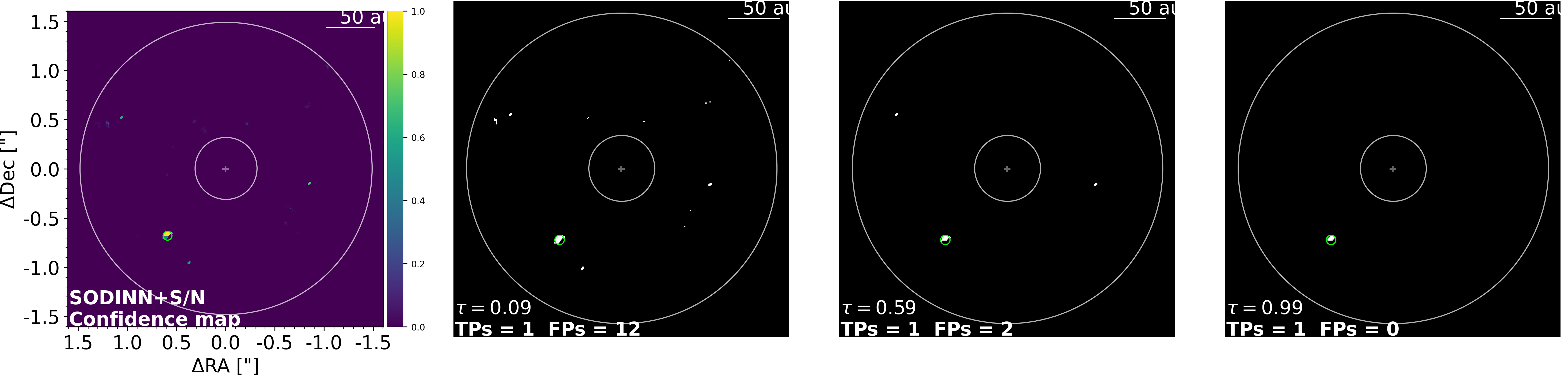}

                \vspace{-0.34cm}
                \includegraphics[width=0.92\textwidth]{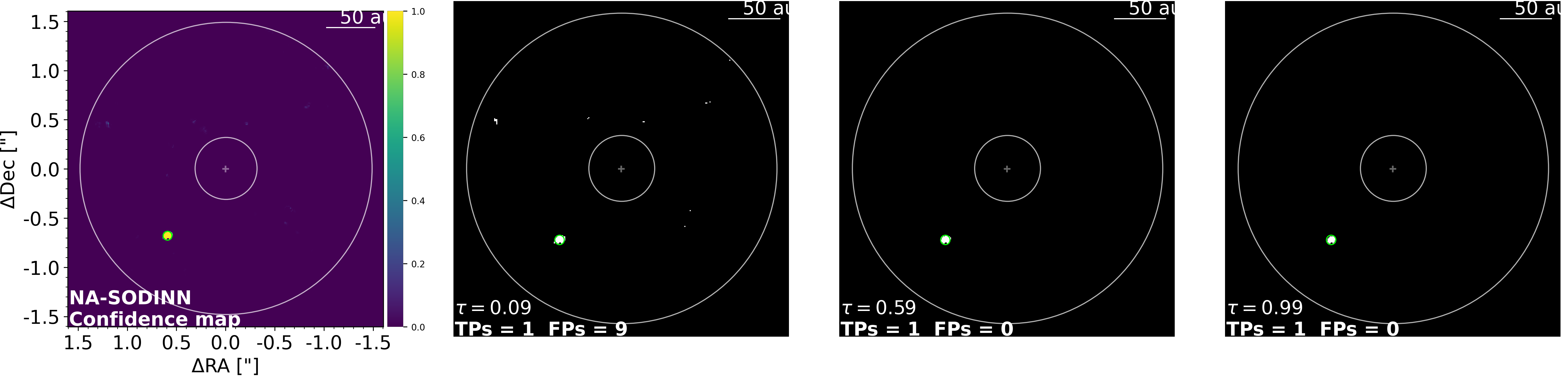}
            
                \caption{Same of Fig.~\ref{fig:maps_sph2_5to7}, but for the regime \textbf{4-16} $\lambda/D$ on \textbf{nrc3}, where a fake companion has been injected with \textbf{S/N=0.84}.}
                \label{fig:maps_nrc3_4to16}
            \end{figure*}

        \section{Details of the EIDC metrics for NA-SODINN} \label{sec:appendixD}
        
        \begin{figure*}
        \begin{tabular}{ccc}
          \includegraphics[width=5.85cm]{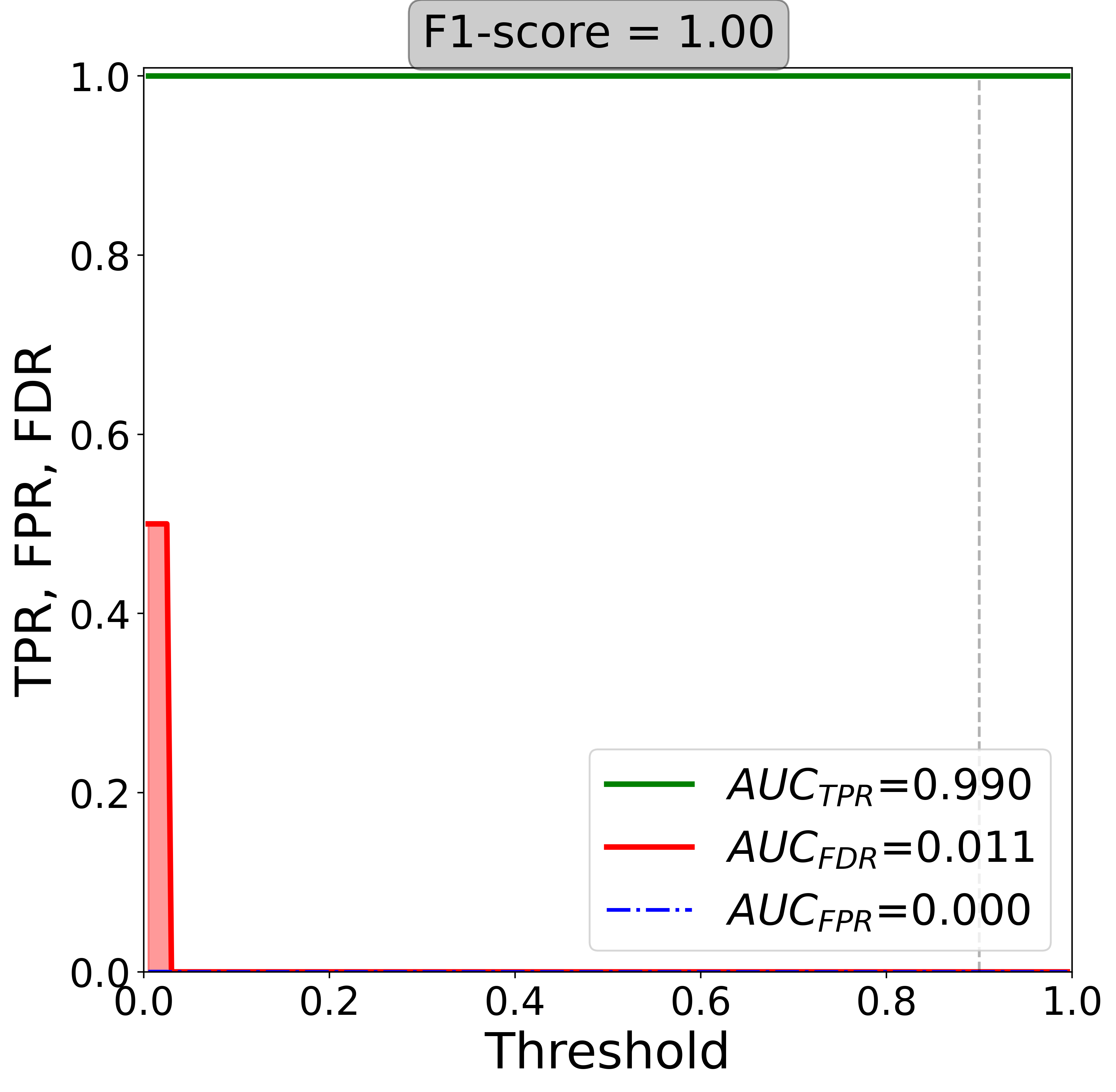} & 
          \includegraphics[width=5.85cm]{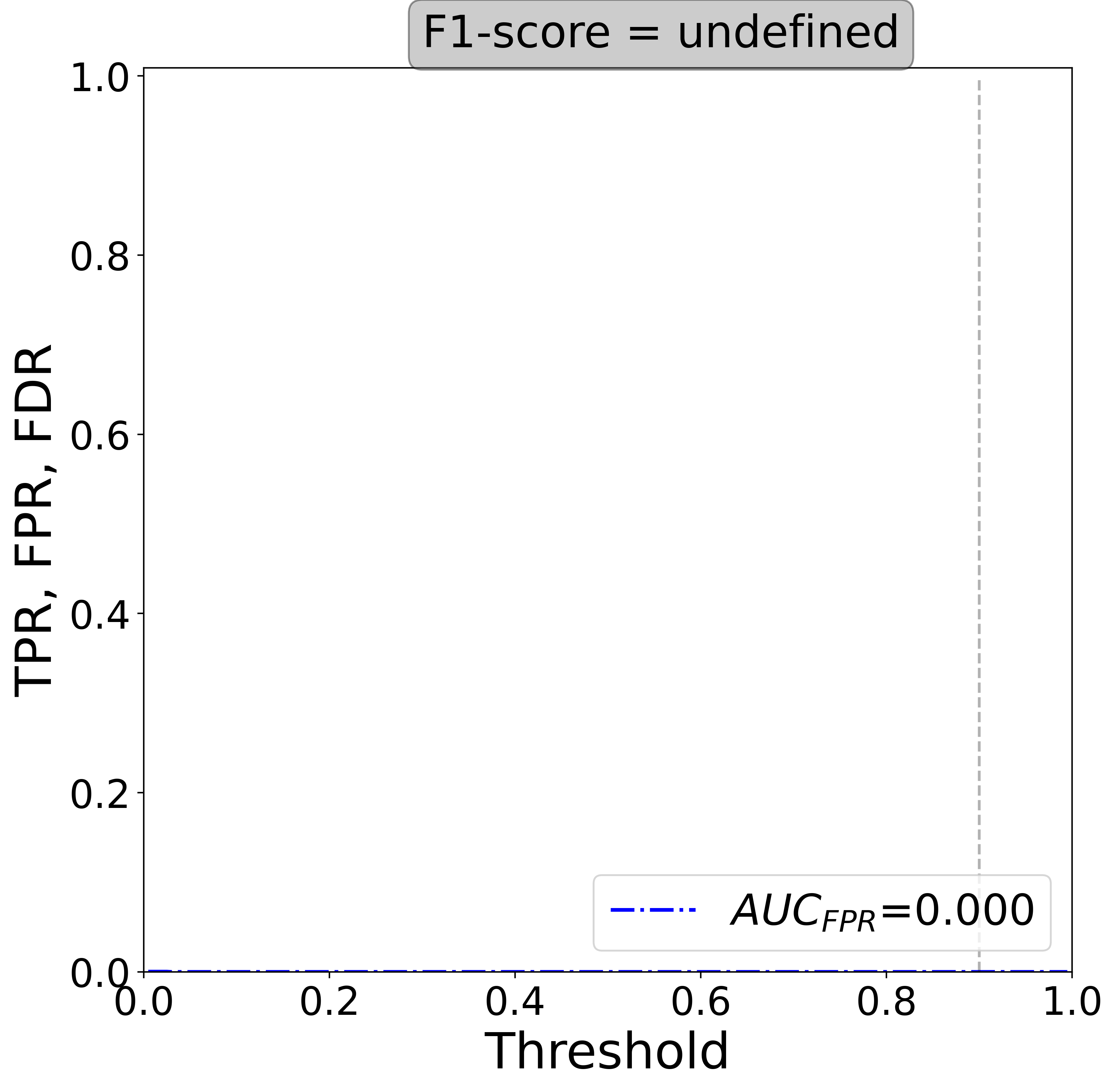} &
          \includegraphics[width=5.85cm]{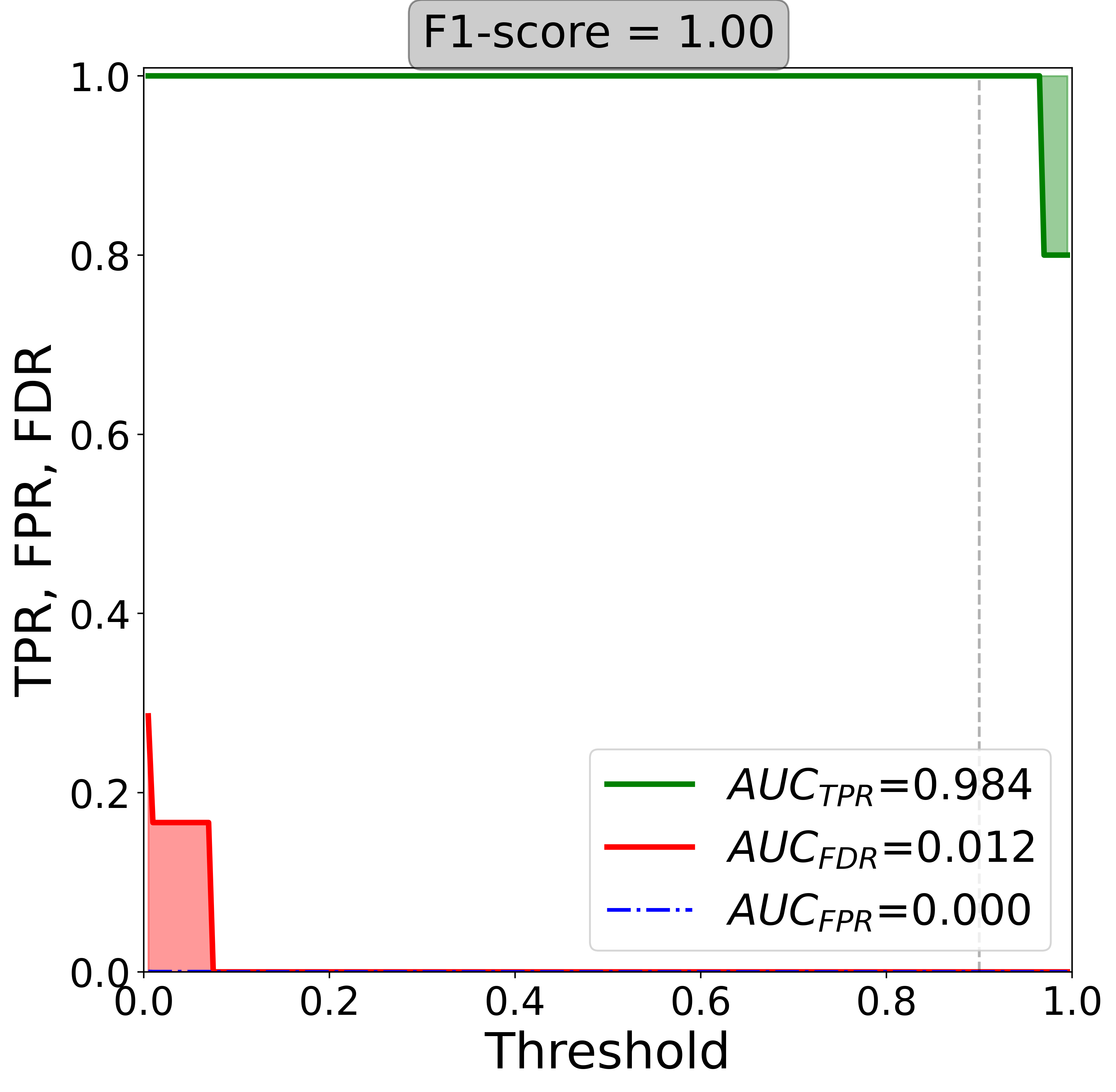} \\
          \textit{sph1} & \textit{sph2} & \textit{sph3} \\[6pt]\\
          \includegraphics[width=5.85cm]{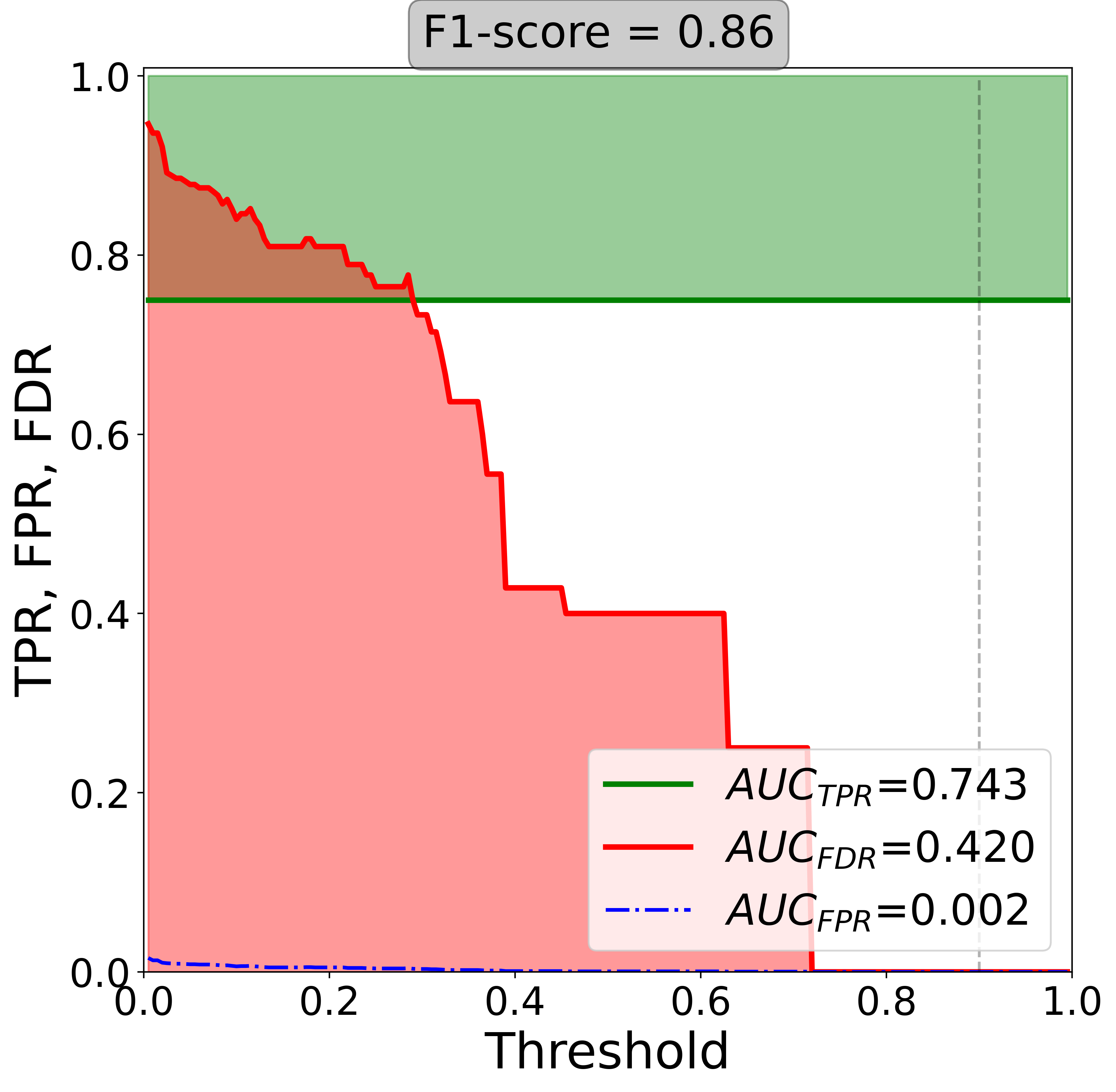} & 
          \includegraphics[width=5.85cm]{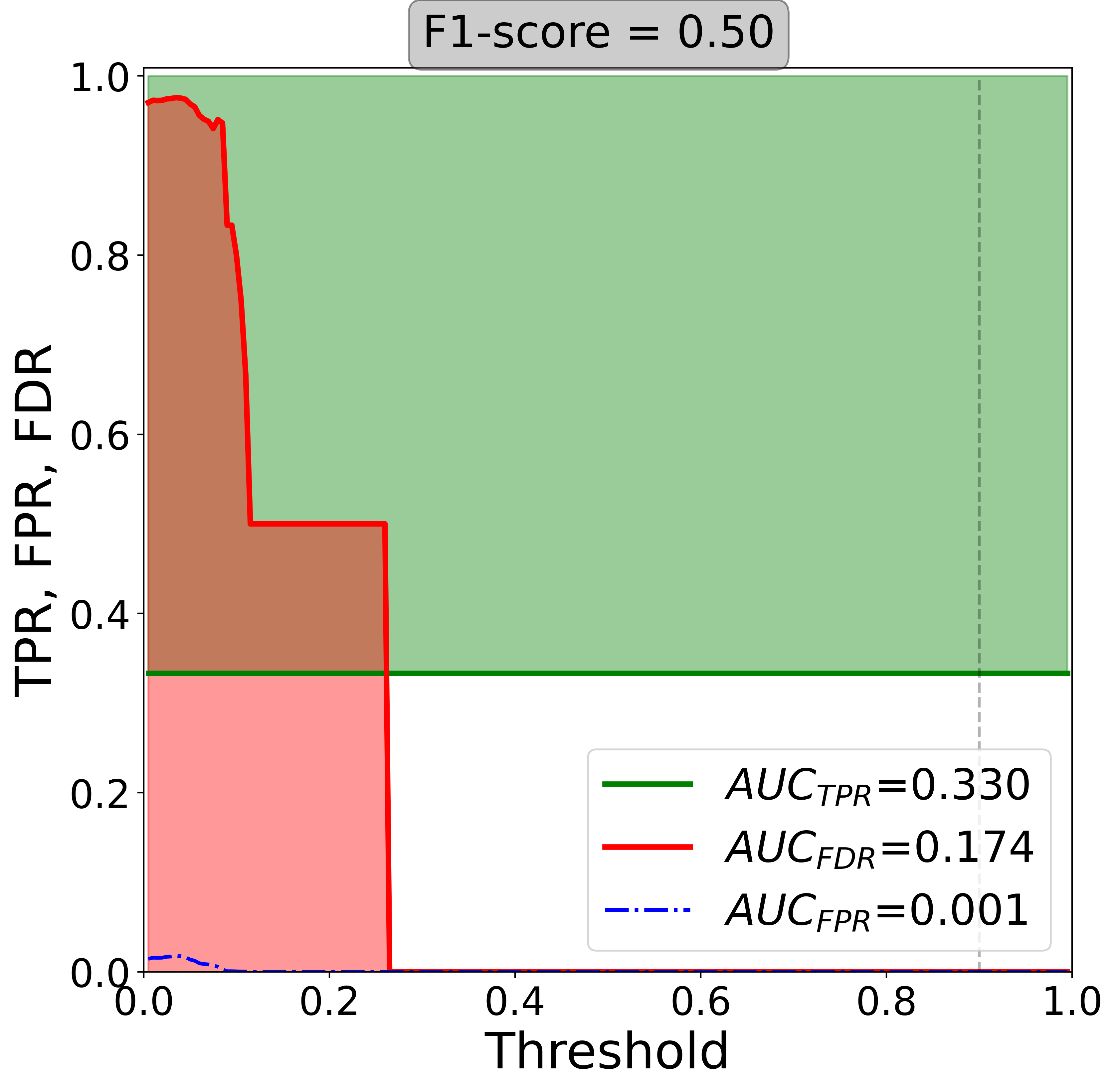} &
          \includegraphics[width=5.85cm]{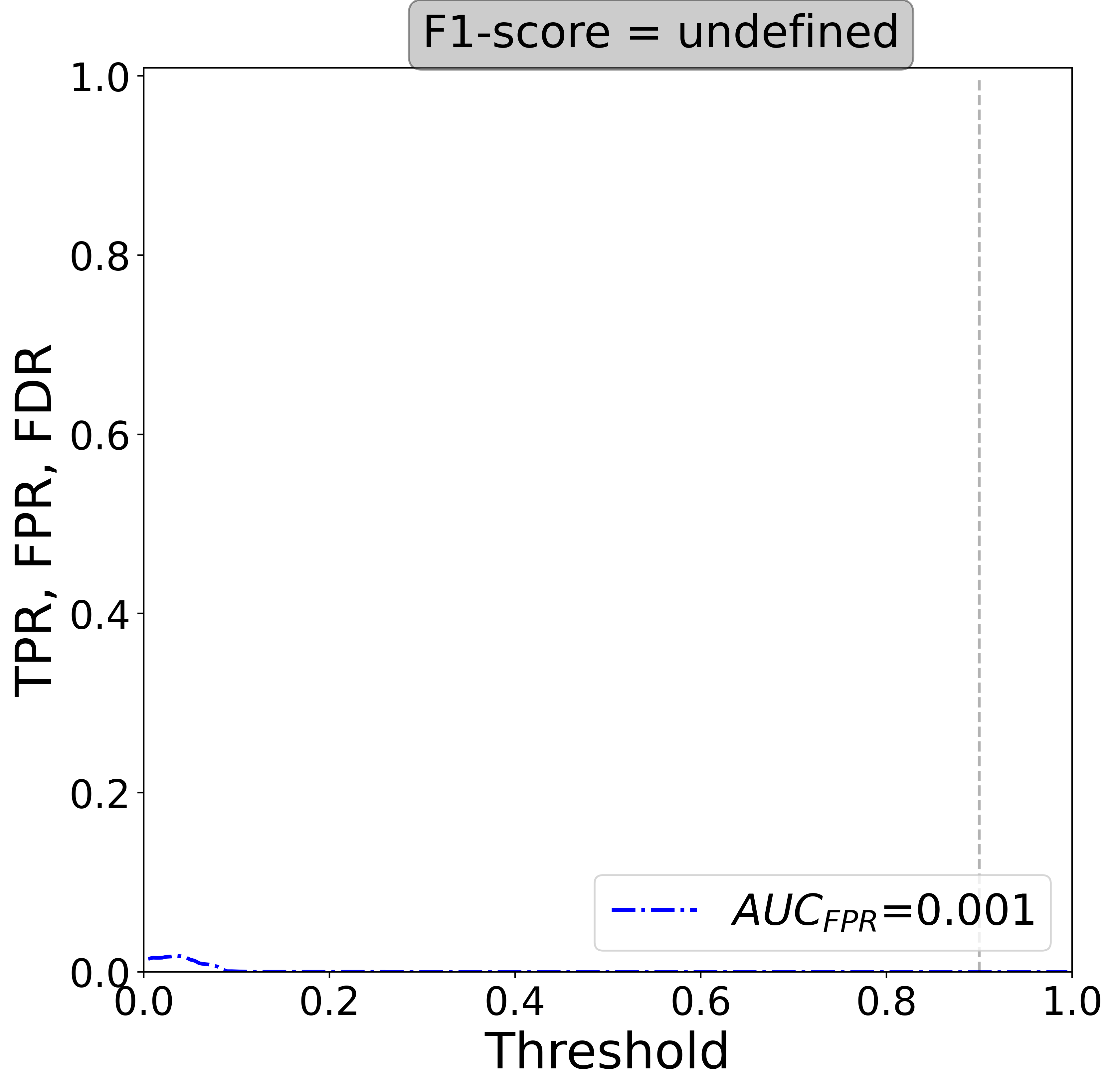} \\
          \textit{nrc1} & \textit{nrc2} & \textit{nrc3} \\[6pt]\\
          \includegraphics[width=5.85cm]{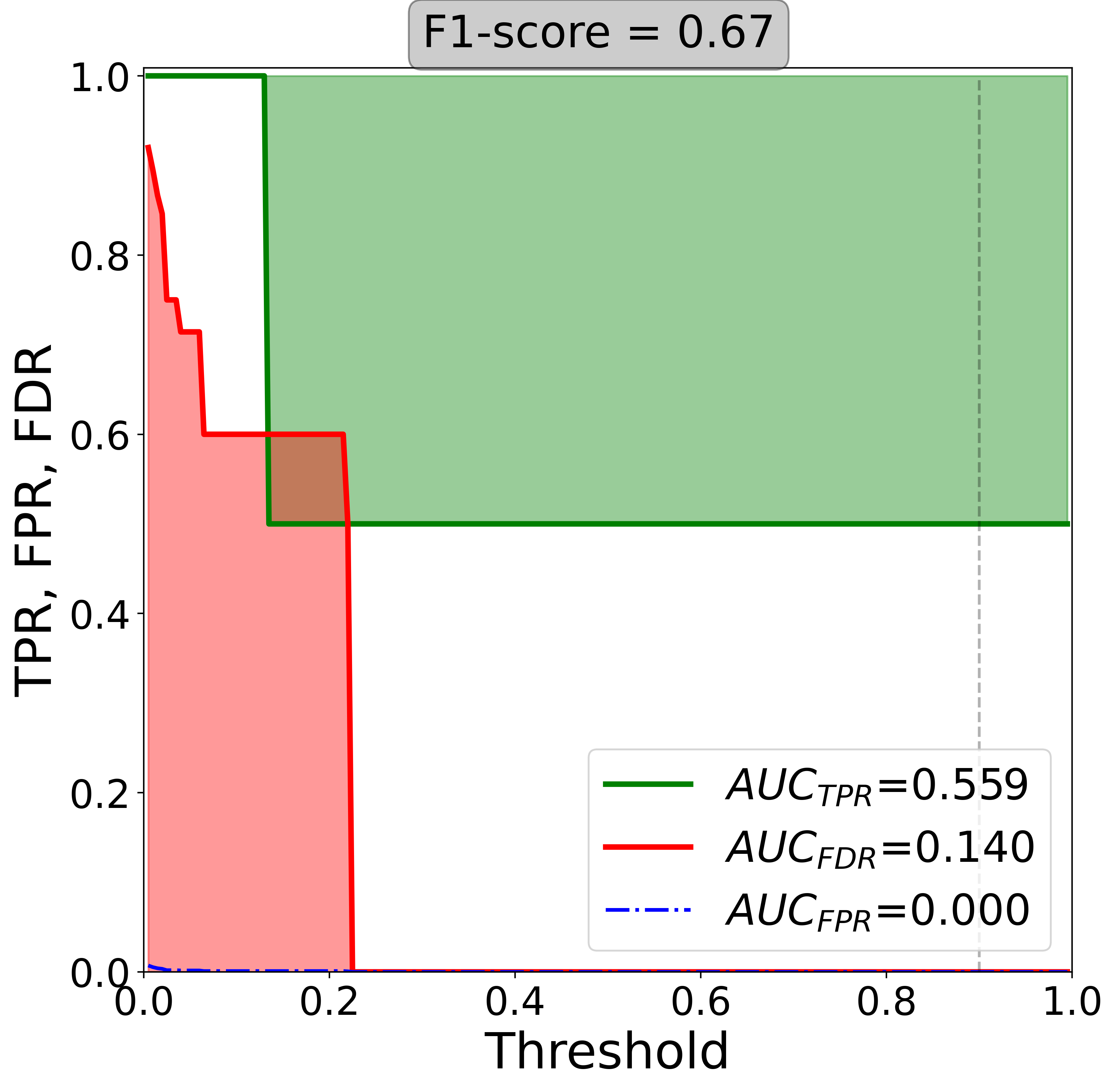} & 
          \includegraphics[width=5.85cm]{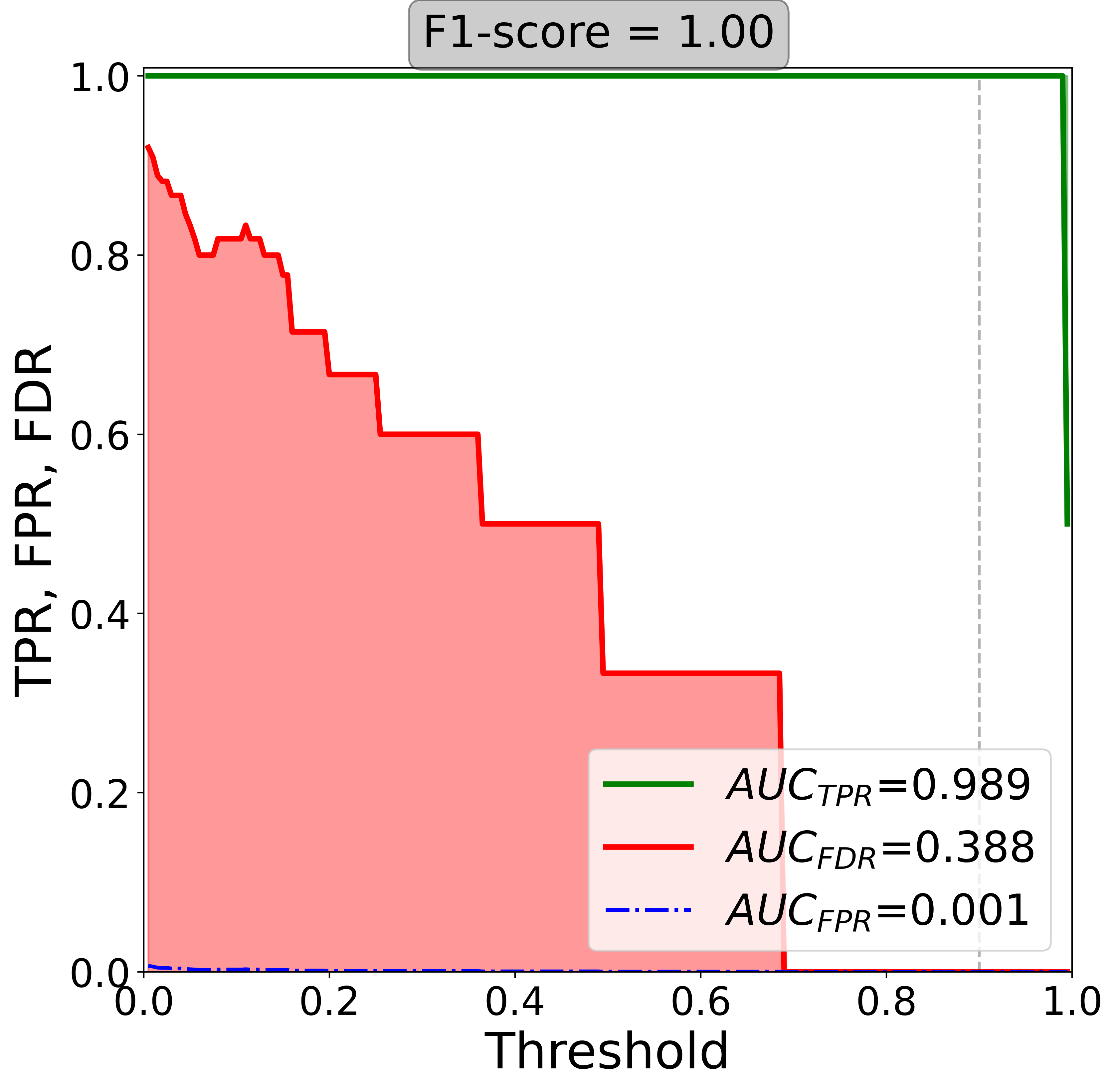} &
          \includegraphics[width=5.85cm]{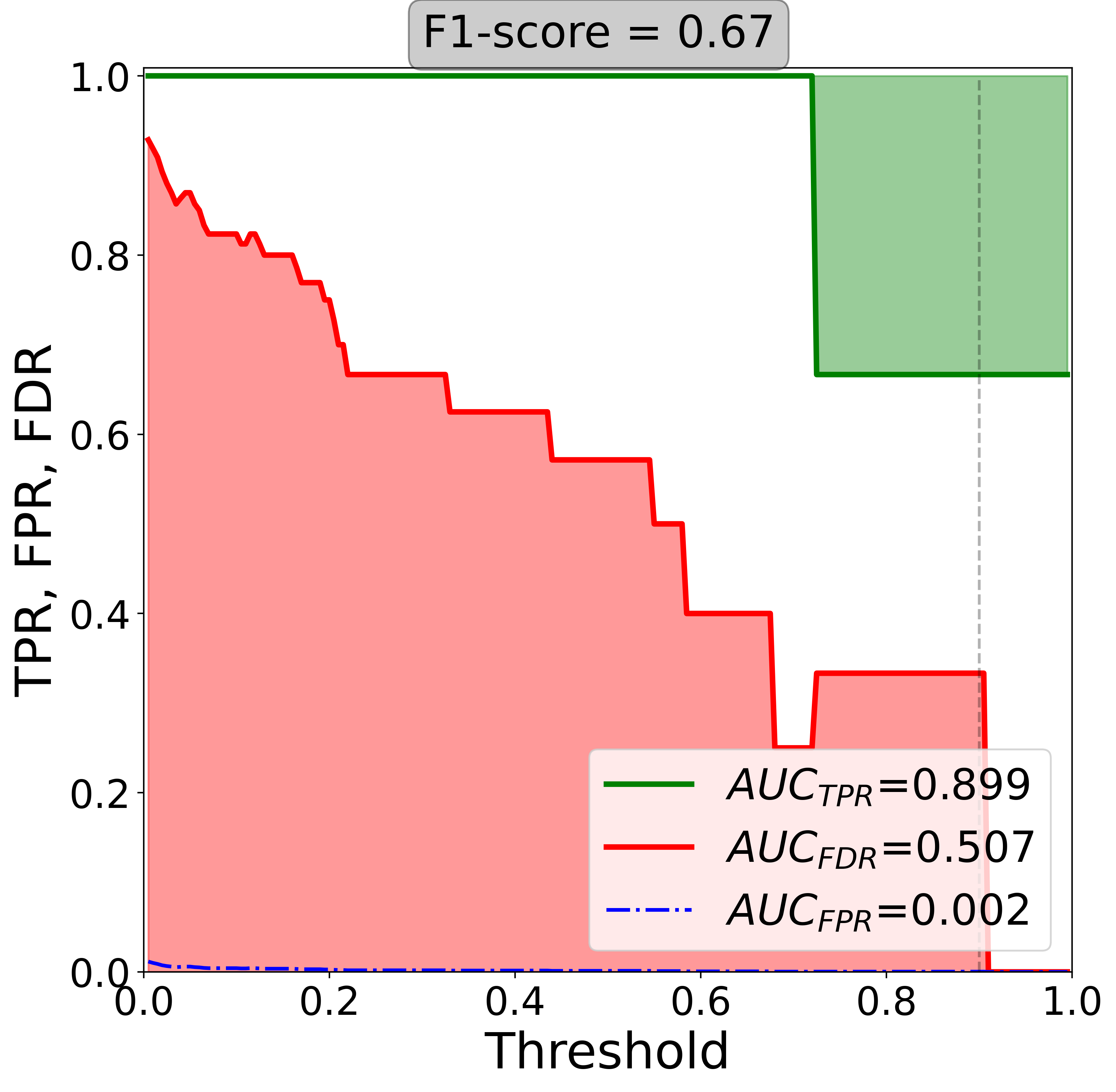} \\
          \textit{lmi1} & \textit{lmi2} & \textit{lmi3} \\[6pt]
        \end{tabular}
        \caption{TPR, FDR, and FPR metrics computed from the confidence maps of Fig.~\ref{fig:eidc_grid} for a range of confidence thresholds varying from zero to one. Their respective AUCs are shown in each legend. The F1-score is computed at the submitted threshold on the challenge $\tau_{sub}=0.9$ (vertical dashed line) and it is shown in the top of each subplot. When the dataset contains injections, TPR and FDR steply decrease with threshold, while FPR decreases monotonically. Thereby, an ideal algorithm would provide a TPR=1, FPR=0 and FDR=0 for any threshold and therefore, an $AUC_{TPR}=1$, $AUC_{FPR}=0$ and $AUC_{FDR}=0$. However, when the dataset does not have injections, the FPR is the only metric that can be defined as it does not depend on TPs.  \label{fig:eidc_grid_metrics}}
        \end{figure*}
    
    \end{appendix}

\end{document}